\documentclass[aps, prd, floats, eqsecnum, superscriptaddress, floatfix, onecolumn, nofootinbib,10pt]{revtex4-2}

\usepackage{graphicx}
\usepackage[format=plain, font=small, labelfont=bf, justification=RaggedRight]{caption}
\usepackage[format=plain, font=small, labelfont=bf, justification=RaggedRight]{subcaption}
\usepackage{xcolor}
\usepackage{latexsym}
\usepackage{slashed}
\usepackage{amssymb}
\usepackage{multirow}
\usepackage{array}
\usepackage{amsmath,amsthm}

\usepackage{lipsum}

\renewcommand{\thesection}{\arabic{section}}
\renewcommand{\thesubsection}{\thesection.\arabic{subsection}}

\makeatletter
\def\p@subsection{}
\def\p@subsubsection{}
\makeatother

\newcommand{\rd}{\mathrm{d}}	
\usepackage{mathrsfs}			

\newcommand{\beq}{\begin{equation}}
\newcommand{\eeq}{\end{equation}}
\newcommand{\bea}{\begin{eqnarray}}
\newcommand{\eea}{\end{eqnarray}}

\newcommand{\pa}{\partial}
\newcommand{\na}{\nabla}
\renewcommand{\ln}{\,\mbox{log}\,}

\renewcommand{\Im}{\,\mbox{Im}\,}

\newcommand{\eq}[1]{(\ref{#1})}
\newcommand{\n}[1]{\label{#1}}

\newcommand{\al}{\alpha}
\newcommand{\be}{\beta}
\newcommand\ga{\gamma}
\newcommand\Ga{\Gamma}

\newcommand\de{\delta}

\renewcommand\th{\theta}

\newcommand\la{\lambda}
\newcommand\La{\Lambda}

\newcommand\si{\sigma}

\newcommand\ph{\varphi}
\newcommand\om{\omega}

\newcommand\Om{\Omega}

\DeclareMathOperator{\cx}{\square}
\DeclareMathOperator{\tri}{\triangle}

\begin{document}

\title{Effective delta sources and Newtonian limit in nonlocal gravity}

\author{Thomas M. Sangy}
\email{thomas.sangy@estudante.ufjf.br}
\affiliation{
Departamento de F\'{\i}sica,  ICE, Universidade Federal de Juiz de Fora,
Juiz de Fora,  36036-900,  MG,  Brazil
}

\author{Nicol\`o Burzill\`a}
\email{burzilla@lnf.infn.it}
\affiliation{
INFN Laboratori Nazionali di Frascati (LNF), Frascati 00044, Roma, Italy
}
\affiliation{
INFN Sezione di Roma Tor Vergata, Roma 00133, Italy
}

\author{Breno L. Giacchini}
\email{breno.giacchini@matfyz.cuni.cz}
\affiliation{
Institute of Theoretical Physics, Faculty of Mathematics and Physics, Charles University, V Hole{\v s}ovi{\v c}k{\'a}ch 2, 180 00 Prague 8, Czech Republic
}

\author{Tib\'{e}rio de Paula Netto}
\email{tiberio.netto@ufjf.br}
\affiliation{
Departamento de F\'{\i}sica,  ICE, Universidade Federal de Juiz de Fora,
Juiz de Fora,  36036-900,  MG,  Brazil
}


\begin{abstract} \noindent
{\bf Abstract}. We investigate the Newtonian limit of a class of nonlocal gravity models with exponential form factors $f_s (\cx) = \exp [(-\cx/\mu_s^2)^{N_s}]$. Our main goal is to identify similarities and differences between models in this family in regard to weak-field solutions. To this end, we use the effective source formalism to compare the related effective delta sources, mass functions, and Newtonian potentials. We obtain a variety of representations for these quantities in terms of series, integrals, and special functions, as well as simple approximations that capture the relevant dependence on the parameters $N_s$ and $\mu_s$---which can be used to explore the weak-field phenomenology of nonlocal gravity. We explain why only for $N_s>1$ the Newtonian potential oscillates and prove that, despite the oscillations, the effective masses are positive. Moreover, we verify that these linearized solutions are regular (without curvature singularities). Finally, we also calculate the form of the leading logarithmic quantum correction to the Newtonian potential in these models. In all our considerations, we assume that $N_s$ is a positive real parameter. The cases of non-integer $N_s$ might be applied beyond nonlocal gravity, in effective approaches to implement quantum corrections in the weak field regime.
\end{abstract}

\maketitle
\noindent


\section{Introduction}
\label{Sec1}

General relativity (GR) accurately describes classical gravity phenomena in a wide range of energies and has been verified in numerous high-precision experiments. Among the most recent ones, we can mention gravitational wave signals from compact binary mergers discovered by the LIGO-Virgo-KAGRA collaboration~\cite{LIGO_VIRGO_KAGRA} and the observation of the Hellings--Downs signature in the stochastic gravitational wave background by pulsar timing array collaborations~\cite{nanograv}. However, despite the fact that GR continues to be one of the most rigorously tested theories in modern physics, there are problems that are still lacking satisfactory descriptions. For example, at galactic and cosmological scales the nature of dark matter and dark energy remains elusive, and the more recent data from the DESI galactic survey~\cite{DESI:2024mwx} suggest that the cosmological constant model ($\Lambda$CDM) appears to be disfavored compared to scenarios with a time-dependent barotropic factor. Such a dynamic dark-energy component would require an extension that goes beyond the simple $\Lambda$CDM framework. If confirmed, the DESI results could add further tension to the $\Lambda$CDM model, enhancing existing discrepancies such as the Hubble tension.

Moreover, in the ultraviolet (UV) regime, GR is plagued by spacetime singularities in important solutions, such as the homogeneous and isotropic, which models the zero-order approximation for the dynamics of our Universe, and inside black holes. This indicates that GR as a classical theory of gravity cannot be expected to remain valid on all scales such as in the vicinity of singularities, where the spacetime curvature inevitably reaches the Planck scale~$\,l_{\rm P}^{-2} \sim M_{\rm P}^2 \approx 10^{19}\, {\rm GeV}$ and the classical description of gravity is supposed to break down. 

From a quantum perspective, GR also suffers from a lack of predictability in the high-energy domain, due to its non-renormalizability~\cite{tHooft:1974toh,Deser:1974cz,Deser:1974xq,GoroffSagnotti}. A natural way to improve the UV behavior of the propagator and the convergence of loop integrals is to extend the Einstein--Hilbert action by including higher-derivative terms, such as $RF_0(\cx)R$ and $C_{\mu\nu\alpha\beta}F_2(\cx) C^{\mu\nu\alpha\beta}$, where $F_{0,2}(\cx)$ are analytic functions of the d'Alembertian $\cx = \nabla_\mu \nabla^\mu$ and $C_{\mu\nu\alpha\beta}$ denotes the Weyl tensor. If $F_{0,2}$ are simply constant parameters, one obtains the well-known fourth-derivative gravity, which is multiplicative renormalizable~\cite{Stelle77}. If, instead, $F_{0,2}(\cx)$ are taken to be non-trivial polynomials, the theory can be made superrenormalizable~\cite{AsoreyLopezShapiro}. Furthermore, even in semiclassical gravity, where gravity is a classical external background field, the renormalization of quantum matter fields in a curved spacetime requires the inclusion of terms quadratic in curvature to the gravitational action~\cite{UtDW} (for an introduction to the subject see, e.g.,~\cite{book,book2}). However, local higher-derivative models suffer from the drawback of introducing ghost-like massive poles in the propagator~\cite{Stelle77,AsoreyLopezShapiro}, which leads to violation of unitarity at the quantum level and might correspond to Ostrogradsky instabilities in the classical theory.

Over the past decades, several proposals have been put forward to reconcile renormalizability with unitarity (see,~e.g.,~\cite{Bender:2007wu,Bender:2008gh,Salvio:2015gsi,Anselmi:2017ygm,Donoghue:2019fcb,ModestoShapiro16,Modesto16}). In particular, the $S$-matrix remains unitary in local higher-derivative theories with actions with six or more derivatives of the metric and such that all the ghost-like poles appear in complex conjugate pairs~\cite{ModestoShapiro16,Modesto16}, provided the Lee--Wick quantization prescription is adopted~\cite{LW1,LW2,CLOP,Anselmi:2017yux,AnselmiPiva2}; for this reason these models are known as Lee--Wick gravity. Another approach consists of avoiding ghost-like degrees of freedom altogether by constructing actions that are non-polynomial in the field derivatives. Specifically, by introducing quadratic-curvature terms containing certain infinite-derivative operators, it is possible to prevent the appearance of extra pathological modes in the physical spectrum while still preserving an improved UV behavior of the propagator. This makes it possible to formulate theories of gravity that are ghost-free at tree-level and renormalizable~\cite{Krasnikov,Kuzmin,Tomboulis,Modesto12,Siegel2,Biswas,Shapiro:2015uxa}. The price to pay is the presence of non-polynomial differential operators, which render the gravitational action explicitly \emph{nonlocal} and the study of classical solutions particularly challenging.

In this work, we will consider the following nonlocal action
\begin{equation}
\n{action}
S = \frac{1}{16 \pi G} \int \rd^4 x \, \sqrt{-g} \left\{
(R - 2\Lambda) 
+ \frac{1}{2} C_{\mu\nu\alpha\beta} F_2(\cx) C^{\mu\nu\alpha\beta} - \frac{1}{6} R F_0(\cx) R 
\right\} + O(R^3)
,
\end{equation}
where $G$ is Newton's constant, $\Lambda$ is the cosmological constant, and $O(R^3)$ collectively denotes terms that are of order at least cubic in curvatures. Also,
\begin{equation}
F_s(\cx) = \frac{f_s(\cx) - 1}{\cx} 
, 
\qquad 
f_s(\cx) = e^{H_s(\cx)}
, 
\qquad s = 0, 2
\end{equation}
are nonlocal form factors with entire functions $H_s(z)$. The exponential operators in the gravitational action are introduced in such a way that the UV behavior of the tree-level (gauge-independent) part of the propagator,
\begin{equation} 
\n{propagator}
G_{\mu\nu,\alpha\beta}(k) = \frac{1}{k^2} \left[ \frac{P^{(2)}_{\mu\nu,\alpha\beta}}{f_2(-k^2)} - \frac{1}{2} \frac{P^{(0-s)}_{\mu\nu,\alpha\beta}}{f_0(-k^2)} \right]
,
\end{equation}
gets modified without introducing ghost degrees of freedom, due to the fact that $f_s(-k^2)$ has no zeros in the entire finite complex plane.\footnote{In \eq{propagator}, $P^{(2,0-s)}_{\mu\nu,\alpha\beta}$ are the usual Barnes--Rivers projectors~\cite{Barnes,Rivers}.} 

The problem of spacetime singularities in the linearized version of the class of theories described by \eqref{action} has been extensively studied in the last fifteen years.\footnote{In contrast, results at nonlinear level are very few---see, for instance,~\cite{Kolar:2023gqi} and references therein.} For instance, it was shown that in the case of the nonlocal model characterized by the exponential form factor
\begin{equation}
\n{exp_form}
f_s(\cx) = e^{ -\cx / \mu_s^2 }
,
\end{equation}
where $\mu_{0,2}$ are massive parameters defining the nonlocality scale~$l_s=1/\mu_{s}$, the linearized Newtonian-limit solution is regular \cite{Tseytlin,Modesto12,Siegel1,Biswas}. In addition, all the linearized curvature invariants that are polynomial in the Riemann tensor, its derivatives, and contractions are finite. This is a consequence of the fact that $f_s(-k^2)=\exp{(k^2/\mu_s^2)}$ grows faster than any polynomial for large values of~$k^2$~\cite{Nos6der}. This observation is a particular example of a more general result on the regularity of Newtonian-limit solutions in higher-derivative local and nonlocal models (see, e.g.,~\cite{BreTibLiv} for a review). Moreover, in the case of the form factor~\eqref{exp_form}, the theory admits nonsingular bouncing cosmological solutions~\cite{Siegel2}, thereby providing a possible resolution of the Big Bang singularity (see also~\cite{Biswas_Bounce}).

In the present work, we focus on the more general family of nonlocal models with the form factor
\begin{equation}
\n{formfactor}
f_s(\cx) = e^{\left( -\cx / \mu_s^2 \right)^{N_s}}
,
\end{equation}
which can be regarded as a generalization of~\eq{exp_form} and of the Krasnikov form factor ($N_s=2$)~\cite{Krasnikov}. Models of this type are studied most often in the particular case $N_0=N_2=N$ and $\mu_0=\mu_2=\mu$, sometimes known as GF$_N$ (for ghost-free theory of type $N$)~\cite{Frolov:2015usa}. The Newtonian-limit solutions in GF$_N$ models were also studied in~\cite{Edholm_NewPot}, where an expression for the Newtonian potential was derived in the form of a power series for an arbitrary integer $N$, and the numerical solution revealed an oscillatory behavior of the potential for $N>1$ (see also~\cite{Perivolaropoulos:2016ucs,Edholm:2017dal,Boos:2018bhd}). A numerical approximation for the potential in the case of large $N$ was obtained in~\cite{Perivolaropoulos:2016ucs}, and in~\cite{Frolov:2015usa} it was shown that the potential and the Newtonian-limit metric are regular at $r=0$ (see also~\cite{Edholm_NewPot,BreTib2}). Nevertheless, it is difficult to perform a detailed analysis of the physical effects of the value of $N$ (or, more generally, $N_0$ and $N_2$) on the solutions.

Here, we revisit the Newtonian limit in models with the form factor~\eq{formfactor} with the goal of showing alternative representations for the solutions, a new approximation for the case of large $N_s$, and related quantities that can be useful for understanding their physical aspects. All this can be done following more recent results on the description of the Newtonian limit of nonlocal gravity in terms of effective sources~\cite{BreTib2,Nos6der,BreTibLiv} and the developments of~\cite{Barvinsky:2019spa} regarding various functions associated with the form factor~\eq{formfactor}. Moreover, we also consider the extension of~\eq{formfactor} to noninteger values of $N_s$; the motivations for this are threefold: First, it corresponds to cases of fractional gravity, which has attracted some attention in recent years (see, e.g.,~\cite{Calcagni:2021aap,Giusti:2020rul}). Second, many of the results derived here using the effective source formalism can be directly applied to scenarios with a minimal length of interaction, like in noncommutative geometry, where non-integer values of $N_s$ are often considered. And finally, from the formal point of view, it is a valid mathematical exercise. 

Last but not least, we consider the effect of the logarithmic one-loop quantum corrections to the Newtonian-limit solutions in these models. This analysis is also done in full generality, and reveals the differences between the classical and one-loop corrected effective sources, masses, and gravitational potentials.

The paper is organized as follows, in Sec.~\ref{Sec2} we consider the Newtonian, weak-field limit to obtain various representations for the effective source, the mass function and the Newtonian potential for any $N_s >0$. In particular, we prove that the source is monotonic if and only if $0<N_s\leqslant 1$, and that the mass function is always positive, for any $N_s>0$. In Sec.~\ref{Sec6} we consider the limit $N_s\to \infty$ of the solutions, which yields an analytic and simple approximation of the potential in the large-$N_s$ regime. In Sec.~\ref{Sec7} we briefly comment on the regularity of curvature invariants. Then, in Sec.~\ref{Sec8} we calculate analytic expressions for the one-loop 1PI quantum correction for the source, mass, and potentials. Finally, in Sec.~\ref{Sec9} we summarize the results and draw our conclusions.\footnote{Throughout the paper we adopt the mostly positive convention for the metric signature $(-,+,+,+)$, with the Ricci tensor defined as $R_{\mu \nu}=R^{\alpha}{}_{\mu \alpha \nu }$, where $R^{\al}{}_{\beta\mu\nu}=\partial_{\mu}\Gamma^{\al}_{\beta \nu}+ \ldots$, and the natural units system $c = \hbar = 1$.}


\section{Newtonian limit}
\label{Sec2}

In the Newtonian limit, we assume the weak-field approximation and consider linear metric fluctuations around the Minkowski spacetime,
\begin{equation}
g_{\mu\nu} = \eta_{\mu\nu} + h_{\mu\nu}
,
\qquad \qquad
| h_{\mu\nu} | \ll 1
.
\end{equation}
Also, matter is assumed to be nonrelativistic, described by the energy-momentum tensor 
\begin{equation}
T_{\mu\nu} = \de^0_\mu \de^0_\nu \rho
,
\end{equation}
where $\rho$ is the mass density. To obtain the linearized field equations, one only needs to consider the terms in the action that are of second order in the perturbation $h_{\mu\nu}$. Hence, the terms $O(R^3)$ are irrelevant in this limit. The bilinear part of the action~\eq{action} takes the form\footnote{We also neglect the cosmological constant $\La$, which is irrelevant in this limit.}
\begin{equation} 
\begin{split}
\n{bili}
S^{(2)} = \frac{1}{32\pi G} & \int  \rd^4 x 
\left\{  
h_{\mu \nu} f_2 ( \cx ) \cx h^{\mu \nu} 
- \frac12 \, h \left[ \frac{f_2 ( \cx ) + 2 f_0 ( \cx )}{3} \right] \cx h 
- h^{\mu\nu} f_2 ( \cx ) \pa_\mu \pa_\la h^\la_\nu 
\right.
\\
&
\left.
+ h^{\mu\nu} \left[ \frac{f_2 ( \cx ) + 2 f_0 ( \cx )}{3} \right]  \pa_\mu \pa_\nu h
+ h^{\mu\nu} \left[ \frac{f_2 ( \cx ) -  f_0 ( \cx )}{3} \right] \frac{\pa_\mu \pa_\nu \pa_\al\pa_\be}{\cx} \, h^{\al\be} 
\right\}
.
\end{split}
\end{equation}
The coupling with matter is introduced through the action 
\begin{equation}
S_{\rm m} =  \frac12 \int \rd^4 x \,\, T^{\mu\nu}  h_{\mu\nu}
,
\end{equation}
where $T_{\mu\nu}$ is the matter energy-momentum tensor. Thus, the principle of least action gives 
\begin{equation} 
\begin{split}
\n{EOM-HD}
\varepsilon^{\mu\nu} \equiv & \,\,  f_2 (\cx) \big(  \cx h^{\mu\nu} - \pa^\mu \pa^\la h^\nu_\la - \pa^\nu \pa^\la h^\mu_\la \big)  
+ \left[ \frac{f_2 ( \cx ) + 2 f_0 ( \cx )}{3} \right]   \left[   \eta^{\mu\nu} \left( \pa^\al \pa^\be h_{\al\be} - \cx h \right)  
+ \pa^\mu \pa^\nu h \right]
\\
&
+ 2 \left[ \frac{f_2 ( \cx ) -  f_0 ( \cx )}{3} \right]  \, \frac{\pa^\mu \pa^\nu \pa^\al \pa^\be}{\cx} \, h_{\al\be}
= - 16 \pi G \, T^{\mu\nu}
.
\end{split}
\end{equation}

Since the principle of superposition holds true in the Newtonian limit, we shall focus on the field generated by a point-like massive source sitting at $r=0$,
\begin{equation}
\rho = M \de(\vec{r})
,
\end{equation}
from which more complicated solutions can be constructed (see, e.g.~\cite{Frolov:Exp,Frolov:Poly,BreTib1,Buoninfante:2022ild}, for some examples). Therefore, we assume a metric ansatz that is static, spherically symmetric, and in isotropic form, 
\begin{equation}
\n{metric-new}
\rd s^2 = -[1 + 2 \ph(r)] \rd t^2 + [1 - 2 \psi(r)] (\rd r^2 + r^2 \rd \Om^2)
,
\qquad \qquad |\ph|,|\psi| \ll 1
,
\end{equation}
where $\rd\Om^2$ is the metric of the unit sphere and $\ph(r)$, $\psi(r)$ are the Newtonian potentials. Substituting~\eq{metric-new} in the field equations~\eq{EOM-HD} and evaluating $\varepsilon^{00}$ and $\varepsilon^\mu{}_\mu$, it follows that the potentials must satisfy
\begin{subequations}
\n{system}
\begin{eqnarray}
\n{eq-ph}
f_2 (\tri) \tri (\ph + \psi) + f_0 (\tri) \tri (2 \psi - \ph) & = & 12 \pi G \rho
,
\\
\n{eq-psi}
f_0 (\tri) \tri (2 \psi - \ph )&  =  &4 \pi G \rho 
.
\end{eqnarray}
\end{subequations}
The above system of differential equations can be decoupled through the introduction of the spin-$s$ potentials~\cite{BreTib1}
\begin{equation}
\chi_2  = \frac{\ph  + \psi }{2}
\qquad
\text{and}
\qquad
\chi_0  = 2 \psi  - \ph
,
\end{equation}
which are defined such that the system~\eq{system} becomes
\begin{equation}
\n{poi-chi-mod}
f_s (\tri) \tri \chi_s = 4 \pi G \rho
, 
\qquad \qquad
s = 0,2
.
\end{equation} 
In other words, they separate the contributions of the spin-0 and spin-2 sectors of the propagator~\eq{propagator} to the solution.
Actually, Eq.~\eq{poi-chi-mod} completely determine the Newtonian-limit solution with a nontrivial source $\rho$.\footnote{Indeed, it can be shown that the solutions of~\eq{poi-chi-mod} also satisfy the other components of the field equations if $\partial_r^2 \left[ f_2(\tri) \chi_2 - f_0(\tri) \chi_0 \right] = 0$. The Newtonian-limit solutions satisfy this condition: Combining $\tri ( - 1/r) = 4 \pi \delta (\vec{r})$ and~\eq{poi-chi-mod} it follows that $f_0(\tri) \chi_0 = f_2(\tri) \chi_2 = - G M/r$ (modulo an irrelevant constant). This fact can be interpreted as if the requirement of matching the singularity of the delta function as $r\to 0$ fixes the $1/r$ (Newtonian) contribution to the potential in the same way in both spin-0 and spin-2 sectors.}

In the literature, Newtonian potentials are typically evaluated in the particular case $f_2 (z) = f_0 (z) \equiv f(z)$ (see, e.g., \cite{Boos:2018bhd,Frolov:2015usa,Biswas,Edholm_NewPot,Modesto12}), which has the advantage of simplifying the coupled system of differential equations~\eq{system} to the generalized Poisson equation $f(\tri) \tri \ph = f(\tri) \tri \psi =4 \pi \rho$. However, by working with the spin-$s$ potentials, one can also obtain a generalized Poisson equation without making any extra restriction in the theory under consideration. Once the solutions for $\chi_s$ are obtained, the original potentials $\ph$ and $\psi$ can be recovered through the inverse transformation
\begin{equation}
\n{pch}
\ph =  \frac43 \chi_2 - \frac13 \chi_0
, 
\qquad  \qquad
\psi = \frac23 \chi_2 + \frac13 \chi_0
.  
\end{equation}
Thus, we can work with the spin-$s$ potentials without loss of generality. For this reason, in what follows we simply call $\chi_s(r)$ the Newtonian potential.

The Eq.~\eq{poi-chi-mod} can be rewritten in an equivalent way by inverting the operator $f_s (\tri)$, namely,
\begin{equation}
\n{poi-chi-eff}
\tri \chi_s = 4 \pi G \rho_s
,
\end{equation}
where $\rho_s$ is an effective (smeared) delta source, defined such that $\rho = f_s (\tri) \rho_s$.\footnote{The conditions that the form factor $f_s (z)$ must satisfy for the existence of the effective source $\rho_s$ are discussed in detail in~\cite{BreTibLiv}; such requirements hold for \eq{formfactor}.} Using the Fourier representation of the delta function, the effective delta source can be expressed as
\begin{equation}
\n{eff-sour-3D}
\rho_s  = M \int \frac{\rd^3 k}{(2\pi)^3} \, \frac{e^{i \vec{k} \cdot \vec{r}}}{f_s(-k^2)}
,
\end{equation}
where $k = |\vec{k}|$. Integration over the angular coordinates yields
\begin{equation}
\n{eff-sour}
\rho_s (r) = \frac{M}{2 \pi^2} \int_0^\infty \rd k \, \frac{k \sin(kr)}{r f_s(-k^2)}
.
\end{equation}

In terms of the mass function
\begin{equation}
\n{massfunction}
M_s(r) = 4\pi \int_0^r \rd r' \, r'^{2} \rho_s (r') 
,
\end{equation}
the solution to Eq.~\eq{poi-chi-eff} can be written as
\begin{equation}
\n{chi_intg}
\chi_s (r) = - \int_\infty^r \rd r' \, g_s (r') 
,
\end{equation}
where 
\begin{equation}
\n{gs_def}
g_s (r) = - \frac{G \, M_s(r)}{r^2}
\end{equation}
is the spin-$s$ contribution to the gravitational field.


\subsection{Effective delta source}
\label{Sec3}

For the form factor~\eq{formfactor} the effective delta source~\eq{eff-sour} can be written as
\begin{equation}
\n{sour-I}
\rho_s (r) =  
\frac{M \mu^3_s}{2 \pi^2} I_{N_s} (\mu_s r) 
,
\end{equation}
where
\begin{equation}
\n{basic-I}
I_N (r) = \int_0^\infty \rd k \, e^{-k^{2N}} \frac{k \sin(k r)}{r}
\end{equation}
is the basic integral to be evaluated. 

There are two cases that are worth mentioning, for which solutions in closed form are known; namely, for $N_s=1$ the effective source has a Gaussian profile,
\begin{equation}
\n{source-gaussian}
\rho_s(r) = \frac{M \mu^3_s}{8 \pi ^{3/2}} e^{-\frac{\mu^2_s r^2}{4} }
,
\end{equation}
while for $N_s = 1/2$,
\begin{equation}
\n{source-nonC}
\rho_s (r) = \frac{M \mu^3_s}{\pi^2 } \, \frac{1}{\left(1+ \mu_s^2 r^2\right)^2}
.
\end{equation}
Both examples of effective sources are commonly used to model noncommutative effects of the spacetime (see, e.g.,~\cite{Nicolini:2005vd,Nozari:2008rc,Liang:2012vx,Kuhfittig:2014iwa}), with the correspondence $\mu_s \to 1/\sqrt{\th}$ between the parameter of nonlocality and the noncommutative parameter $\th$. In fact, the considerations presented here have applications beyond nonlocal gravity, and they can be used to study new effects from noncommutative geometry and models with minimal length.  

Now, consider~\eq{basic-I} with an arbitrary $N$. Applying the change of integration variable $k = t^{1/(2N)}$, we obtain 
\begin{equation}
\n{IN-int}
I_N(r) = \frac{1}{2 N r} 
\int_0^\infty \rd t  \, e^{-t} \, t^{\frac{1}{N}-1}\,  \sin ( t^{\frac{1}{2N}} r ) 
.
\end{equation}
For $N > 1/2$ this integral admits a representation in power series. Indeed, using the Taylor series for the sine function, it follows
\begin{equation}
\n{IN-series}
I_N(r) = \frac{1}{2 N} \sum_{\ell=0}^\infty \frac{(-1)^\ell}{(2\ell + 1)!} \,  r^{2 \ell}
\int_0^\infty \rd t  \, e^{-t} \,  t^{\frac{2\ell+3}{2N} -1} 
,
\end{equation}
since the integration and the summation commute if $N> 1/2$. The above integral can be evaluated by means of the Gamma function, $\Ga (z) = \int_0^\infty \rd t \,  e^{-t} t^{z-1}$. Consequently,
\begin{equation}
\n{I-series}
I_N(r) = \frac{1}{2 N} \sum_{\ell=0}^\infty \frac{(-1)^\ell}{(2\ell + 1)!} \, 
\Ga \left( \frac{2\ell+3}{2N} \right) r^{2 \ell}
.
\end{equation}

The infinite sum~\eq{I-series} can be rewritten in terms of many different special functions. For example, it has been shown that for $N_s \in \lbrace 2, 3 \rbrace$ the effective source is equivalent to a finite sum of generalized hypergeometric functions~\cite{Boos:2018bhd}, and it can also be expressed in terms of a single Meijer $G$-function for $N_s \in \mathbb{N}$. We leave the details of such cumbersome representations for Appendices~\ref{ApA} and~\ref{ApB}, respectively. Here, instead, we express it in a more compact and useful form. To this end, we shall use the generalized exponential function (GEF) 
\begin{equation}   
\n{GEF}
\mathcal{E}_{\nu, \alpha} (z) = \frac{1}{\nu} \sum_{\ell=0}^{\infty} \frac{\Gamma \left( \frac{\ell + \alpha}{\nu} \right)}{\Gamma ( \ell + \alpha )} \frac{z^{\ell}}{\ell!}
,
\end{equation}
as defined in~\cite{Barvinsky:2019spa}. Note that ${\cal E}_{1,\al} (z) = \exp(z)$. For $\nu > 1/2$, the power series~\eq{GEF} is absolutely convergent for all $z \in \mathbb{C}$, whereas if $\nu < \frac{1}{2}$, it diverges for all $\mathbb{C} \backslash \left\{ 0 \right\}$ \cite{Barvinsky:2019spa}. Moreover, if $\nu = \frac{1}{2}$, the series~\eq{GEF} converges for all $z \in \mathbb{C}$ such that $\left| z \right| < \frac{1}{4}$, and it can be summed analytically \cite{Barvinsky:2019spa}:
\begin{equation} 
\n{gef1/2}
\mathcal{E}_{\frac{1}{2}, \alpha} ( z ) = \frac{4^\alpha}{\sqrt{\pi}} \sum\limits_{m=0}^\infty \Gamma\left(\alpha+\frac{1}{2}+m\right) \frac{(4z)^m}{m!} = \frac{4^{\alpha} \Gamma \left( \alpha + \frac{1}{2} \right)}{\sqrt{\pi}} ( 1 - 4 z )^{- \alpha - \frac{1}{2}}
.
\end{equation}

Comparing~\eq{GEF} with~\eq{I-series}, it is direct to see that
\begin{equation}
\n{I-GEF}
I_N(r) = \frac{\sqrt{\pi}}{4} \, {\cal E}_{N,\frac32} \left(- \frac{r^2}{4} \right)
.
\end{equation}
This relation is also valid for the critical value $N=1/2$, as it can be explicitly verified by comparing Eqs.~\eq{sour-I},~\eq{source-nonC} and~\eq{gef1/2} (with $\alpha = 3/2$ and $z = -r^2/4$). In particular,  $\mathcal{E}_{\frac{1}{2}, \frac{3}{2}} ( z )$ is well defined for $z<0$, which is the relevant domain here. For the sake of completeness, in Appendix~\ref{ApC} we present a detailed analysis of the convergence of the series~\eq{I-series}.

More generally, the solution~\eq{I-GEF} in terms of the GEF holds even for $0 < N < 1/2$; the main difference is that in this case the GEF is non-analytic, being defined through its integral representation~\cite{Barvinsky:2019spa}
\begin{equation} 
\n{GEF-int}
{\cal E}_{\nu,\alpha}(z) = \frac{1}{\nu} \int\limits_0^\infty \rd \mu\, \mu^{\alpha/\nu-1} e^{-\mu}\, {\cal C}_{\alpha-1}(z\mu^{1/\nu})
,
\end{equation}
where $\mathcal{C}_{n} ( z )$ is the Bessel--Clifford function. Indeed, for $\alpha = 3/2$ and $z = -r^2/4$, using the identities
\begin{equation} 
\n{IC}
{\cal C}_n (z) = z^{- \frac{n}{2}} {\cal I}_n (2 \sqrt{z})
, 
\qquad 
{\cal I}_{1/2} (z) = \sqrt{\frac{2}{\pi z} } \sinh z
,
\end{equation}
where ${\cal I}_n (z)$ is the modified Bessel function of the first kind, we get 
\begin{equation}
\n{GEF-32-int}
{\cal E}_{N,\frac32} \left( - \frac{r^2}{4} \right) = \frac{2}{\sqrt{\pi} N r} 
\int_0^\infty \rd t  \, e^{-t} \, t^{\frac{1}{N}-1}\,  \sin ( t^{\frac{1}{2N}} r ) 
.
\end{equation}
Then, the comparison of the above equation with~\eq{IN-int} shows that Eq.~\eq{I-GEF} its true even for $0< N < 1/2$.  

Therefore, from~\eq{sour-I} it is direct to see that
\begin{equation}
\n{sour-GEF}
\rho_s (r) = \frac{M \mu^3_s}{8 \pi^{3/2}} \, \mathcal{E}_{N_s, \frac{3}{2}} \left(-\frac{\mu^2_s r^2}{4} \right)
.
\end{equation}
That is, for all $N_s > 0$ the effective source is given by a ``generalized Gaussian function", which also provides the extension of the source for non-integer values of $N_s$.\footnote{In principle, models with $N <0$ can also be studied in the effective source formalism starting with Eq.~\eq{basic-I}. However, in this case the form factor~\eq{formfactor} goes to a constant in the UV and vanishes in the IR, corresponding to IR modifications of GR rather than a UV completion. Therefore, the effective source, mass function, and potentials possibly have a very different qualitative behavior in comparison to the models with $N>0$. We shall not discuss the case $N<0$ here, with the exception of some comments in Sec.~\ref{Sec7}.} 

For $N_s \geqslant 1/2$ the effective source has the power series representation 
\begin{equation}
\n{sour-series}
\rho_s (r) = 
\frac{M \mu_s^3}{4 \pi^2 N_s} \sum_{\ell=0}^\infty \frac{(-1)^\ell}{(2\ell + 1)!} \, 
\Ga \left( \frac{2\ell+3}{2N_s} \right) (\mu_s r)^{2 \ell}
.
\end{equation}
It is also worth recalling that for $\nu > 1/2$ the GEF is a particular case of the Fox--Wright $\Psi$-function~\cite{Barvinsky:2019spa}
\begin{equation}
\begin{split}
\n{GEF-Psi}
\mathcal{E}_{\nu, \alpha} ( z ) = \frac{1}{\nu} {}_{1} \Psi_{1} \biggl[
\left( \frac{\alpha}{\nu}, \frac{1}{\nu} \right) ; 
( \alpha, 1 );
z
\biggr]
,
\end{split}
\end{equation}
which can be expressed as a Fox $H$-function \cite{mathai},
\begin{equation}
\begin{split}
\mathcal{E}_{\nu, \alpha} ( z ) = \frac{1}{\nu} H^{1,1}_{1,2} \biggl[
\begin{array}{c}
\left( 1- \frac{\alpha}{\nu}, \frac{1}{\nu} \right) \\
(0,1), ( 1-\alpha, 1 )
\end{array}
\bigg \vert
-z
\biggr]
.
\end{split}
\end{equation}
Therefore, the effective source can also be expressed for $N_s > 1/2$ as
\begin{equation}
\n{source-H}
\rho_s (r) = \frac{M \mu^3_s}{8 \pi^{3/2} N_s}  H^{1,1}_{1,2} \Biggl[
\begin{array}{c}
\left(\frac{2N_s-3}{2N_s}, \frac{1}{N_s} \right) \\
(0,1), \left( -\frac{1}{2}, 1 \right)
\end{array}
\bigg \vert
\frac{\mu^2_s r^2}{4}
\Biggr]
.
\end{equation}

\begin{figure}[t]
\centering
\begin{subfigure}{.5\textwidth}
\centering
\includegraphics[width=7.28cm]{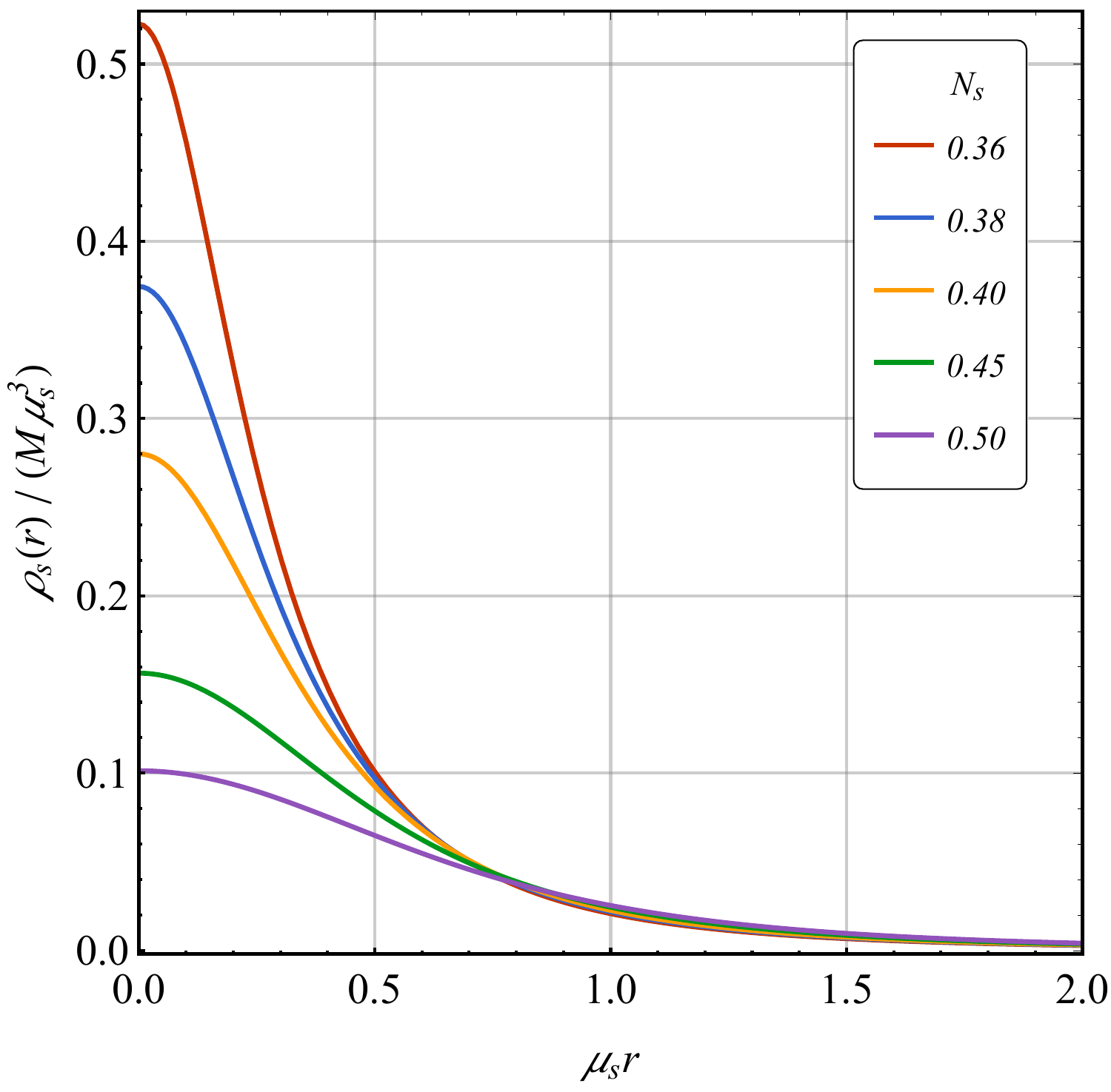}
\end{subfigure}%
\begin{subfigure}{.5\textwidth}
\centering
\includegraphics[width=7.5cm]{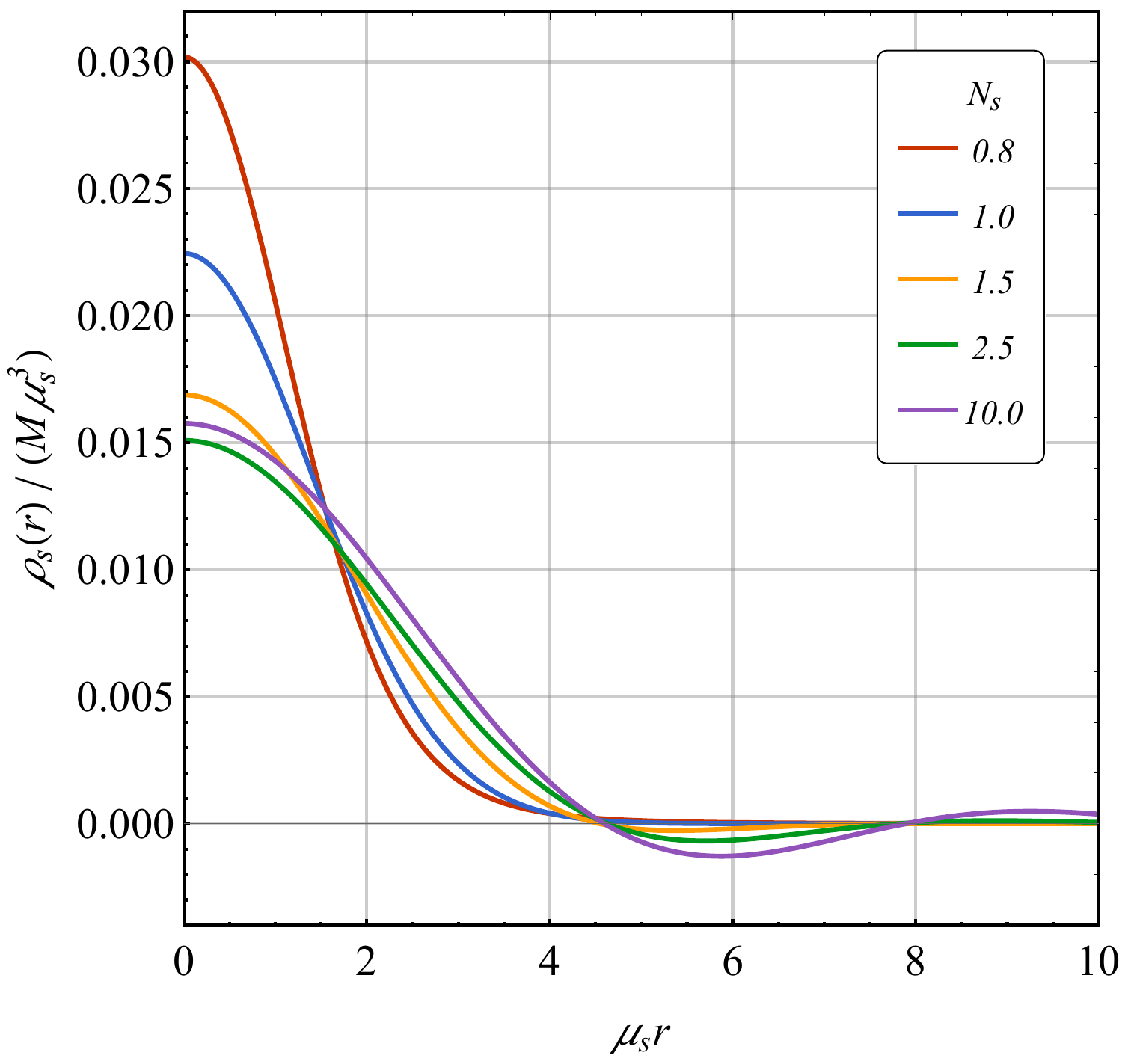}
\end{subfigure}
\caption{Graph of the source $\rho_s(r)/(M\mu_s^3)$ as a function of $\mu_s r$ for different values of $N_s$. The oscillations are present only for $N_s > 1$. The value of $\rho_s (0)$ grows very fast as $N_s \to 0$.}
\label{Fig1}
\end{figure}

In Fig.~\ref{Fig1} we plot $\rho_s(r)$ for some values of $N_s$, to illustrate some important features of the effective delta sources. First, note that for all $N_s > 0$,
\begin{equation}
\n{rho_0}
\lim_{r \to 0} \rho_s (r) = {\rm max} \, [\rho_s (r)] = \frac{M \mu_s^3}{4 \pi ^2 N_s} \, \Gamma \left(\frac{3}{2 N_s}\right)
\qquad 
{\rm and}
\qquad 
\lim_{r \to \infty} \rho_s (r) = 0
.
\end{equation}
As proved in~\cite{BreTib2} (see also~\cite{BreTibLiv}), the effective source achieves its maximum at $r=0$ if the form factor is such that $f_s(-k^2) \sim k^4$ (or faster) for $k$ large enough; the specific value at this maximum (first equality) follows from~\eq{sour-GEF}. On the other hand, the second property follows from the fact that the form factor is analytic and $f_s(0) = 1$. Note that ${\rm max} \, [\rho_s (r)]\to\infty$ in the limit $N_s\to0$, as in this case $f_s\to e$ and the source tends to a delta.\footnote{This limit corresponds to the form factor $F_s(\Box)=\tfrac{e-1}{\Box}$, which provides an IR modification of GR, as considered in~\cite{Gorbar:2002pw,Deser:2007jk,Ferreira:2013tqn}.}

Second, for $N_s > 1$ the effective delta source possesses spatial oscillations and can assume negative values. These oscillations are similar to the ones present in Lee--Wick theories of gravity~\cite{Accioly:2016qeb,Newton-BLG,Burzilla:2023xdd} and cause the gravitational potential also to oscillate, which may have interesting applications, as discussed in~\cite{Accioly:2016qeb,Accioly:2016etf,BreTib1,Boos:2018bhd,BreLuc,Krishak:2020opb,Antoniou:2017mhs,Perivolaropoulos:2016ucs}. From a mathematical perspective, the absence of oscillations in the effective delta source for $0 < N_s \leqslant 1$ can be explained by the fact that $\rho_s$ is given in terms of the Fourier transform $\ga_q$ of the function $\exp (-|z|^{2q})$, 
and $\ga_q$ is positive only in the interval $0< q \leqslant 1$~\cite{koldobsky2005}. Therefore, we have the following theorem. 

\vskip 4mm 
\noindent
\textbf{Theorem 1.} The effective smeared delta source $\rho_s (r)$ is strictly positive for $0 < N_s \leqslant 1$.
\vskip 4mm 
\noindent
\textit{Proof.} 
For $N_s=1$, the effective source can be evaluated in closed form, Eq.~\eq{source-gaussian}, which is clearly positive. Now, for $0<N_s<1$ and $z\in [0,\infty)$, note that the function $\exp ( -z^{2N_s} )$ is positive and completely monotonic, i.e., it is infinitely differentiable on $(0,\infty)$, its first derivative is negative, and the signs of its successive derivatives alternate. Then, it follows from Bernstein's theorem that for every $z \in [0,\infty)$ it can be expressed as a Laplace transform of a positive measure ${\cal M}_{N_s}$, namely,
\begin{equation} 
\n{gen-gauss-lap}
e^{-(k/\mu_s)^{2N_s}} = e^{-[ (k/\mu_s)^2 ]^{N_s} } = \int_{0}^{\infty} \rd t \, {\cal M}_{N_s} ( t ) \,  e^{-t ( k / \mu_s )^2}
.
\end{equation}
Using~\eq{gen-gauss-lap}, the effective source can be expressed as 
\begin{equation} 
\begin{split}
\rho_{s} ( r ) 
&= \frac{M}{2 \pi^2 r} \int_0^\infty \rd k \, k \sin(kr) \, e^{- ( k/ \mu_s )^{2 N_s}}
= \frac{M}{2 \pi^2 r} \int_{0}^{\infty} \rd t \, {\cal M}_{N_s} ( t ) \, \int_0^\infty \rd k \, k \sin(kr) \,  e^{-t ( k / \mu_s )^2}
\\
& \qquad \qquad \qquad \qquad \qquad
= \frac{M\mu_s^3}{8 \pi^{3/2}} \int_{0}^{\infty} \rd t \, {\cal M}_{N_s} ( t ) \, t^{-3/2} e^{- \frac{\mu^2_s r^2}{4 t}}
.
\end{split}
\end{equation}
Thus, since ${\cal M}_{N_s} ( t )$ is a positive measure and the remaining integrand is strictly positive, $\rho_s ( r ) > 0$ for all values of~$r>0$. 
\qed

\vskip 4mm 

It is instructive to note that the above proof fails for $N_s > 1$ because the function $\exp ( -z^{2N_s} )$ is \emph{not} completely monotonic. Indeed, there is always a point $z=\big(1-\tfrac{1}{2 N_s}\big)^{1/(2 N_s)}$ where its second derivative changes sign.


\subsection{Mass function}
\label{Sec4}

For the particular cases of $N_s = 1$ and $N_s = 1/2$, one can directly integrate~\eq{massfunction} using the sources~\eq{source-gaussian} and~\eq{source-nonC} to find~\cite{BreTibLiv,Nozari:2008rc,Liang:2012vx,Kuhfittig:2014iwa}, 
\begin{equation}
\n{MexpBox}
M_s(r) = M \left[ {\rm erf} \left( \frac{\mu_s r}{2} \right) - \frac{\mu_s r}{\sqrt{\pi}}  e^{-\frac{\mu^2_s r^2}{4} } \right]
\end{equation}
for $N_s=1$, where 
\begin{equation}
{\rm erf} (x) = \frac{2}{\sqrt{\pi}} \int_0^x \rd t \, e^{-t^2} 
\end{equation}
is the error function, and
\begin{equation}
M_s ( r )  
= \frac{2M}{\pi} \left[ \arctan ( \mu_s r ) -\frac{\mu_s r}{1 + \mu_s^{2} r^{2}} \right]
,
\end{equation}
for $N_s = 1/2$.

In order to find a compact expression for the mass function that holds for $N_s >0$ we can once again use the GEF~\eq{GEF}. We start by substituting~\eq{sour-GEF} into the definition of the  mass function~\eq{massfunction} and then employ the Feynman trick of integration to eliminate the factor $r^2$ in the integrand of~\eq{massfunction}. More precisely, using the derivative formula for the GEF,
\begin{equation}
\n{deGEF}
\frac{\rd}{\rd z} {\cal E}_{\nu,\al} (z) = {\cal E}_{\nu,\al+1} (z),
\end{equation}
the integrand of~\eq{massfunction} with~\eq{sour-GEF} can be rewritten as
\begin{equation}
r^2 \rho_s(r) = - \frac{M \mu_s^3}{8 \pi^{3/2}} \frac{\pa}{\pa \la_s} \left[ {\cal E}_{N_s,1/2} (- \la_s r^2 ) \right]
,
\end{equation}
where we defined the Feynman parameter $\la_s = \mu_s^2/4$. Thence, defining the ``generalized error function'',
\begin{equation}
\n{G-ERROR}
{\cal E}{\rm{rf}}_{\nu,\al} (x) = \frac{2}{\sqrt{\pi}} \int_0^x \rd t \, {\cal E}_{\nu,\al}(-t^2)
,
\end{equation}
it follows 
\begin{eqnarray}
\n{mass-GEF}
M_s (r) & = & - \frac{M \mu^3}{4} \frac{\pa}{\pa \la_s} \left[ \frac{1}{\sqrt{\la_s}} \, {\cal E}{\rm{rf}}_{N_s,1/2} (\sqrt{\la_s} r ) \right]_{\la_s = \mu_s^2/4}
\nonumber
\\
& = & M \left[ {\cal E}{\rm{rf}}_{N_s,1/2} \left( \frac{\mu_s r}{2} \right) - \frac{\mu_s r}{\sqrt{\pi}} \, {\cal E}_{N_s,1/2} \left(- \frac{\mu_s^2 r^2}{4} \right) \right]
,
\end{eqnarray}
that generalizes~\eq{MexpBox} for $N_s >0$.

For $N_s \geqslant1/2$ one can also obtain a power-series representation of the mass function  by directly integrating [according to Eq.~\eq{massfunction}] each term of~\eq{sour-series}. The result is 
\begin{equation}
\n{mass_series}
M_s(r)  =   \frac{M }{ \pi N_s } \sum_{\ell=0}^\infty \frac{(-1)^\ell}{(2\ell + 1)!} \, 
\Ga \left( \frac{2\ell+3}{2 N_s} \right) \frac{(\mu_s r)^{2 \ell + 3}}{2l + 3}.
\end{equation}
Another useful representation valid for $1/2 < N_s < \infty$, is given in terms of the $H$-function. Indeed, the identities~\cite{mathai}
\begin{equation}
\n{H1}
\int_{0}^{t} \rd x \ x^{\rho - 1} ( t - x )^{\sigma - 1} H^{m, n}_{p, q}
\biggl[
\begin{array}{c}
( a_{p}, A_{p} ) \\
( b_{q}, B_{q} )
\end{array}
\bigg \vert
b x^{k}
\biggr]
=
t^{\rho + \sigma - 1} \Gamma ( \sigma ) 
\, H^{m, n+1}_{p+1, q+1} \biggl[
\begin{array}{c}
(1-\rho, k), (a_{p}, A_{p})\\
(b_{q}, B_{q}), (1-\rho-\sigma, k)
\end{array}
\bigg \vert
b t^{k} \biggr],
\end{equation}
and
\begin{equation}
\n{H2}
z^\si
H_{p,q}^{m,n} 
\left[
\begin{array}{c}
(a_p, A_p) \\
(b_q, B_q)
\end{array}
\bigg \vert z
\right] 
= 
H_{p,q}^{m,n} 
\left[
\begin{array}{c}
(a_p + \sigma A_p, A_p) \\
(b_q + \sigma B_q, B_q)
\end{array}
\bigg \vert z
\right],
\qquad z \in \mathbb{C},
\end{equation}
together with Eqs.~\eq{massfunction} and \eq{source-H} give
\begin{equation}
\n{mass-function-H}
M_s(r) 
=\frac{4 M}{\sqrt{\pi} N_s} H^{1,2}_{2,3} \biggl[
\begin{array}{c}
(1,2), \left( 1, \frac{1}{N_s} \right) \\
(\frac{3}{2}, 1), \left( 1, 1 \right), ( 0, 2 )
\end{array}
\bigg \vert
\frac{\mu_s^{2} r^{2}}{4}
\biggr]
.
\end{equation}

\begin{figure}[t]
\centering
\begin{subfigure}{.5\textwidth}
\centering
\includegraphics[width=7.55cm]{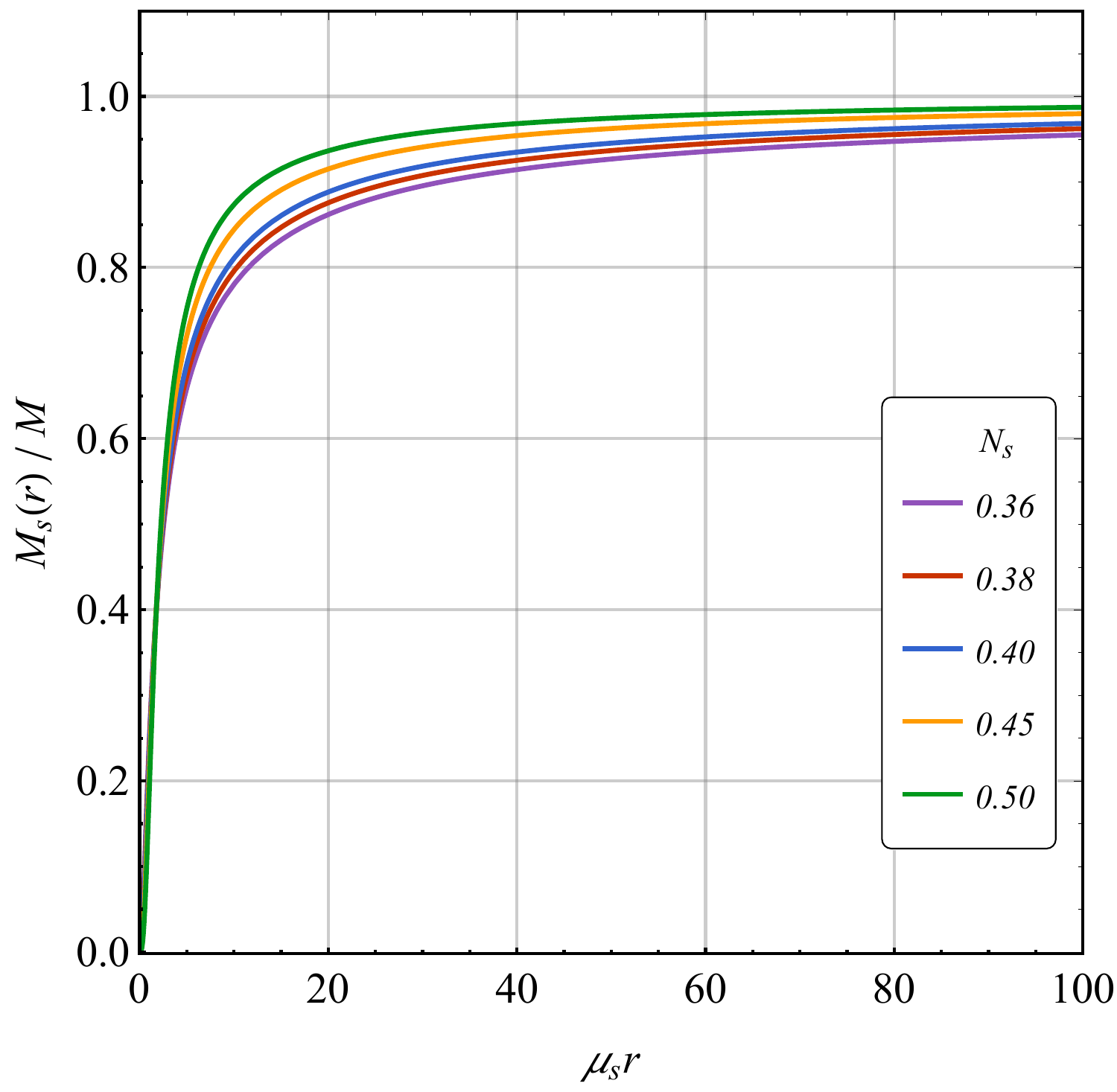}
\end{subfigure}%
\begin{subfigure}{.5\textwidth}
\centering
\includegraphics[width=7.5cm]{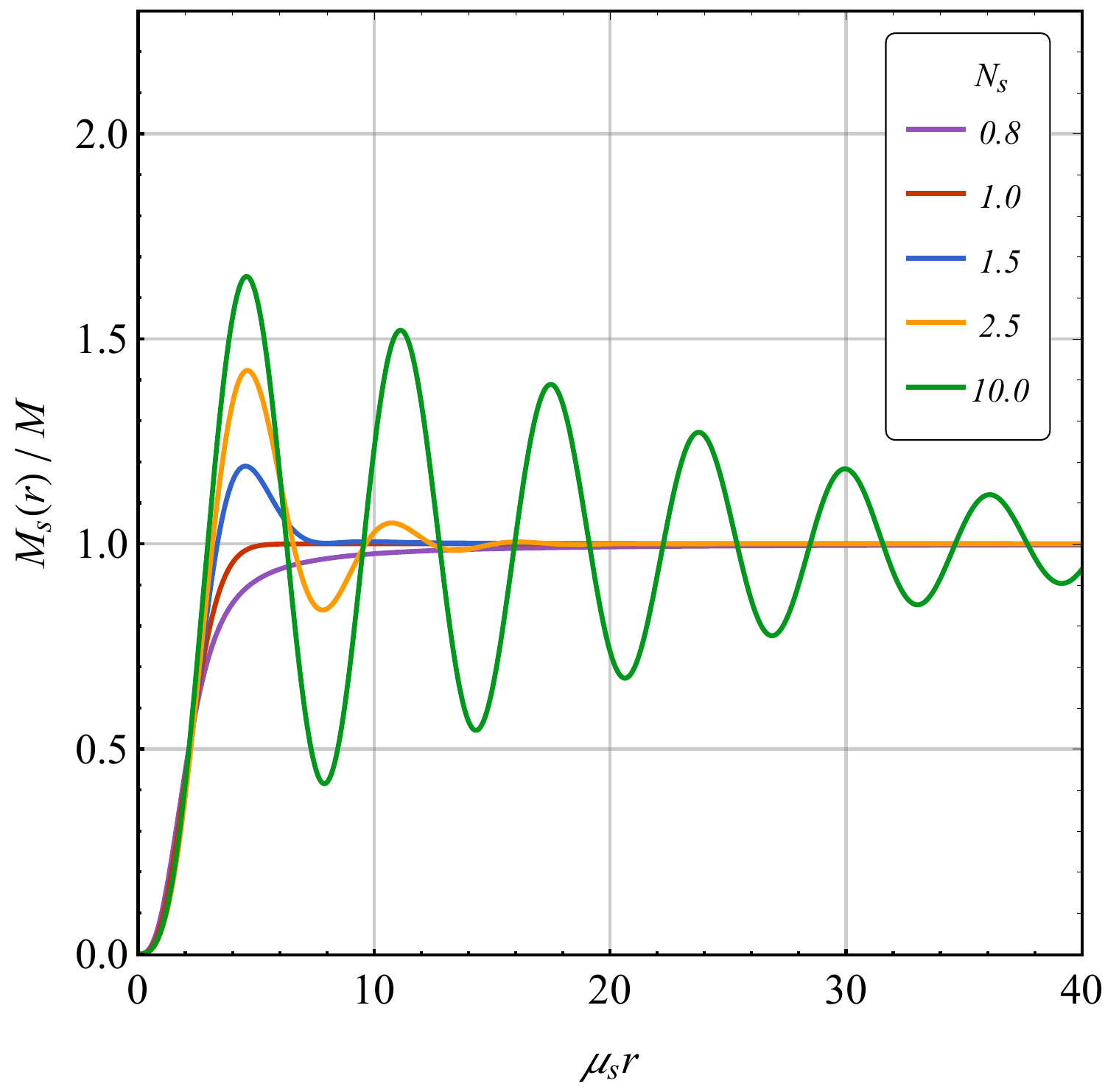}
\end{subfigure}
\caption{Plot of the dimensionless mass function $M_s(r)/M$ as a function of $\mu_s r$ for different values of $N_s$. The oscillations are present only for $N_s > 1$. For any value of $N_s$ we can observe the asymptotic limits: For $\mu_s r \ll 1$, $M_s(r) \sim r^3$, while  for $\mu_s r \gg 1$, $M_s (r) \sim M$.}
\label{Fig2}
\end{figure}

Fig.~\ref{Fig2} displays the graphs of the mass function for selected values of $N_s$. The plots illustrate the following general properties of the function $M_s (r)$:

\vskip 4mm
\noindent
\textit{1. Positivity.} 
Although the effective source $\rho_s(r)$ may assume negative values, we have the following result:
\vskip 4mm 
\noindent
\textbf{Theorem 2.} The effective mass function $M_s(r)$ is positive for any $N_s>0$.
\vskip 4mm 
\noindent
The proof is carried out in Appendix~\ref{ApD}, and involves expressing $M_s(r)$ as the integral of a positive function.

\begin{figure}[t]
\includegraphics[width=7.5cm]{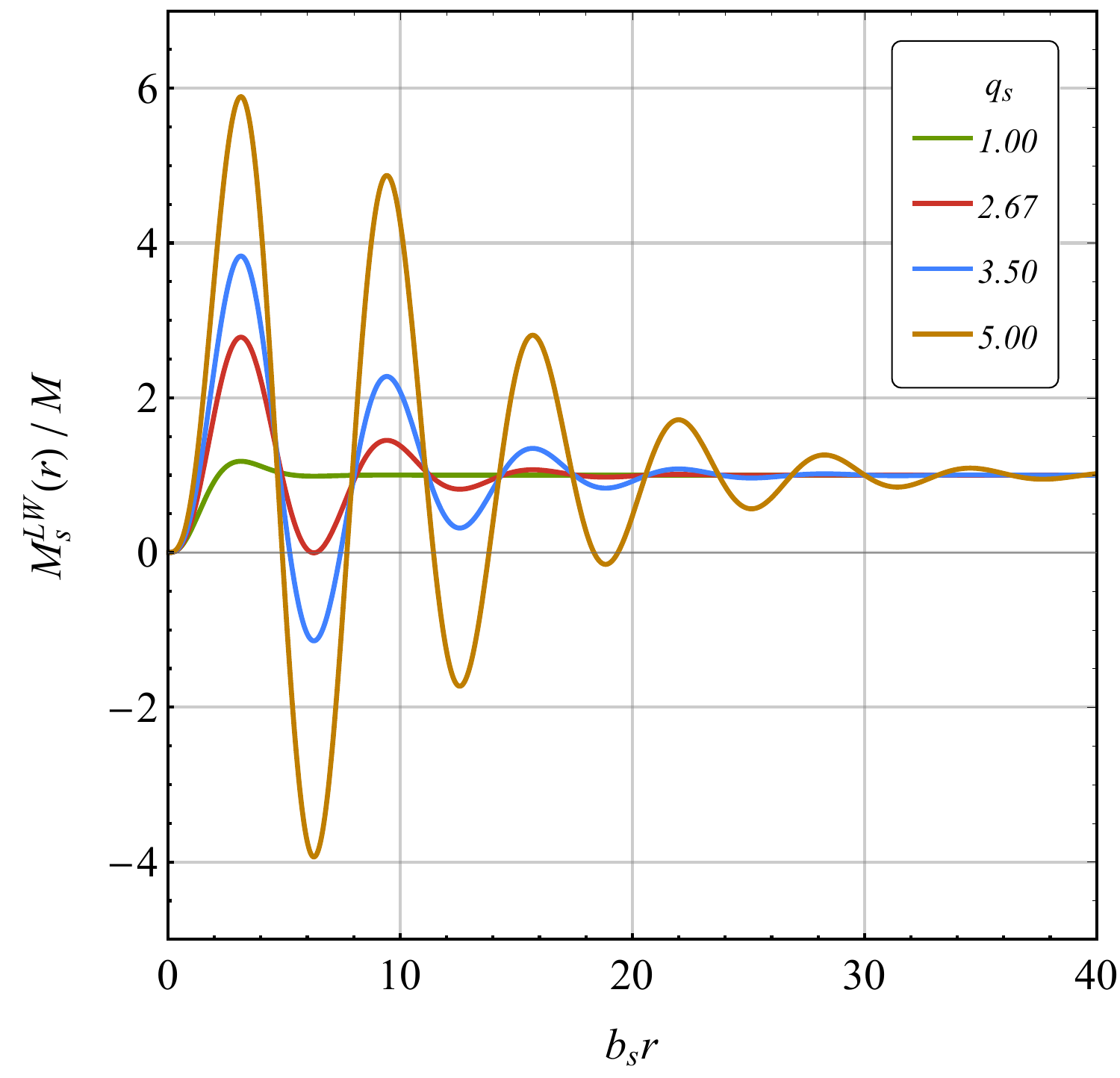}
\caption{Plot of the dimensionless mass function~\eq{meff-lw} for Lee--Wick gravity, $M_s^{\text{LW}}(r)/M$, as a function of $b_s r$ for different values of $q_s = b_s/a_s$. ${M_s^{\text{LW}}}(r)$ assumes negative values for $q_s>2.67$.}
\label{Fig3}
\end{figure}

From the physical point of view, Theorem 2 can be regarded as a consequence of the fact that the model does not have massive poles with tachyonic behavior in the propagator, which could generate a net repulsive spin-$s$ gravitational force. Indeed, since $M_s(r) \geqslant 0$, a test particle of mass $m$ experiences a spin-$s$ component of the gravitational force given by $m g_s(r) = - m M_s(r)/r^2$ which is always attractive.\footnote{Of course, like in generic higher-derivative gravity models, when the contributions of the spin-0 and spin-2 sectors are combined, via~\eq{pch}, it can happen that there exist regions where the net resultant force $-\vec{\na} \ph$ is repulsive, depending on the particular values of $N_{0,2}$ and $\mu_{0,2}$. In the case of GF$_N$ models, however, we can guarantee that the force is always attractive, as $\ph=\chi_2=\chi_0$ and the combined effective mass function $M(r)=M_0(r)=M_2(r)$ is positive.}
This situation is to be contrasted, e.g., with the case of Lee--Wick gravity, for which the imaginary part of the massive poles has a tachyonic character and, if it dominates, the mass function can become negative. For example, the effective mass function in the six-derivative Lee--Wick local model with mass $\mu_s = a_s+ib_s$ is given by~\cite{Burzilla:2023xdd} 
\begin{equation}
\n{meff-lw}
M_{s}^{\text{LW}}(r)  =  M-\frac{M}{2 a_s b_s } e^{-a_s r} \Big\{ b_s \left[  2a_s +(a_s^2+b_s^2)r \right] \cos(b_s r) 
+ \left[ a_s^2-b_s^2 +a_s(a_s^2+b_s^2)r \right]  \sin(b_s r) \Big\}
,
\end{equation}
where $a_s$ and $b_s$ are model-dependent parameters. There exist regions where $M_s^{\rm LW}(r)<0$, if $q_s \equiv b_s/a_s > 2.67$ (see Fig.~\ref{Fig3}), showing the dominance of the repulsive force of the tachyonic part. (It is worth mentioning that real ghosts are also associated with repulsive interactions; however, the tug of war of ghosts and healthy modes cannot cause a net repulsive force in the spin-0 or spin-2 sector unless complex modes are present. This was explicitly verified for sixth-derivative gravity~\cite{Accioly:2016qeb,NosG}, and is conjectured to occur in higher-order models as well.)

\vskip 4mm
\noindent
\textit{2. Asymptotic behavior.} The behavior of the effective mass function for small and large values of $r$ is
\begin{equation}
\n{mass-prop-1}
\lim_{r \to 0} M_s (r) = 0
,
\qquad \qquad
\lim_{r \to \infty} M_s (r) = M
.
\end{equation}
The first property follows from the fact that the total mass $M$ is effectively smeared rather than concentrated in a point-like source. In this way, the effective mass inside a sphere with vanishing diameter is also vanishing. Specifically, one can show that $\lim_{r \to 0} \rho_s (r) = {\rm max} \, [\rho_s(r)]$, which applied to~\eq{massfunction} yields
\begin{equation}
\n{mass-rto0}
M_s (r)  \underset{\mu_s r \ll 1}{\sim} \frac{4\pi}{3} \, {\rm max} \,  [\rho_s(r)] r^3  
.
\end{equation}
The second expression in~\eq{mass-prop-1} is related to the suppression of the form factor in the IR, namely $f_s (0) = 1$, so that the total mass $M$ is recovered as $r \to \infty$. In fact, for a form factor that is continuous and with $f_{s} ( 0 ) \ne 0$, 
\begin{equation}
\begin{split}
\n{proveM}
\lim_{r \to \infty} M_s ( r ) 
&= M \int_{\mathbb{R}^3} \rd^3 r^{\prime} \int_{\mathbb{R}^3} \frac{\rd^3 k}{( 2 \pi )^3} \frac{e^{i \vec{k} \cdot \vec{r}\,^{\prime}}}{f_{s} ( - k^2 )}
\\
&= M \int \frac{\rd^3 k}{( 2 \pi )^3} \frac{1}{f_{s} ( -k^2 )} \int_{\mathbb{R}^3} \rd^3 r^{\prime} e^{i \vec{k} \cdot \vec{r} \,^{\prime}}
\\
&= M \int_{\mathbb{R}^3} \frac{\rd^3 k}{( 2 \pi )^3} \frac{1}{f_{s} ( -k^2 )} \, \delta ( \vec{k} )=
\frac{M}{f_{s} ( 0 )}
.
\end{split}
\end{equation}
Physically, this means that for $r \gg 1/\mu_s$ we approach the standard limit of linearized GR.

\begin{table}[t]
\begin{tabular}{|c|c|c|c|c|c|}
\hline
$Y(\delta)$ & 6 & 7 & 8 & 9 & 10 \\ 
\hline
$\delta $ 
&
$2.48 \times 10^{-2} $ 
& 
$1.22 \times 10^{-2} $
& 
$5. 96 \times 10^{-3}$  
&
$2.88 \times 10^{-3} $ 
&
$1.38 \times 10^{-3} $ 
\\
\hline
\end{tabular}
\caption{Some values for the multiplicative factor in Eq.~\eq{r-estrela} and the related maximum deviation $\de$ of the effective mass function with respect to its asymptotic value $M$. That is, if $r>Y(\de) N_s/\mu_s$, then $\vert M_s(r)/M-1\vert<\de$.}
\label{Table-Y}
\end{table}

\vskip 4mm
\noindent
\textit{3. Oscillation and local maxima.} For $N_s>1$, there are regions where $M_s(r) > M$; this is a consequence of the existence of regions where $\rho_s(r) < 0$. Together with Eq.~\eq{mass-prop-1}, this implies that the mass function oscillates and has local maxima. The local maxima of $M_s (r)$ constitute a monotonically decreasing positive sequence, while the sequence of minima is monotonically increasing. In addition, the position of the first peak increases with $N_s$. The asymptotic behavior of $M_s(r)$ for large values of $\mu_sr$ is provided by Eq.~\eq{pacote} in Appendix~\ref{ApE}, which explicitly shows how the mass function oscillates in this regime.

It is natural to wonder at which scale the oscillations are suppressed and the mass function becomes sufficiently close to $M$. This problem can be formulated as follows: Given an arbitrarily small $\delta>0$, find the value $r_*$ such that  
\begin{equation}
r>r_* \quad \Longrightarrow \quad
\left| \frac{M_s(r)}{M} - 1 \right| < \delta 
.
\end{equation}
Figure~\ref{Fig2} suggests that, given a $\delta$, $r_*$ increases with $N_s$. An educated guess for the form of this dependence is a linear relation with $N_s$,
\begin{equation}
\n{r-estrela}
r_* = Y(\delta) \, \frac{N_s}{\mu_s}
,
\end{equation}
where $\delta$ can be regarded as the maximum deviation from the asymptotic value $M$ and $Y(\delta)$ is a function that reproduces the dependence on $\de$. Indeed, by fixing $Y$ and numerically calculating $M_s (r_*)$ [using the integral representation~\eq{mass-integral}] for several thousands of values of $N_s$ in the range $[ 1, 1000]$, we checked that~\eq{r-estrela} is indeed a good approximation for  $N_s \geqslant 2 $ (and with some small deviations in the interval $1 < N_s < 2$). Additionally, in Appendix~\ref{ApE} we provide an alternative explanation of the linear relation~\eq{r-estrela}, by obtaining a function that describes the attenuation of the mass function's oscillations. In Table~\ref{Table-Y} we list some values of the pair $(\de,Y)$. These values were obtained using Eq.~\eq{envFunc} and verified by numerically integrating~\eq{mass-integral}. For example, for $Y=6$ it follows that if $r>6N_s/\mu_s$ the maximum relative deviation of $M_s(r)$ with respect to $M$ is $2.48\%$, while for $Y=10$ if $r>10N_s/\mu_s$ it is $0.138 \%$.

From a physical point of view, $r_*$ gives the scale beyond which the nonlocal effects to the gravitational force become negligible (as defined by $\delta$). Indeed, since $M(r) \approx M$ for $r>r_*$, a test particle of mass $m$ at a distance $r > r_*$ feels a gravitational force of magnitude [see~\eq{gs_def}]
\begin{equation}
|m g_s (r > r_*)| \sim  \frac{GmM}{r^2}
.
\end{equation} 
The fact that the range of the nonlocal interaction grows with $N_s$ and is not only $1/\mu_s$ is an interesting property of the models with form factor~\eq{formfactor}. In principle, by choosing $N_s$ large enough it is possible to make nonlocal effects relevant at  arbitrarily large scales, with physical consequences that will be discussed in a separate publication.


\subsection{Newtonian Potential}
\label{Sec5}

The effective delta source for $N_s = 1$ has the Gaussian profile~\eq{source-gaussian}, while for other values of $N_s$, as shown in Sec.~\ref{Sec3}, $\rho_s(r)$ can be expressed in terms of ``generalized Gaussian functions''. The Newtonian potential for $N_s=1$ has the well-known solution found by Tysetlin~\cite{Tseytlin} in the context of string theory (see also~\cite{Siegel2,Biswas,Modesto12}, for nonlocal gravity considerations), 
given by the error function,
\begin{equation}
\n{chi-error}
\chi_s (r) = - \frac{GM}{r} \, {\rm erf} \left( \frac{\mu_s r}{2} \right)
.
\end{equation}
Similarly to the case of the effective source, for general $N_s$ the spin-$s$ potential is given in terms of ``generalized error functions''. 

In order to show this, note that Eqs.~\eq{deGEF} and~\eq{mass-GEF} yield the identity 
\begin{equation}
\frac{M_s (r)}{r^2} = \frac{\rd}{\rd r} \left[ - \frac{M}{r} {\cal E}{\rm{rf}}_{N_s,1/2} \left( \frac{\mu_s r}{2} \right) \right]
. 
\end{equation}
Therefore, the result of the integral \eq{chi_intg} is simply
\begin{equation}
\n{chi-GERROR}
\chi_s (r) = - \frac{GM}{r} \, {\cal E}{\rm{rf}}_{N_s,1/2} \left( \frac{\mu_s r}{2} \right)
,
\end{equation}
which generalizes the solution for the potential~\eq{chi-error} for any $N_s >0$. It also implies that the potential is even in $r$ and finite at $r=0$; indeed, 
\begin{equation}
\n{chi_0}
\chi_s (0) = - \frac{GM\mu_s}{\pi N_s} \Ga \left( \frac{1}{2N_s} \right) 
.
\end{equation}
As expected, the potential diverges in the limit $N_s\to 0$, in which case the form factor $f_s$ becomes constant and its overall effect is to redefine Newton's constant $G$.

Furthermore, using the power series representation of the GEF~\eq{GEF} and the definition of the generalized error function~\eq{G-ERROR} for $N_s \geqslant 1/2$, one obtains the power-series solution for the potential:
\begin{equation}
\n{chi-series}
\chi_s (r) = -\frac{ G M \mu_s}{\pi N_s} \sum_{\ell=0}^\infty \frac{(-1)^\ell}{(2\ell + 1)!} \, \Ga \left( \frac{2\ell+1}{2N_s} \right) (\mu_s  r)^{2 \ell}.
\end{equation}
This expression reproduces the solution found in~\cite{Edholm_NewPot} for the potential in the particular case in which $\ph = \psi = \chi_2 = \chi_0$. 

\begin{figure}[t]
\centering
\begin{subfigure}{.5\textwidth}
\centering
\includegraphics[width=7.5cm]{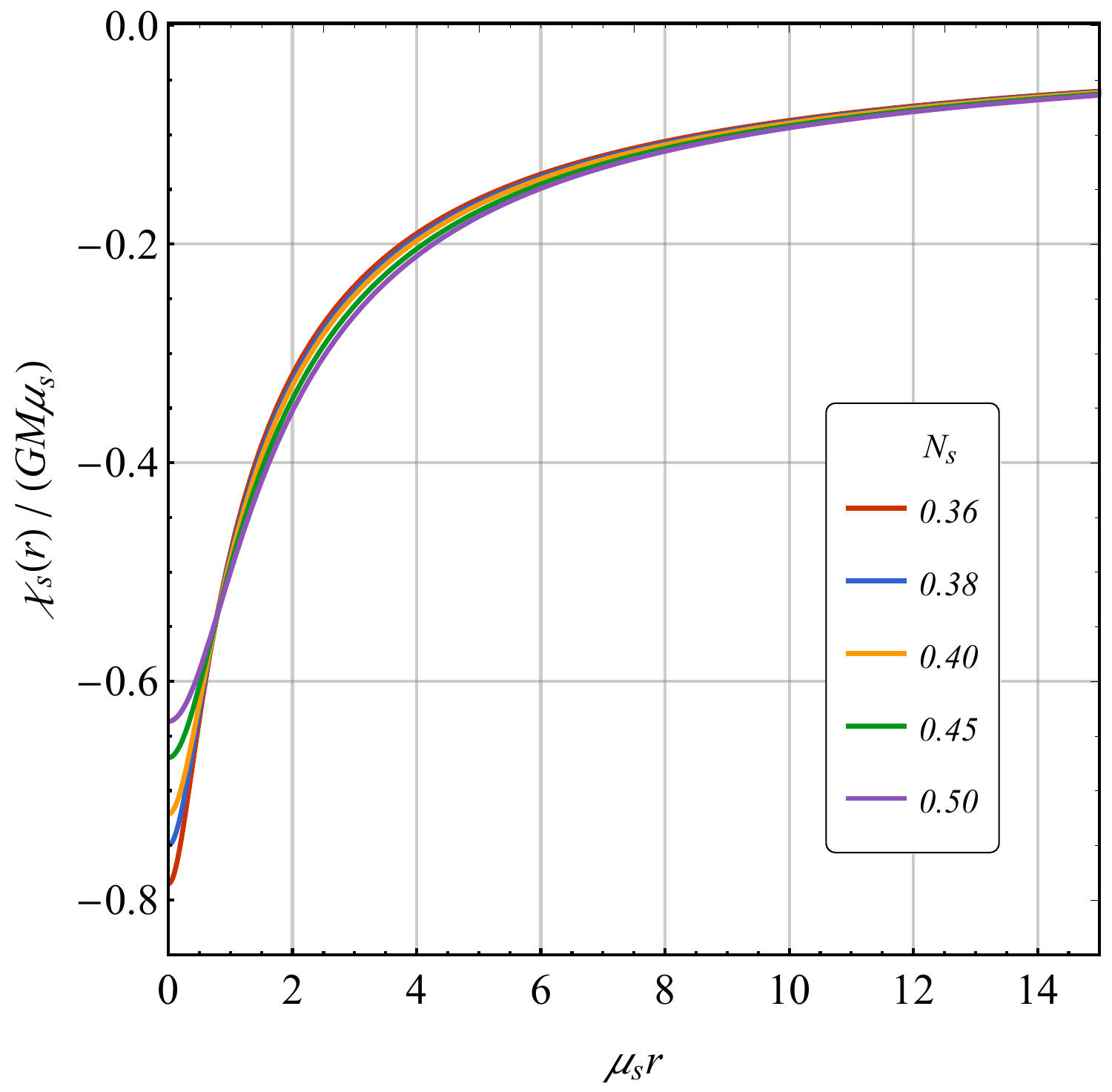}
\end{subfigure}%
\begin{subfigure}{.5\textwidth}
\centering
\includegraphics[width=7.5cm]{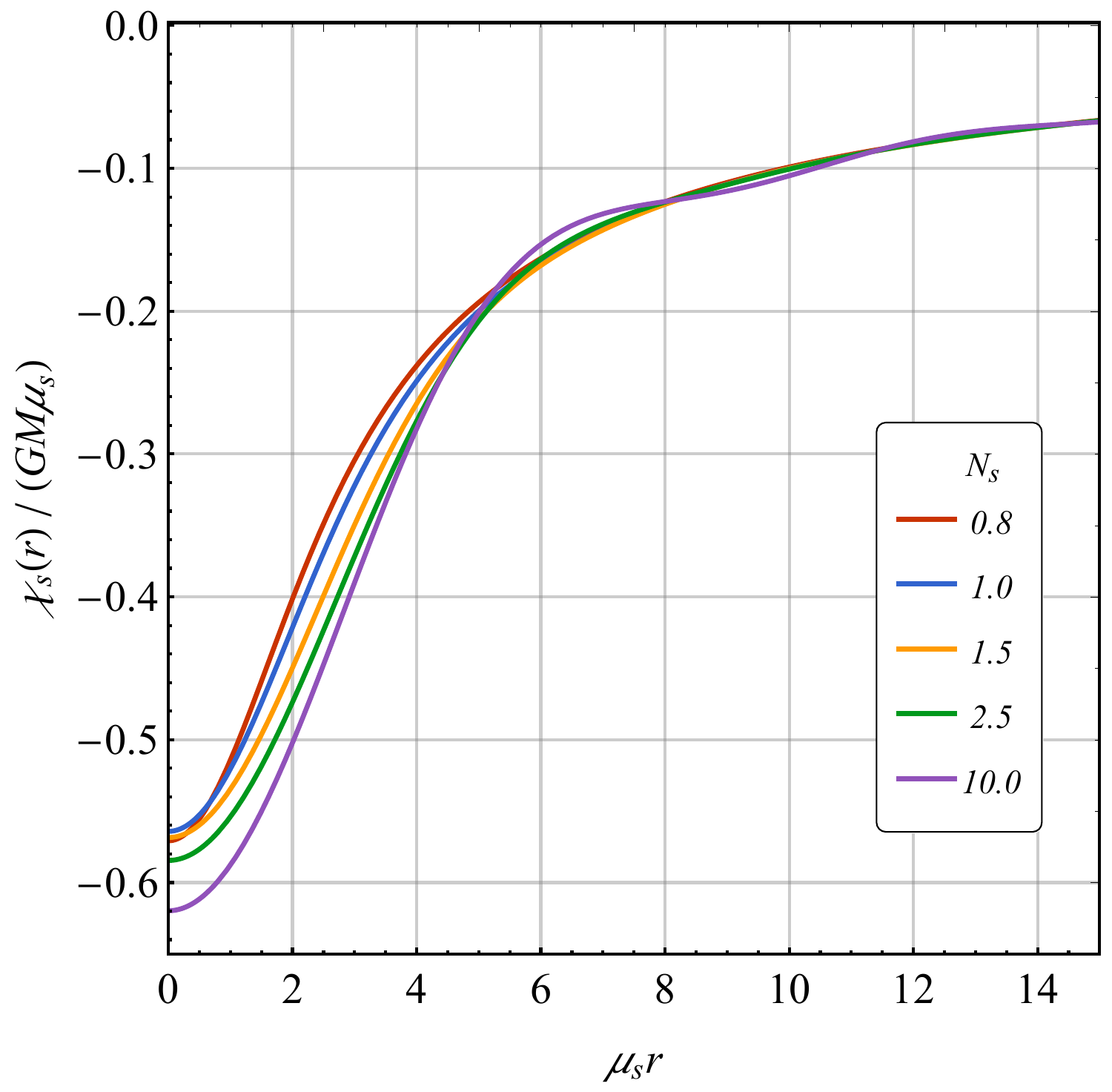}
\end{subfigure}
\caption{ Plot of $\chi_s(r)/(G M \mu_s)$ as a function of $\mu_s r$ for different values of $N_s$. The oscillations are present only for $N_s > 1$. Note that for any $N_s >0$ we have $\chi'_s(0)=0$.}
\label{Fig4}
\end{figure}

In Fig.~\ref{Fig4} we plot~\eq{chi-GERROR} for several values of $N_s$. The following properties can be observed: For $N_s>1$ the potential oscillates, $\chi_s (r) \sim -1/r$ for large~$r$, and it is regular at $r=0$. Moreover, $\chi'_s(0) = 0$. As we discuss in Sec.~\ref{Sec7}, the last property is closely related with the regularization of the linearized curvature invariants at $r=0$.

Finally, some general comments about the potential are in order:
\vskip 4mm
\noindent
1) The solution for the potential, valid for $N_s > 1/2$, in terms of the Fox $H$-function,
\begin{equation}
\chi_s ( r ) 
= -\frac{2 G M}{\sqrt{\pi} r} 
H^{1,2}_{2,3} \biggl[
\begin{array}{c}
(1, 2), \big( 1, \frac{1}{N_s} \big) \\
(1/2, 1), \left( 1, 1\right), ( 0, 2 )
\end{array}
\bigg \vert
\frac{\mu_s^2 r^2}{4} 
\biggr]
,
\end{equation}
can be directly obtained using Eqs.~\eq{chi_intg},~\eq{H1},~\eq{H2} and~\eq{mass-function-H}. Also, for $N_s = 1/2$ the potential can be written in closed form:
\begin{equation}
\chi_s ( r ) = - \frac{2GM}{\pi r} \arctan ( \mu_s r )
.
\end{equation}

\vskip 4mm
\noindent
2) The power series representation~\eq{chi-series} is also easily found without using effective sources but directly applying the Fourier transform method. Specifically, starting with the solution of~\eq{poi-chi-mod} in the form
\begin{equation}
\n{chi-fo}
\chi_s (r) = -\frac{2GM}{\pi r} \int_0^\infty \rd k \, \frac{\sin (kr)}{k f_s(-k^2)}
,
\end{equation}
and then applying a procedure similar to the one used at the beginning of Sec.~\ref{Sec3}.

\vskip 4mm
\noindent
3) The potential $\chi_s(r)$ can also be obtained using the heat kernel method~\cite{Frolov:Exp,Frolov:Poly} (see also~\cite{Buoninfante:2022ild}). In fact, the effective source $\rho_s (r)$ is closely related to the heat kernel of the operator ${\tri^N}$. As shown in~\cite{Barvinsky:2019spa}, this heat kernel in a $D$-dimensional spacetime is given by the GEF,
\begin{equation}
\n{HK}
\langle \vec{r} \,\big| e^{-s \tri^N} \big|  \vec{r} \,' \rangle = K_{N,D} (| \vec{r} - \vec{r} \,'| , s) = \frac{1}{(4 \pi s^{1/N})^{D/2}} \, \, {\cal E}_{N,D/2} \left(- \frac{|\vec{r}-\vec{r} \,'|^2}{4 s^{1/N}} \right)
. 
\end{equation}
The comparison of~\eq{HK} and~\eq{sour-GEF} reveals that 
\begin{equation}
\n{sour-HK}
\rho_s (r) = M K_{N_s,3} \left( r , 1/\mu_s^{2N_s}\right)
,
\end{equation}
and the heat-kernel representation of the potential reads
\begin{equation}
\n{chi-HK}
\chi_s (r) = -\frac{GM}{N_s} \, \int_0^\infty \frac{\rd s'}{s'} \, \theta \left( s' - \frac{1}{\mu_s^{2N_s}} \right) K_{N_s,1} (r, s')
,
\end{equation}
where $\theta (x)$ is the Heaviside step function. This is a generalization of the formula for $N_s = 1$ obtained in~\cite{Frolov:Exp}. Finally, the change of integration variable to $t = r^2/4 s'^{1/N_s}$ in~\eq{chi-HK} results in the generalized error function defined in~\eq{G-ERROR}, leading to~\eq{chi-GERROR}.


\section{The limit $N_s \to \infty$}
\label{Sec6}

In the previous section, we obtained different ways of expressing the effective source, mass function, and Newtonian potential. Nevertheless, because of their somewhat complicated expressions, it is difficult to grasp intuition about the way oscillations occur and how the weak-field solutions change as the value of $N_s$ increases. This can be partially remedied by using the approximation~\eq{pacote} for the mass function, or by studying the limiting functions as $N_s \to \infty$. For the Newtonian potential, this formal (but useful) limit was considered numerically in~\cite{Perivolaropoulos:2016ucs} (see also~\cite{Edholm:2017dal}), resulting in an analytical approximation. Here, we show how our results can lead to a better understanding of the $N_s \to \infty$ limit, even providing exact and closed-form expressions.

First, notice that for $N_s \to \infty$ the form factor~\eq{formfactor} converges to the rectangle (pulse) function~\cite{Edholm:2017dal}:
\begin{equation}
\n{fInfinity}
\lim_{N_s \to \infty} f_s (-k^2) = \lim_{N_s \to \infty} e^{-(k^2/\mu_s^2)^{N_s} } = {\rm rect} (k^2/\mu_s^2)
.
\end{equation}
Since the limiting function is not continuous, the convergence is not uniform. For this reason, the direct use of~\eq{fInfinity} in Fourier integral representations, such as~\eq{eff-sour} and~\eq{chi-fo}, might be problematic. Thus, in order to study the $N_s \to \infty$ limit, let us start with the power series representations obtained in the previous sections.

Using the property $z \Ga(z) = \Ga (z+1)$ of the Gamma function, the effective source~\eq{sour-series} can be cast as
\begin{equation}
\rho_{s} (r) = \frac{M \mu_s^3}{2 \pi^2 } \sum_{\ell=0}^\infty \frac{(-1)^\ell}{(2\ell + 3) (2\ell + 1)!} \, 
\Ga \left( 1+ \frac{2\ell+3}{2N_s} \right) (\mu_s r)^{2 \ell}
.
\end{equation}
Now, $N_s \to \infty$ yields
\begin{equation}
\rho^{\infty}_{s} (r) \equiv
 \lim_{N_s \to \infty}  \rho_s (r) = 
\frac{M \mu_s^3}{2 \pi^2 } \sum_{\ell=0}^\infty (-1)^\ell \frac{1}{(2\ell + 3) (2\ell + 1)!} \, 
(\mu_s r)^{2 \ell}
.
\end{equation}
A simple summation index reshuffling and manipulation with factorials allow us to rewrite the above equation as
\begin{equation}
\rho^{\infty}_{s} (r) 
= \frac{M}{2 \pi^2 r^3} \left[ 
\sum_{\ell=0}^\infty \frac{(-1)^{\ell}}{(2\ell + 1)!} (\mu_s r)^{2\ell+1}
- (\mu_s r) \sum_{\ell=0}^\infty \frac{(-1)^{\ell}}{(2\ell)!} (\mu_s r)^{2\ell}
\right]
,
\end{equation}
whence
\begin{equation}
\n{sour-inf}
\rho^{\infty}_{s} (r) = \frac{M}{2 \pi^2 r^3} \left[ \sin (\mu_s r) - \mu_s r \cos (\mu_s r) \right]
.
\end{equation}
This closed-form solution can also be directly obtained from the source representation~\eq{sour-GEF} in terms of the GEF, since~\cite{Barvinsky:2019spa}
\begin{equation}
\n{GEF-infinity}
\mathcal{E}_{\infty, \alpha} ( z ) \equiv \lim_{\nu\to \infty} \mathcal{E}_{\nu, \alpha} ( z ) = \mathcal{C}_{\alpha} ( z ) 
,
\end{equation}
where $\mathcal{C}_{\alpha} ( z )$ is the Bessel--Clifford function, and
\begin{equation}
\n{JC}
{\cal J}_\al (z) = \left( \frac{z}{2} \right)^\al \mathcal{C}_{\alpha} \left( - \frac{z^2}{4} \right)
,
\qquad
{\cal J}_{3/2} (z) = \sqrt{\frac{2}{\pi z}} \left( \sin z - \frac{\cos z}{z} \right)
,
\end{equation}
where ${\cal J}_\al (z)$ is the Bessel function of the first kind. Therefore, given~\eq{GEF-infinity}, one can say that the solutions obtained expressed in terms of the GEF also hold for $N_s \to \infty$.

\begin{figure}[t]
\centering
\begin{subfigure}{.5\textwidth}
\centering
\includegraphics[width=7.28cm]{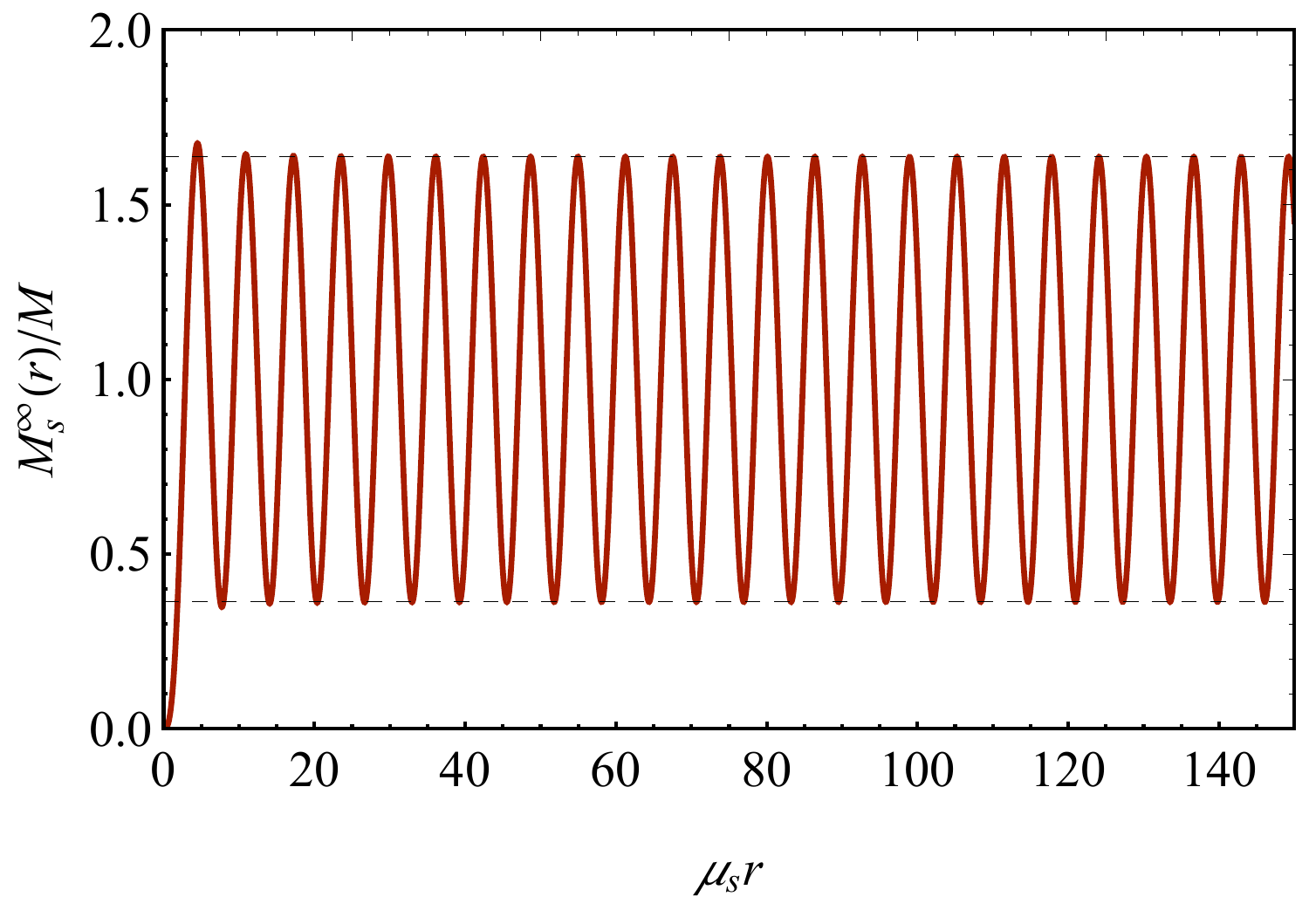}
\caption{}
\label{Fig5}
\end{subfigure}%
\begin{subfigure}{.5\textwidth}
\centering
\includegraphics[width=7.5cm]{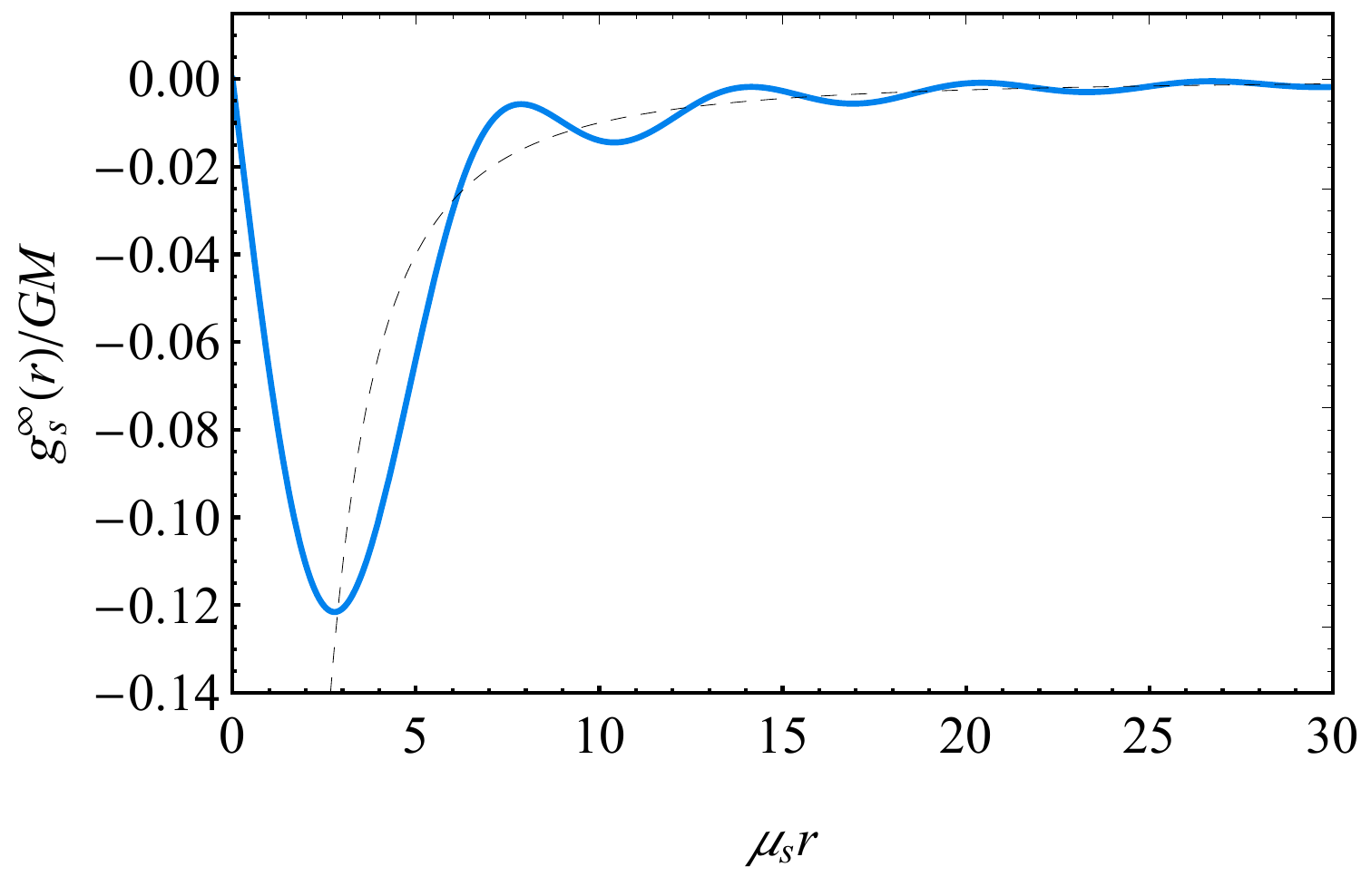}
\caption{}
\label{Fig6}
\end{subfigure}
\caption{ {\bf (a)} $M_s^{\infty}(r)/M$ as a function of $\mu_s r$, the horizontal dashed lines corresponds the value $(1 \pm 2/\pi) = (0.363,1.637)$. {\bf (b)} Plot of the gravitational field $g_s^{\infty}(r)/GM$ as a function of $\mu_s r$. The dashed lines correspond to the standard $-1/r^2$ gravitational field. Although the mass function $M_s^\infty(r)$ does not have a definite limit when $r \gg  1/\mu_s$, the gravitational field~$g_s^\infty(r) \to 0$ owning to the $r^{-2}$ damping factor in Eq.~\eq{gs_def}. Furthermore, the typical Newtonian singularity at $r=0$ is absent.}
\end{figure}

In what concerns the effective mass function, the substitution of \eq{sour-inf} into~\eq{massfunction} leads to
\begin{equation}
\n{mass-inf}
M_s^{\infty} (r) = \frac{2M}{\pi} \left[\text{Si}\,  (\mu_s r) - \sin (\mu_s r) \right] 
,
\end{equation}
where
\begin{equation}
\label{Si(z)}
{\rm Si} \, (z) = \int_0^z \rd t \, \frac{\sin t}{t}
\end{equation}
is the sine integral. The limiting function~\eq{mass-inf} has an interesting property: it violates the relation $\lim_{r \to \infty} M_s(r) = M$, valid for finite values of $N_s$. Indeed, since $\lim_{z \to \infty} {\rm Si} \, (z) = \pi/2$, the dimensionless mass function $M_s^{\infty} (r)/M$ tends to an oscillation in the range $\left[ 1 - 2/\pi, 1 + 2/\pi  \right]$ for large values of the argument $\mu_s r$, see Fig.~\ref{Fig5}. As explained in Sec.~\ref{Sec4}, for finite values of $N_s$, for any given $\de>0$ there exists $r_*$ such that $r>r_*$ implies $\vert M_s (r)/ M -1\vert < \de$; moreover, $r_*$ grows with $N_s$ [see discussion involving Eq.~\eq{r-estrela}]. Thus, $M_s (r) \approx  M$ is unattainable in the limit $N_s \to \infty$, for $r_*$ diverges. This gives the physical interpretation of why $M_s^{\infty} (r)$ does not approach $M$ for big enough~$r$. From a mathematical point of view, one of the assumptions underlying the proof of~\eq{proveM} is violated, for~\eq{fInfinity} is discontinuous at $r=0$.

Although $M_s^{\infty}(r)$ does not have a defined limit when $r \to \infty$, both the gravitational field $g_s (r)$ in Eq.~\eq{gs_def} (see Fig.~\ref{Fig6}) and the potential $\chi_s(s)$ have the correct limit even when $N_s \to \infty$, approaching the conventional behavior of Newtonian gravity. In fact, taking~\eq{mass-inf} into~\eq{chi_intg} it follows
\begin{equation}
\n{chi_in}
\chi_s^{\infty} (r) = - \frac{2GM}{\pi r} \, {\rm Si} \, (\mu_s r)
.
\end{equation}
Therefore, 
\begin{equation}
\chi_s^{\infty} (0) = - \frac{2GM \mu_s}{\pi}
,
\qquad \qquad
\chi_s^{\infty} (r) \underset{r \to \infty}{\sim }  - \frac{GM}{r}
.
\end{equation}

\begin{figure}[t] 
\includegraphics[width=7.5cm]{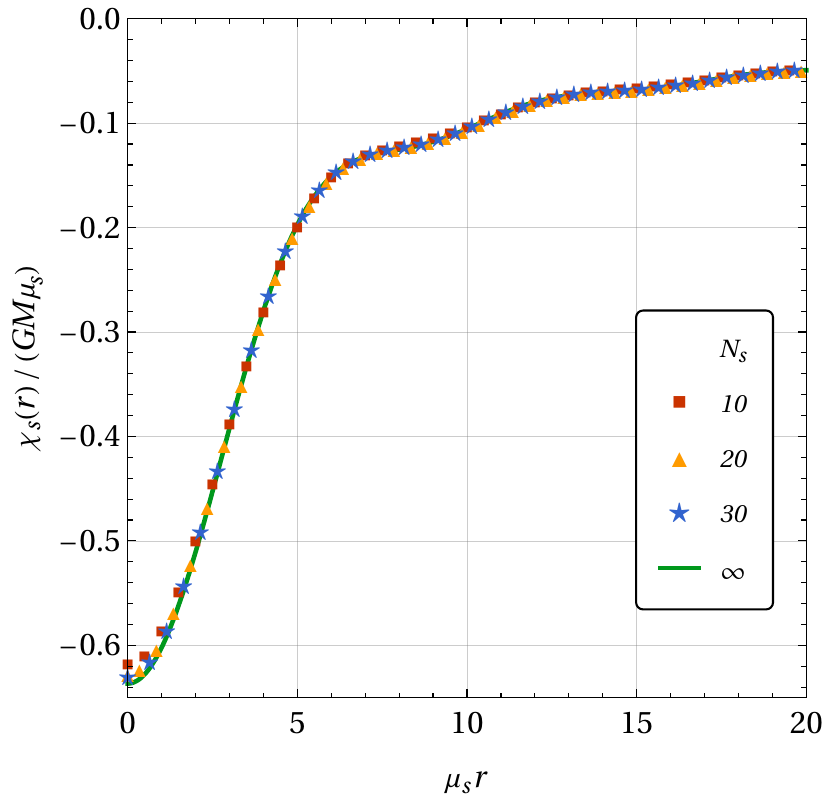}
\caption{Comparison between $\chi_s^\infty(r)$ and $\chi_s(r)$ for $N_s \in \{ 10,20,30 \}$.}
\label{Fig7}
\end{figure}

\begin{figure}[t] 
\includegraphics[width=7.5cm]{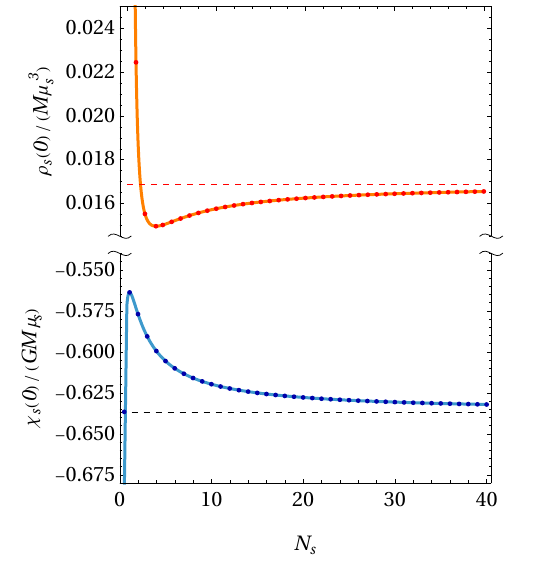}
\caption{Graph of $\rho_s(0)/(M \mu_s^3)$ and $\chi_s(0)/(GM\mu_s)$ as a function of $N_s$. The dots represents $N_s = 1/2$ and integer values of~$N_s$. The horizontal dashed lines correspond to the values $\rho^{\infty}_s (0)/(M \mu_s^3) = 1/(6\pi^2) = 0.0168$ and $\chi^{\infty}_s (0)/(GM\mu_s) = - 2/\pi = - 0.636$, which represent, respectively, an upper bound for $N_s \geqslant 3/2$ and a lower bound for $N_s \geqslant 1/2$. }
\label{Fig8}
\end{figure}

In Fig.~\ref{Fig7} we plot the comparison of Eqs.~\eq{chi_in} and~\eq{chi-GERROR} for some values of $N_s$. It suggests that increasing $N_s$ beyond $N_s=10$ does not significantly modify (especially for large $\mu_s r$) the behavior of the potential, which approaches~$\chi_s^\infty$. In Fig.~\ref{Fig8} we plot the value of the maximum ${\rm max} \, [  \rho_s (r) ] = \rho_s(0)$ [Eq.~\eq{rho_0}] of the effective delta source and the value of $\chi_s(0)$ [Eq.~\eq{chi_0}] for different values of $N_s$ compared to $\rho^{\infty}_s (0)$ and $\chi^{\infty}_s (0)$, respectively. The points where the solid curves cross the horizontal dashed lines correspond to $N_s = 3/2$ for the effective source and $N_s = 1/2$ for the Newtonian potential. Therefore, $\rho^{\infty}_s (0)$ is actually the upper bound to the effective delta sources for $N_s \geqslant 3/2$, while $\chi^{\infty}_s (0)$ is a lower bound to the potential for $N_s \geqslant 1/2$. 

\begin{figure}[t] 
\includegraphics[width=7.5cm]{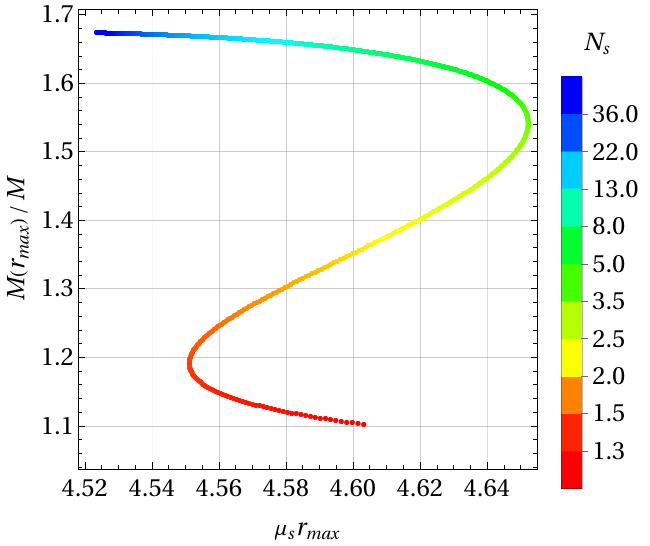}
\caption{Absolute maximum of the dimensionless mass function $M_s(r)/M$ for $N_s > 1$. The horizontal axes represent the position of the mass function's first peak in the coordinates $\mu_s r_\text{max}$, while the vertical axes is $M_s(r_\text{max})/M$. The colors gives the values of $N_s$. Note that $\mu_s r_\text{max}$ increases for $1.505 < N_s < 3.766$, and decreases otherwise.}
\label{Fig9}
\end{figure}

Also, in Fig.~\ref{Fig9} we plot the mass function's first peak (its global maximum) as a function of its position. The value ${\rm max} \, [M_s^\infty(r)/M] = 1.675$ occurs at the position $\mu_s r_{\rm max} = 4.493$ (see also Fig.~\ref{Fig5}) and represents an upper bound for theories with $N_s > 1$ (recall that for $N_s \leqslant 1$ the mass function does not oscillate and has no global maximum, but tends to the supremum $M$ as $r\to\infty$). For $N_s > 3.766$, the position of the first peak decreases when $N_s$ increases.

\begin{figure}[t]
\centering
\begin{subfigure}{.5\textwidth}
\centering
\includegraphics[width=7.5cm]{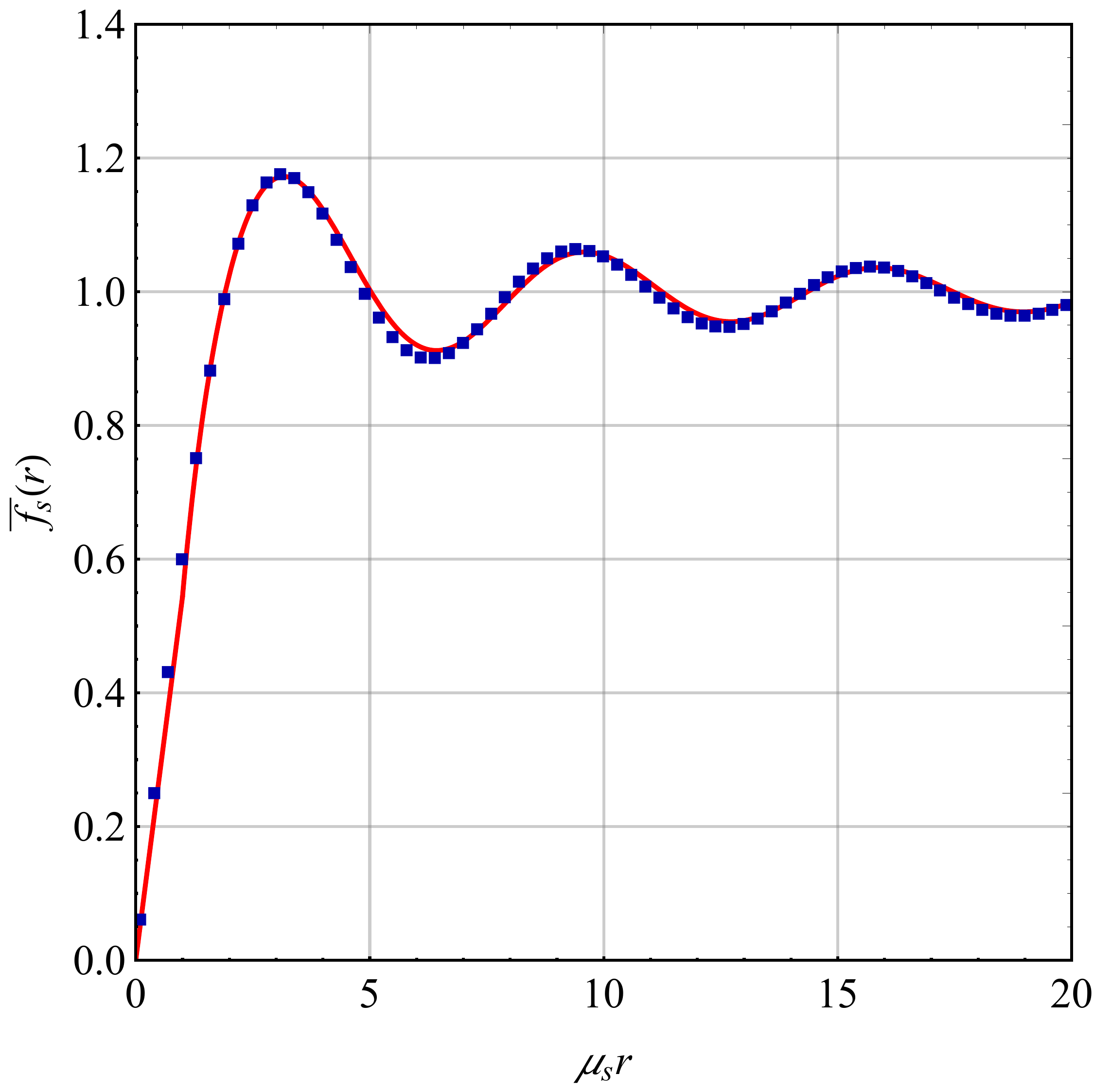}
\caption{}
\end{subfigure}%
\begin{subfigure}{.5\textwidth}
\centering
\includegraphics[width=7.5cm]{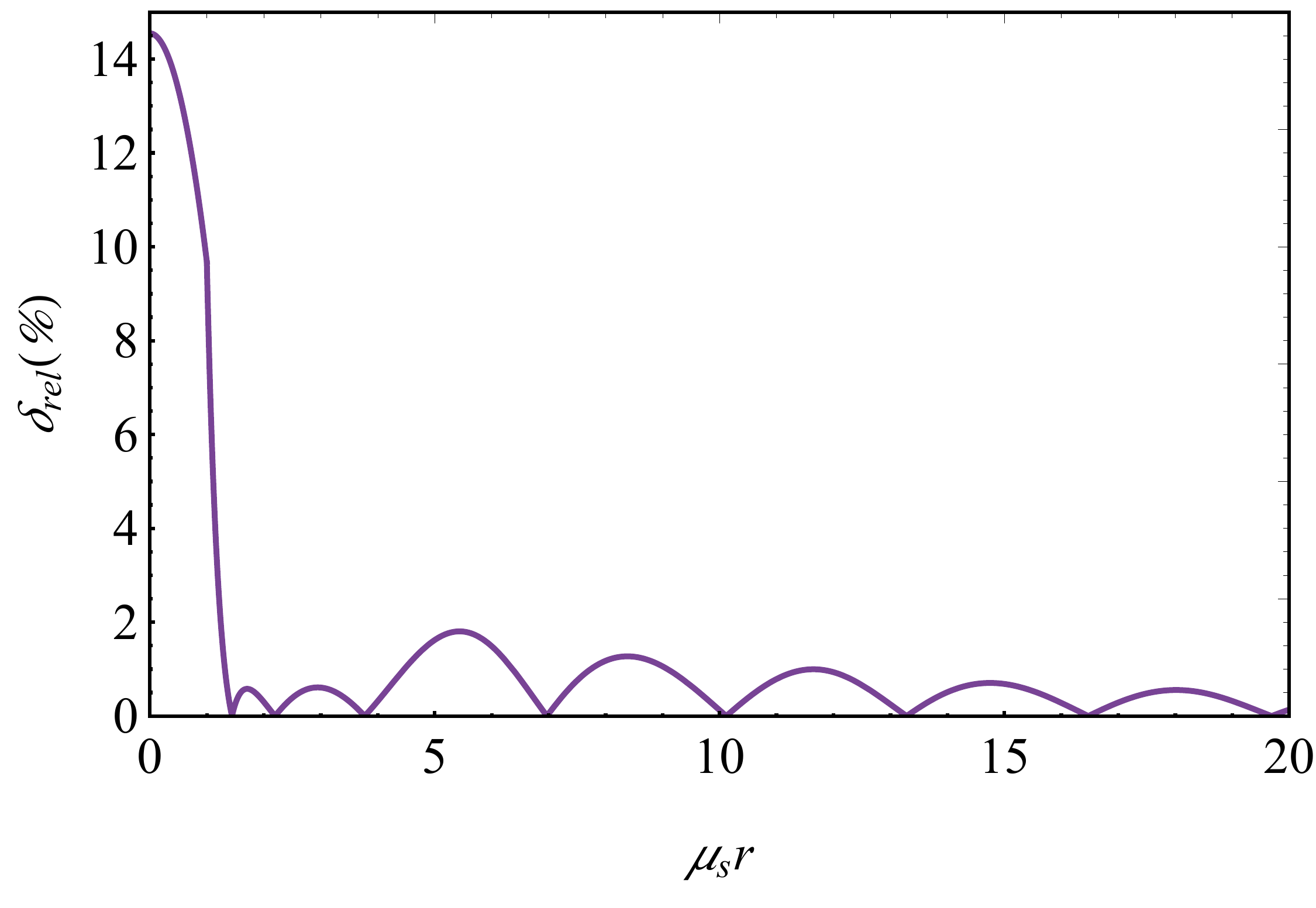}
\caption{}
\end{subfigure}
\caption{{\bf (a)} Plot of the approximation~\eq{h_ap_peri} for $\bar{f}_s(r)$ and the exact solution given by Eq.~\eq{f_exato} (blue squares). {\bf (b)} Relative error between~\eq{h_ap_peri} and~\eq{f_exato}.}
\label{Fig10}
\end{figure}

\begin{figure}[t] 
\includegraphics[width=7.5cm]{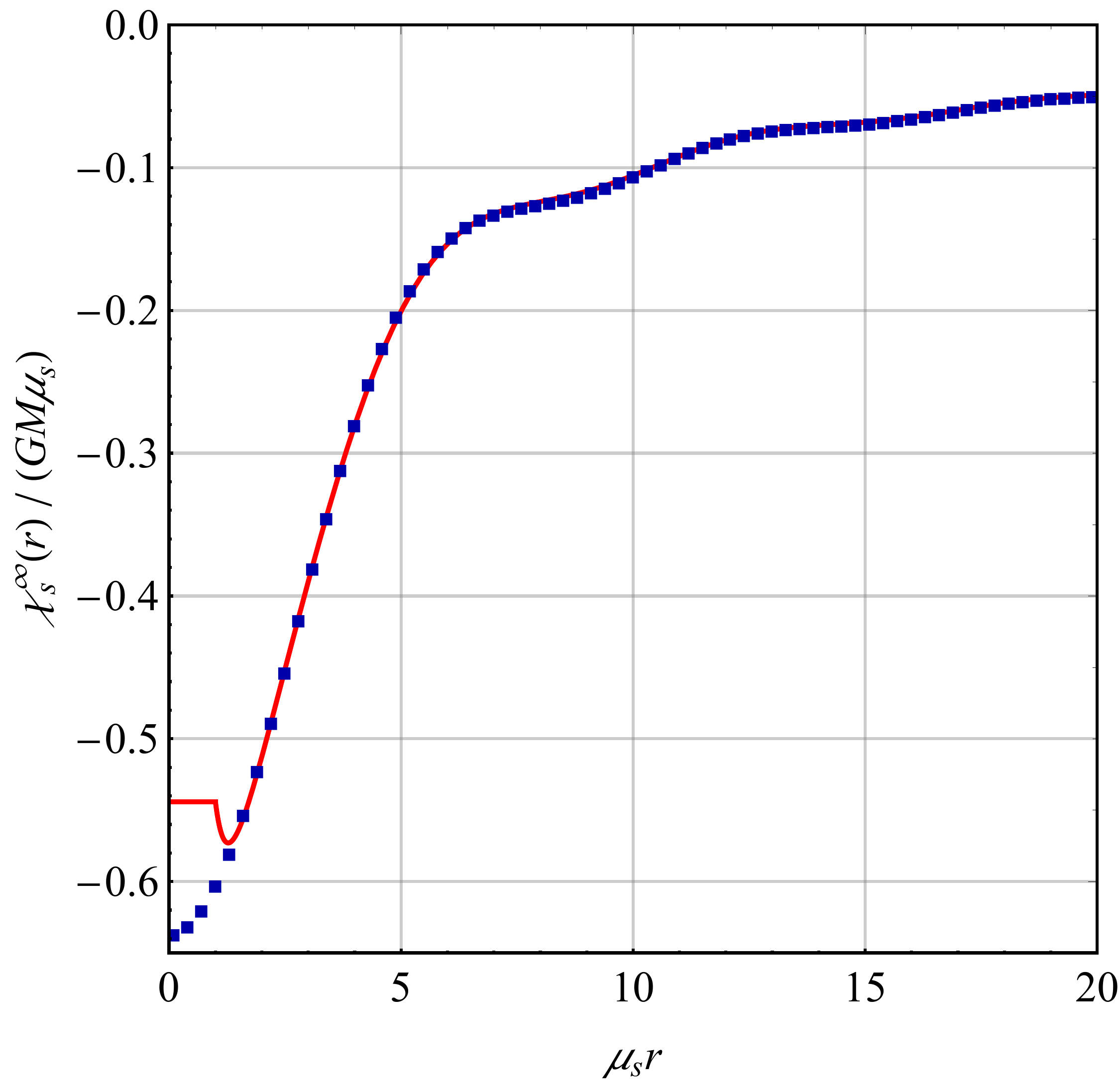}
\caption{Plot of $\chi_s^\infty(r)$. Red line: approximated solution, Eq.~\eq{chi_ap}. Blue squares: exact solution, Eq.~\eq{chi_in}.}
\label{Fig11}
\end{figure}

It is interesting to compare our results with the approximation for large $N_s$ obtained in~\cite{Perivolaropoulos:2016ucs}. Using a parametrization of the potential in the form\footnote{Note that the function $\bar{f}_s(r)$ here is not the effective mass function $M_s(r)$ defined in~\eq{massfunction}. In fact, the effective mass $M_s(r)$ enters equation~\eq{gs_def} for the gravitational field, not the one for the potential.}
\begin{equation}
\n{chi_ap}
\chi_s^\infty (r) = - \frac{GM }{r} \, \bar{f}_s(r)
,
\end{equation}
the following approximation was proposed:
\begin{equation}
\n{h_ap_peri}
\bar{f}_s(r) \approx
\begin{cases}
\alpha_1 {\mu_s r} ,  & \text{if }\quad 0<{\mu_s r}<1
, 
\\
1 + \alpha_2 \frac{ \cos({\th_0 \mu_s r} + \th_1)}{{\mu_s r}} , &\text{if } \quad 1< {\mu_s r} 
,
\end{cases}
\end{equation}
where $\al_1 = 0.544$, $\al_2 = 0.572$, $\th_0 = 1.000$ and $\th_1 = 0.885 \pi$, which was obtained by assuming the specific ansatz~\eq{h_ap_peri}, solving for the parameters $\alpha_{2}$, $\th_0$ and $\th_1$ that provided the best fit to the potential for $N_s=20$ in the interval $1 \leqslant \mu_s r \leqslant 20$, and making a linear approximation for $0 \leqslant \mu_s r \leqslant 1$. Although this approximation seems to work with a relative error\footnote{The relative error $\delta_{\rm rel}$ between an approximated function $\Phi_\text{app.}$ and the exact one is defined as $\delta_{\rm rel} = \vert \Phi_\text{exact} - \Phi_\text{app.}\vert/\Phi_\text{exact}$.} smaller than 0.6\% for $\mu_s r > 15$ with respect to the exact function 
\begin{equation}
\n{f_exato}
\bar{f}_s(r) = \frac{2 \, {\rm Si} (\mu_s r)}{\pi},
\end{equation} 
the errors for small $r$ (where the approximation is linear) can be as large as 14\% (see Fig.~\ref{Fig10}). Furthermore, in Fig.~\ref{Fig11} we compare the approximation for $\chi_s^\infty (r)$ from Eq.~\eq{h_ap_peri}, and its exact solution~\eq{chi_in}. There is a clear mismatch for $\mu_s r < 1$ due the fact that the linear approximation of $\bar{f}_s(r)$ in this range makes the potential to be constant. The relative errors of $\chi_s^\infty(r)$ and $\bar{f}_s(r)$ are basically the same, with a maximum of 14.55\% of relative error for the potential at $r=0$.

\begin{figure}[t] 
\includegraphics[width=7.5cm]{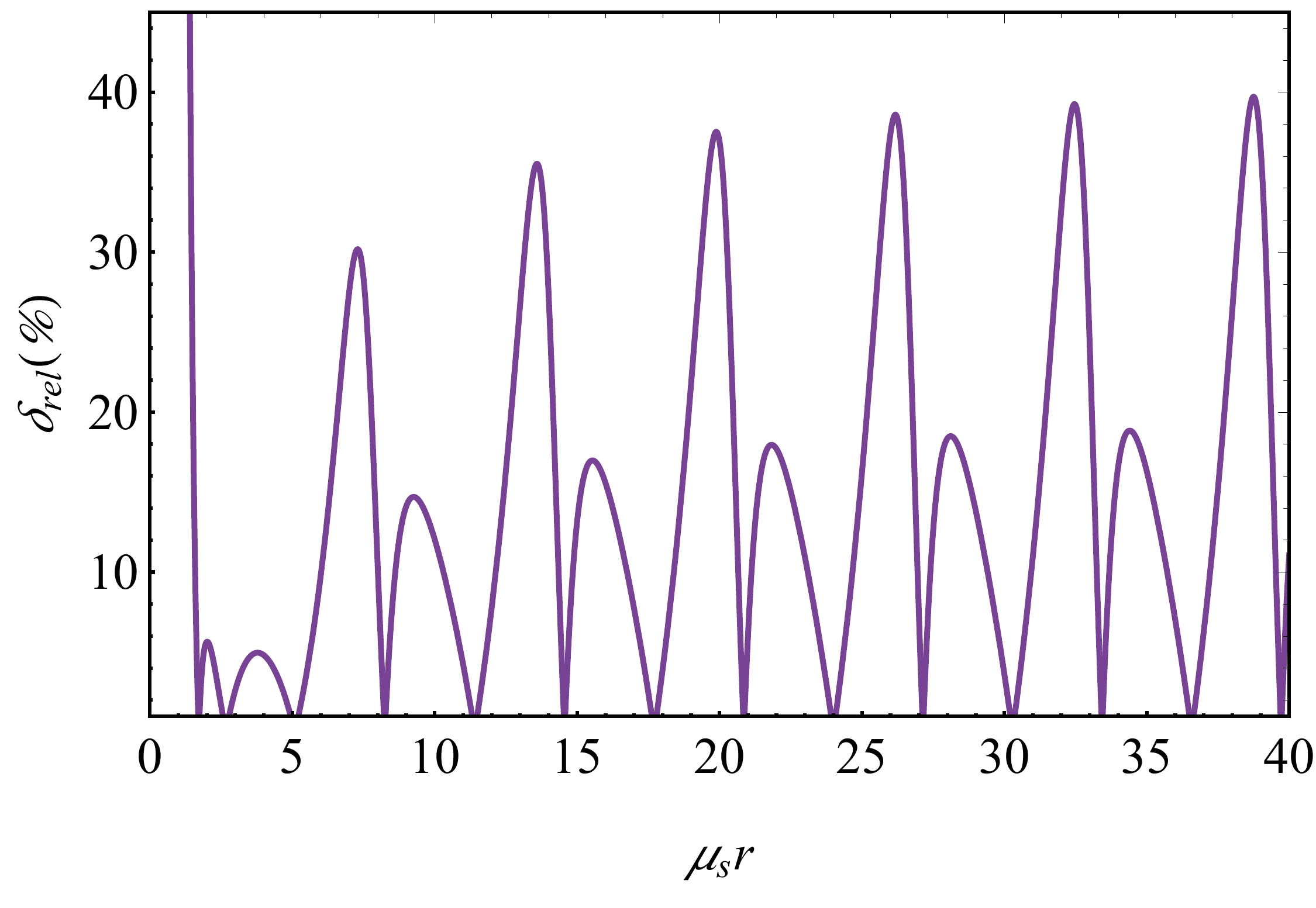}
\caption{Relative error of $\chi_s'(r)$ obtained through the derivative of~\eq{chi_in} (exact expression) and~\eq{chi_ap} (approximation). The error is of $100\%$ for $0\leqslant \mu_s r <1$, as $\chi'_s =0$ for the approximation in this interval. The discontinuity at $\mu_sr=1$ occurs because the approximation is not smooth at that point. The amplitude of variation of the relative error has peaks of around $40\%$, which do not seem to decrease even for $\mu_s r \gg 1$.}
\label{Fig12}
\end{figure}

The differences are more striking when one takes the approximation beyond the estimate of the value of the potential and checks whether it can reproduce other features, such as the characteristic oscillations and their inflection points. For example, the relative errors for $\chi_s'$ derived from~\eq{h_ap_peri} seem to oscillate in the range 0\%--40\%, see Fig.~\ref{Fig12}, with an average relative error of about 17\% that remains approximately constant. This approximation can also be very sensitive to the values of its parameters. For instance, if the parameter $\th_0$ is increased by $0.010$, the relative error for $\chi_s^\prime$ can be as large as 300\% even for large values of $\mu_s r$, as the oscillations in the derivative of~\eq{chi_in} and~\eq{chi_ap} become completely out of phase.\footnote{This is not a fortuitous observation, as the number $\th_0=1.010$ was actually considered in~\cite{Perivolaropoulos:2016ucs} (see Supplemental Material) as the unrounded best fit for the potential with $N_s=20$.} Since the derivative of the potential is related to gravitational force, the modeling of laboratory experiments of Newton's law (e.g., with torsion balances) can be significantly affected by such errors.

\begin{figure}[t]
\centering
\begin{subfigure}{.5\textwidth}
\centering
\includegraphics[width=7.5cm]{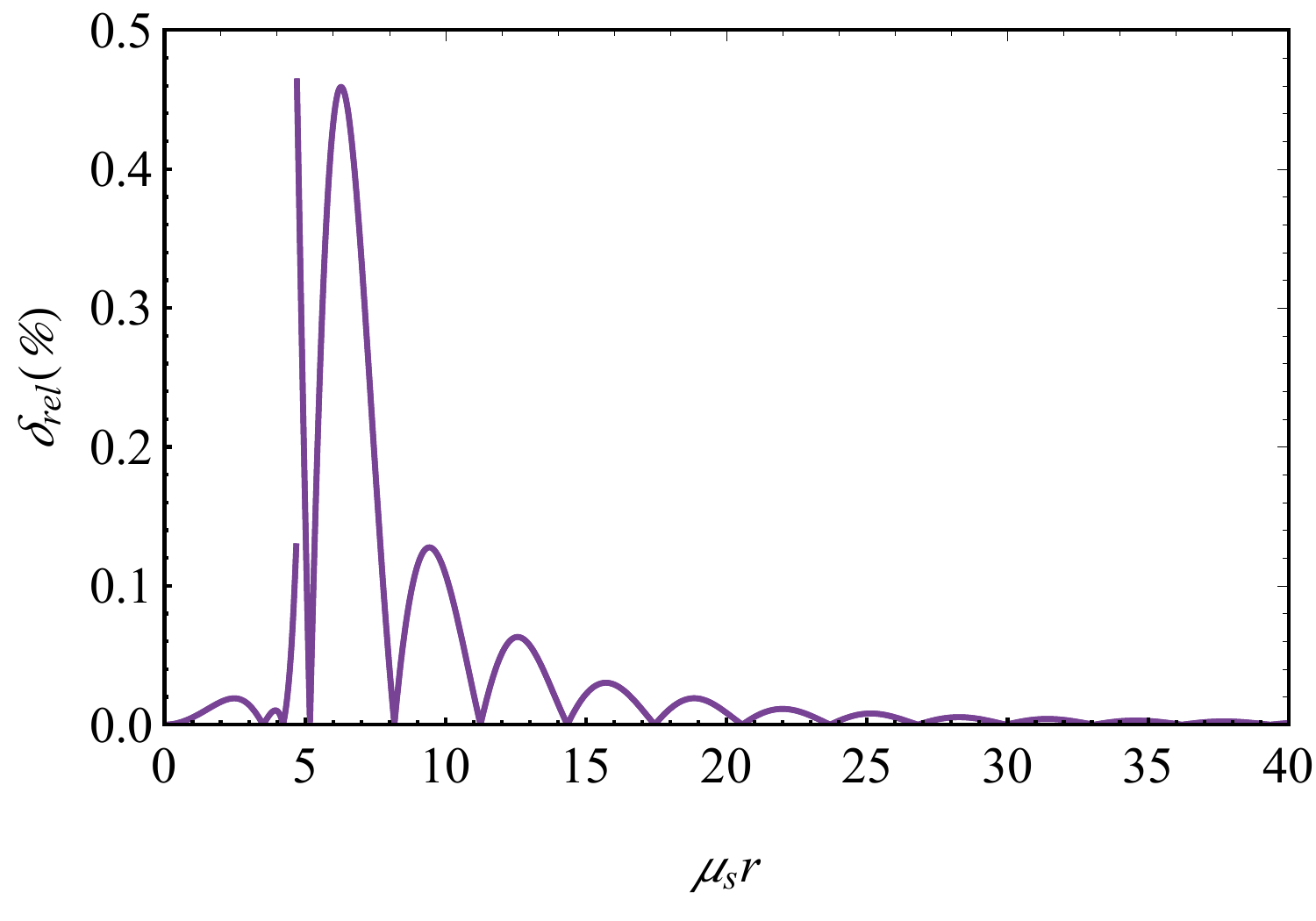}
\caption{}
\end{subfigure}%
\begin{subfigure}{.5\textwidth}
\centering
\includegraphics[width=7.5cm]{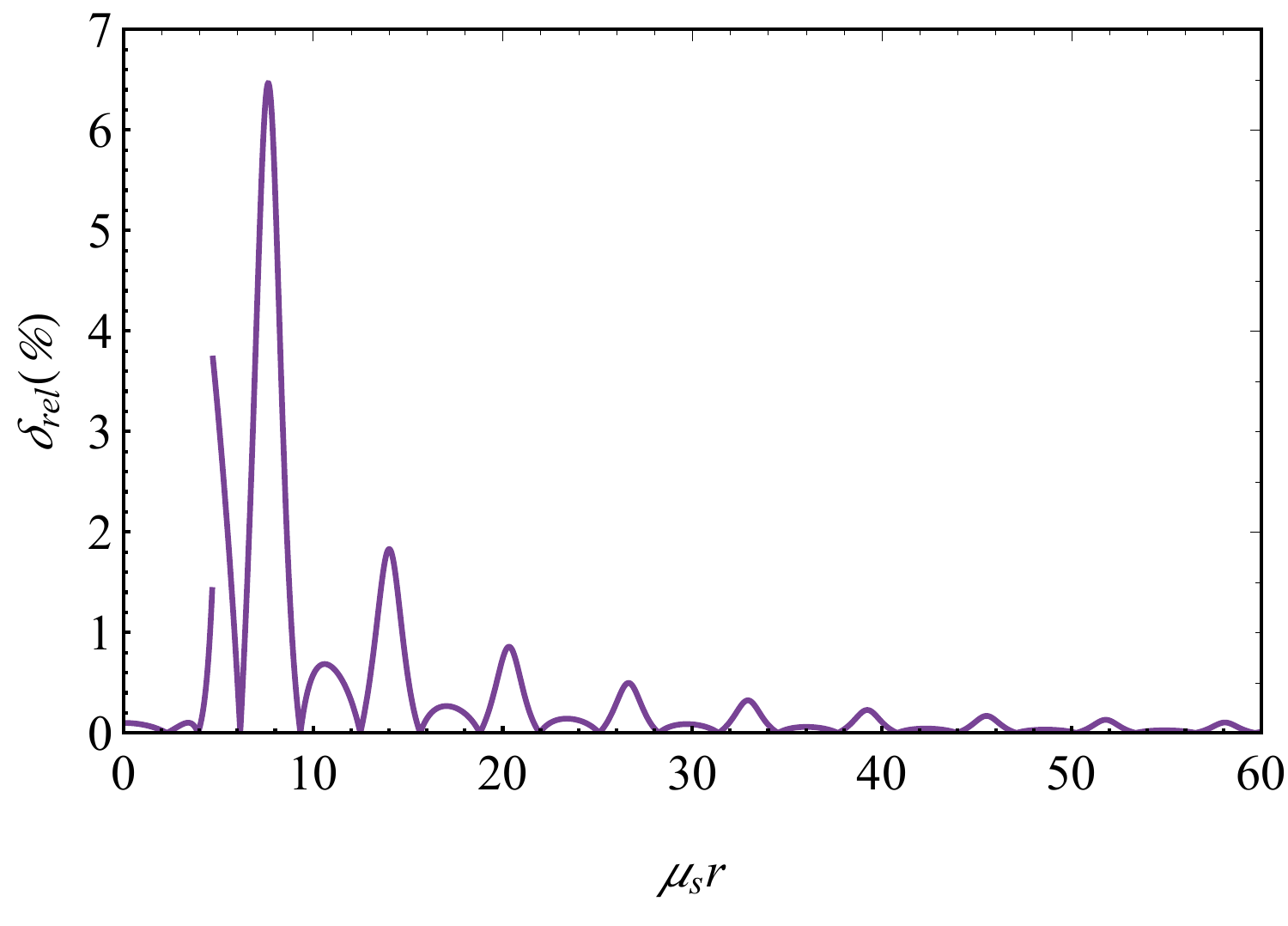}
\caption{}
\end{subfigure}
\caption{Plot of relative error obtained comparing the exact solution, Eq.~\eq{chi_in} and the new approximation, Eq.~\eq{chi_ap2}. {\bf (a)}~Relative error for the potential $\chi_s^\infty (r)$.  {\bf (b)} Relative error for the derivative of the potential $\chi'_s {}^\infty(r)$. In both cases the amplitude of the relative errors approaches zero for $\mu_s r \gg 1$.}
\label{Fig13}
\end{figure}

Knowing the exact solution, we can derive more accurate closed-form approximations to the potential that might simplify phenomenological applications. For example, in the regime $\mu_s r \gg 1$ one can use~\cite{NIST}
\begin{equation}
\n{apprSi}
{\rm Si} \, (z) = \frac{\pi}{2} - \frac{\cos z}{z} \left[ 1 + O(z^{-2}) \right]  - \frac{\sin z}{z^2} \left[ 1 + O(z^{-2}) \right]
, 
\qquad z \to \infty 
.
\end{equation}
The truncation  involving only the terms explicitly written in~\eq{apprSi}, namely,
\begin{equation}
\n{chi_ap_nossa}
\chi_s^\infty(r) \approx - \frac{GM}{r} \left[  1 - \frac{2\cos (\mu_s r)}{\pi \mu_s r} - \frac{2\sin (\mu_s r)}{\pi (\mu_s r)^2} \right]  
, 
\qquad \mu_s r \gg 1
,
\end{equation}
gives a relative error smaller than 2\% of the exact potential already for $\mu_s r > 2.1$, and smaller than 0.5\% for $\mu_s r > 4.68$. An approximation for small values of $\mu_s r$ can be obtained from the Taylor series of ${\rm Si}(z)$ or, better, with a Pad\`e approximant. For example,
\begin{equation}
\n{apprSiPade}
{\rm Si} \, (z) \approx z \frac{1+ \sum \limits_{i=1}^4 a_i z^{2i}}{1+\sum \limits_{i=1}^3 b_i z^{2i}} 
\end{equation}
(with coefficients $a_i$ and $b_i$ shown in Table~\ref{Bre-Pade}) yields an approximation for the potential with relative error smaller than 1\% for $\mu_s r < 5.3$. Combining these two approximations, we obtain
\begin{equation}
\n{chi_ap2}
\chi_s^\infty(r) \approx - \frac{GM}{r} \times
\begin{cases}
\frac{2\mu_s r}{\pi} \frac{1 + a_1 (\mu_s r)^{2} + a_2 (\mu_s r)^{4}+ a_3 (\mu_s r)^{6}+ a_4 (\mu_s r)^{8}}{1 + b_1 (\mu_s r)^{2} + b_2 (\mu_s r)^{4}+ b_3 (\mu_s r)^{6}}  
,  
& \text{if }\quad 0<{\mu_s r}<4.7
, 
\\
1 - \frac{2\cos (\mu_s r)}{\pi \mu_s r} - \frac{2\sin (\mu_s r)}{\pi (\mu_s r)^2} 
, 
&\text{if } \quad {\mu_s r} > 4.7 
,
\end{cases}
\end{equation}
with a relative error smaller than 0.5\% for all values of $r$. Moreover, the approximation~\eq{chi_ap2} reproduces $\chi_s'$ with a maximal relative error of 6.5\% (for $\mu_s r = 7.6$) that decreases to less than 1\% for $\mu_s r > 14.7$; although the error oscillates, its amplitude tends to zero, see~Fig.~\ref{Fig13}.

\begin{table}[t]
\begin{tabular}{|c|c|c|c|c|c|c|}
\hline
$a_1$ & $a_2$ & $a_3$ & $a_4$ & $b_1$ & $b_2$ & $b_3$ \\ 
\hline
$-4.543
\times 10^{-2}$ &
$1.154
\times 10^{-3}$ & 
$-1.410
\times 10^{-5}$  & 
$9.432
\times 10^{-8}$  &
$1.011
\times 10^{-2}$ &
$4.991
\times 10^{-5}$ &
$1.565
\times 10^{-7}$ \\
\hline
\end{tabular}
\caption{Values of the coefficients of the Pad\`e approximant~\eq{apprSiPade}~\cite{Rowe:2014cza}.}
\label{Bre-Pade}
\end{table}

\section{Curvature regularity}
\label{Sec7}

One of the well-understood aspects of Newtonian-limit solutions in generic higher-derivative and nonlocal gravity is the issue regarding the resolution of the singularity at $r=0$ (see, e.g.~\cite{BreTibLiv} and references therein). In particular, since for $N_s >0$ the form factors considered here grow faster than any polynomial, there are general theorems that guarantee that the Newtonian potential is bounded, as well as all the curvature invariants that are polynomial in the Riemann tensor and its derivatives~\cite{Nos6der}. A sufficient condition for the regularity of the curvature-derivative invariants is that the potential is bounded and is an even analytic function of $r$ (around $r=0$)~\cite{Nos6der}---which is the case of~\eq{chi-series}, valid for $N_s \geqslant 1/2$. For the sake of completeness, in this short section we revisit the topic of singularity resolution. Also, we refer the interested reader to the works~\cite{Frolov:2015usa,Edholm_NewPot,BreTib2}, where aspects of this problem were considered for the first time in the context of GF$_N$ gravity models for $N \in \mathbb{N}$.

The regularity of the curvature invariants can be studied in a unified way by noticing that any Riemann-polynomial curvature invariant can be written as contractions of the Riemann tensor components $R^{\mu\nu} {}_{\al\be}$ using only Kronecker~deltas~\cite{Bronnikov:2012wsj}. For the metric~\eq{metric-new}, $R^{\mu\nu} {}_{\al\be}$ has four independent components, namely,
\begin{subequations}
\n{oskas}
\begin{align}
&
K_{1} \equiv R^{t r}{}_{tr} = \ph''( r ) = \frac43 \chi_2''(r) - \frac13 \chi''_0 (r) 
, 
\\ 
&
K_{2} \equiv R^{t \th}{}_{t \th} = \frac{\ph' ( r )}{r} = \frac43 \, \frac{\chi'_2 (r)}{r} - \frac13 \, \frac{\chi'_0 (r)}{r}
,
\\
&
K_{3} \equiv R^{r \th}{}_{r \th} = - \psi''(r) - \frac{\psi'(r)}{r} = -  \frac23 \, \left[ \chi''_2 (r) + \frac{\chi'_2 (r)}{r} \right] - \frac13 \, \left[ \chi''_0(r) + \frac{ \chi'_0(r)}{r} \right]
,
\\ 
&
K_{4} \equiv R^{\theta \phi}{}_{\theta \phi} = -\frac{\psi' ( r )}{r} = -  \frac23 \, \frac{\chi'_2 (r)}{r} - \frac13 \, \frac{ \chi'_0(r)}{r}
.
\end{align}
\end{subequations}
For example, the linearized Kretschmann scalar can be expressed as 
\begin{equation}
\n{Kre}
K \equiv R_{\al\beta \la \tau} R^{\al\be\la\tau} = 4 (K_1^2 + 2 K_2^2 + 2 K_3^2 +  K_4^2)
.
\end{equation}
If all $K_{i}$ are regular, then all curvature invariants constructed from the contractions of the Riemann tensor are bounded as well. 

This is clearly the case for $N_s \geqslant 1/2$: Since the potential $\chi_s(r)$ is bounded and is an analytic even function of $r$, it follows trivially from~\eq{chi-series} and the Taylor's theorem that 
\begin{equation}
\n{chi-d0}
\frac{\rd^{2\ell-1} }{\rd r^{2\ell-1} }\chi_s ( r ) \big|_{r=0} = 0
, 
\qquad \qquad 
\frac{\rd^{2\ell}}{\rd r^{2\ell}} \chi_s ( r ) \big|_{r=0} = - \frac{G M \mu_s^{2\ell+1}}{\pi N_{s}}  \frac{(-1)^\ell}{2\ell+1}  \Gamma \left( \frac{2\ell+1}{2N_s} \right)
.
\end{equation}
Therefore, all quantities $K_{1,\ldots,4}$ are bounded and even. Moreover, according to the general theorem\footnote{For a recent generalization to the nonlinear regime, see~\cite{Antonelli:2025zxh}.} proven in~\cite{Nos6der}, since all odd derivatives of $\chi_s (r)$ vanish at $r=0$, while the even ones are finite, not only all Riemann-polynomial curvature invariants are regular but also the scalars constructed with covariant derivatives of the curvatures, such as $\cx^\ell R$, $ (\na_\al \na_\be \na_\ga \na_\de R_{\mu\nu\rho\si})^2$, $R_{\mu\nu\al\be} \cx^\ell  R^{\mu\nu\al\be}$ ($\ell \in \mathbb{N}$) and so on. 
 
It is more subtle to explicitly check the regularity of the curvature invariants for $0<N_s < 1/2$ because the potential~\eq{chi-GERROR} is non-analytic (see, e.g., the discussion in~\cite{Burzilla:2020bkx,Nos6der}). In fact, in this case, the GEF is defined through~\eq{GEF-32-int} that yields the integral representation~\eq{chi-integral} of the potential. This expression can still be used to evaluate the derivatives of the potential as $r\to0$ since the smoothness of the effective source allows one to differentiate  and pass the limit inside the integral. By doing so, it is possible to prove that Eq.~\eq{chi-d0} also holds in this case. This result can be traced back to the fact that the series~\eq{I-series} converges at $r=0$ even for $0 < N_s < 1/2$, see the discussion in Appendix~\ref{ApC}. Hence, for all $N_s > 0$ the nonlocal theory with form factor~\eq{formfactor} has a complete regularization of Newtonian singularities. As discussed in~\cite{Nos6der,BreTibLiv}, this physical behavior is explained because, in the UV, the propagator~\eq{propagator} is suppressed faster than any polynomial.

\begin{figure}[t] 
\includegraphics[width=7.5cm]{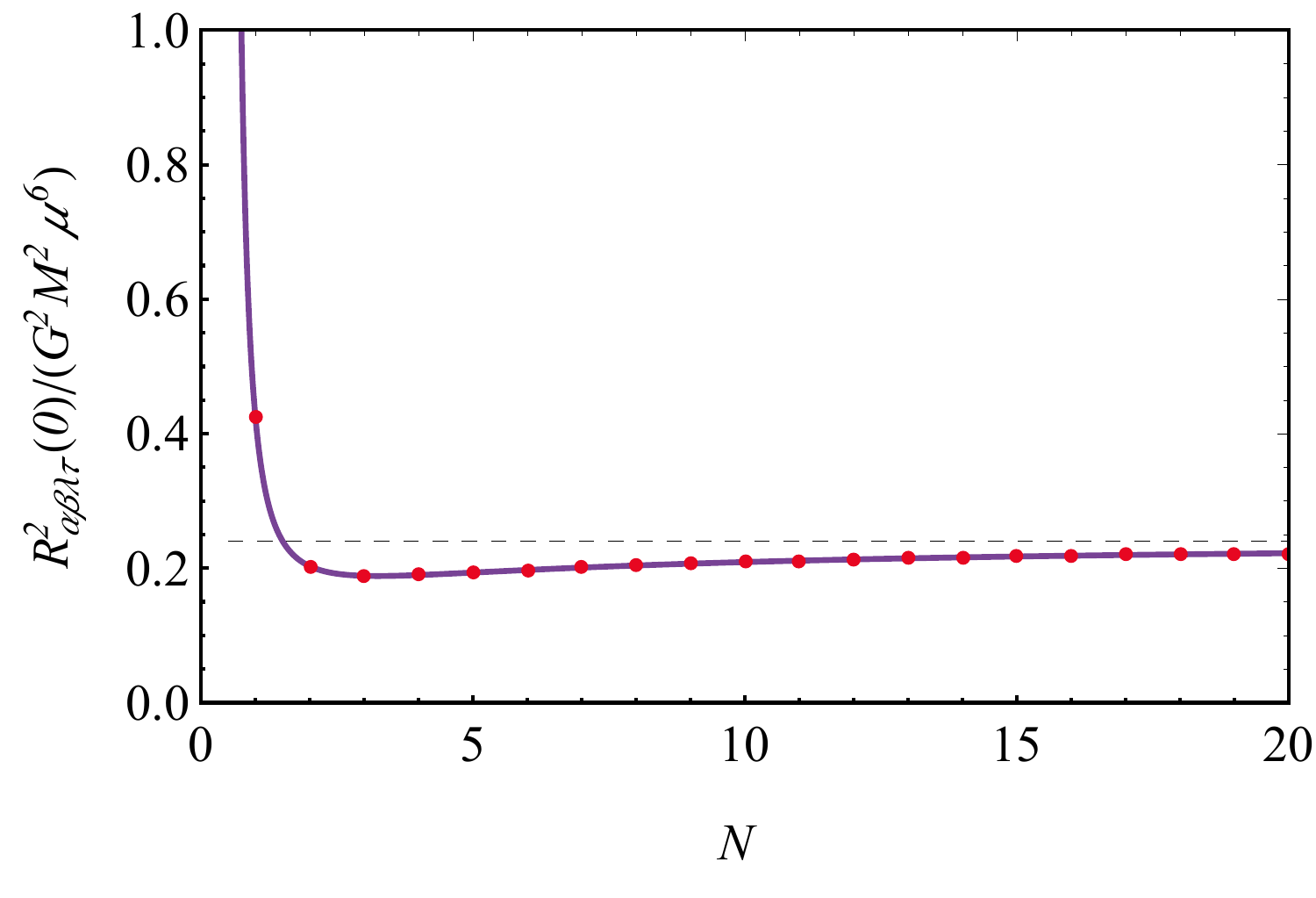}
\caption{Plot of the value dimensionless Kretschmann invariant $K(0)/(G^2M^2\mu^6)$ at $r=0$ as a function of $N$, for the particular case where $\mu_2 = \mu_0 \equiv \mu$ and $N_2 = N_0 \equiv N$. The dashed line corresponds to the value when $N \to \infty$, $K^\infty  (0)$, which serves as an upper bound for high values of $N$. The dots represent the integer values of $N$. Notice that when $N = 3/2$, $K (0) = K^\infty  (0)$.}
\label{Fig14}
\end{figure}

As an explicit example, let us comment on the Kretschmann invariant at $r=0$. To simplify the discussion, consider the case of the GF$_N$ model with $\mu_2 = \mu_0 \equiv \mu$ and $N_2 = N_0 \equiv N$, so that $\ph = \psi = \chi_2 = \chi_0$. Hence, using~\eq{Kre} and~\eq{chi-d0} we obtain
\begin{equation}
\n{Kre0}
\frac{K(0)}{G^2 M^2 \mu^6} = \frac{16 }{3 \pi ^2 N^2} \, \Gamma \left(\frac{3}{2 N}\right)^2
,
\qquad 
N > 0
,
\end{equation}
whereas for $N \to \infty$ we have
\begin{equation}
\n{KreInf}
\frac{K^\infty(0)}{G^2 M^2 \mu^6} = \frac{64 }{27 \pi ^2}
.
\end{equation}
In Fig.~\ref{Fig14} we plot the value of the Kretschmann scalar at $r=0$ as a function of $N$. Interestingly, the plot shows that $K^\infty(0)$ works as an upper bound for this scalar in all models with $N \geqslant 3/2$; for $N_s = 3/2$ the Eqs.~\eq{Kre0} and~\eq{KreInf} give exactly the same value.

Finally, to highlight the difference between nonlocal modifications in the UV and the IR, let us briefly comment on the more exotic scenario with $N_s < 0$, mentioned in footnote 5. In this case, the form factor~\eq{formfactor} becomes
\begin{equation}
f_s (-k^2) = e^{(\mu_s^2/k^2)^{\vert N_s\vert} },
\end{equation}
which tends to a constant when $k \to \infty$. Therefore, the UV behavior of the propagator~\eq{propagator} is not improved (compared to GR), which means, according to the theorems of~\cite{Nos6der}, that there can be curvature singularities. As an explicit example, let us consider the case of $N_s = - 1$, which was proposed in~\cite{Conroy:2014eja}. The direct integration of~\eq{chi-fo} then yields
\begin{equation}
\begin{split}
\chi_s (r) &= -\frac{2 G M }{\pi  r} \left[\frac{\pi}{2} \, {}_0 F_2\left(-;\frac{1}{2},1;\frac{ \mu_s^2 r^2}{4}\right)-\sqrt{\pi} \mu_s  r \, {}_0 F_2\left(-;\frac{3}{2},\frac{3}{2};\frac{\mu_s^2 r^2 }{4}\right)\right]
\\
& = -\frac{G M}{r} + \frac{2 G M \mu_s}{\sqrt{\pi }} -\frac{G  M \mu_s^2}{2} \, r 
+ O(r^2)
.
\end{split}
\end{equation}
Thus, the potential is not bounded for $N_s = - 1$, the terms in~\eq{oskas} diverge, and there is a curvature singularity at $r=0$.


\section{One-loop logarithmic quantum corrections to the Newtonian limit}
\label{Sec8}

In this section, we investigate general properties of the Newtonian limit of nonlocal gravity including one-loop quantum corrections. Typically, quantum corrections have a universal structure in both the ultraviolet (UV) and infrared (IR) regimes in the form of leading logarithms (see the discussion in~\cite{dePaulaNetto:2021axj} and references therein),
\begin{equation}
\n{EA}
\begin{split}
{\Ga}^{(1)} = & \,\, \frac{1}{16 \pi G} \int \rd^4 x \sqrt{-g} 
\left\{  
(R - 2 \La) + \frac12 C_{\mu\nu\al\be} F_2 (\cx) C^{\mu\nu\al\be} - \frac16 R F_0 (\cx) R 
\right.
\\
&
\left.
\hspace{1cm}
-  \frac{\beta_2}{2} C_{\mu\nu\al\be} \log (-\cx/\mu_R^2) C^{\mu\nu\al\be}
+ \frac{\beta_0}{6} R \log (-\cx/\mu_R^2) R
\right\} + O(R^3)
,
\end{split}
\end{equation}
where $\mu_R$ is a mass scale and, for the sake of generality, in the following considerations we shall treat $\be_{0,2}$ as arbitrary parameters.\footnote{The values of $\beta_{2,0}$ may change in the UV and IR limits due to the decoupling theorem of Applequist--Carazzone~\cite{AC}. At intermediate scales, the one-loop quantum corrections are given by complicated form factors that depend on the mass of matter fields~\cite{Gorbar:2002pw, Gorbar:2003yt, Franchino-Vinas:2018gzr, Franchino-Vinas:2019upg}. Nonetheless, the contributions of massive fields get suppressed in the IR, in such a way that only logarithmic contributions of massless fields remain in this regime.}    

To derive the classical Newtonian-limit field equations from the action~\eq{EA}, one can apply the same procedure of Sec.~\ref{Sec2}, but now with the replacement $f_s(\cx) \to f_s(\cx) - \beta_s \log(-\cx/\mu_R^2)$. The result is that, for the action~\eq{EA}, the equations for the spin-$s$ potentials have the form 
\begin{equation}
\n{poi-chi-mod-quantum}
f_s (\tri) \tri \chi_s -  \beta_s \log (-\tri / \mu_R^2) \tri^2 \chi_s = 4 \pi G \rho
, 
\qquad \qquad
s = 0,2
.
\end{equation} 
To solve this equation we perform the loop expansion of the potential,
\begin{equation}
\chi_s = \chi_s^{(0)} + \chi_s^{(1)} + O(\hbar^2),
\end{equation}
where $\chi_s^{(l)}$ is of the order $O(\hbar^l)$. Since $\beta_s = O (\hbar)$, the equations at zero and first orders in $\hbar$ read
\begin{align}
&
\n{Q1}
f_s (\tri) \tri \chi_s^{(0)} = 4 \pi G \rho, 
\\
&
\n{Q2}
f_s (\tri) \tri \chi_s^{(1)} = \beta_s \log ( -\tri/ \mu_R^2) \tri^2 \chi_s^{(0)}
.
\end{align}
Equation~\eq{Q1} defines the classical part of the potential, and it is nothing else but~\eq{poi-chi-mod}, which was already solved in Sec.~\ref{Sec2}. On the other hand, Eq.~\eq{Q2} represents the one-loop quantum correction to the potential, which is the main subject of this section. 

Following~\cite{Nos6der}, one can solve the system \eq{Q1}, \eq{Q2} by means of the Fourier transform method. This procedure gives the solution for the one-loop quantum correction to the potential in the form
\begin{equation}
\n{chi-Q-int}
\chi_s^{(1)} (r) = \frac{2 \beta_s GM}{\pi r} \int_0^\infty \rd k \, \frac{k\sin(kr) \log(k^2 /\mu_R^2 )}{[f_s(-k^2)]^2}
.
\end{equation}
Alternatively, one could study the problem using the effective source formalism of Sec.~\ref{Sec2}. In fact, Eq.~\eq{Q2} can be recast as
\begin{equation}
\tri \chi_s^{(1)} = 4 \pi G \rho_s^{(1)}
,
\end{equation}
where
\begin{equation}
\n{source-Q-int}
\rho_s^{(1)} (r) = - \frac{\beta_s M}{2\pi^2 r} \int_0^\infty \rd k \, \frac{k^3\sin(kr) \log(k^2 /\mu_R^2 )}{[f_s(-k^2)]^2}
\end{equation}
is the one-loop quantum correction to the effective source. The one-loop quantum correction to the mass function can then be obtained from~\eq{source-Q-int} through the relation
\begin{equation}
M_s^{(1)}(r) = 4 \pi \int_0^r \rd x \, x^2  \rho_s^{(1)} (x)
.
\end{equation} 
Or,  using~\eq{chi-Q-int},
\begin{equation}
\n{M-de-quant}
M_s^{(1)} (r) = \frac{r^2}{G} \frac{\rd}{\rd r} \chi_s^{(1)} (r)
.
\end{equation} 

Since the integrals in~\eq{chi-Q-int} and~\eq{source-Q-int} are very similar, let us define the master integral  
\begin{equation}
\n{I-quantum}
I^{(1)}_{N,\mu} ( r,n ) = \int_{0}^{\infty} \rd k \, k^n \, e^{-2 ( k/ \mu )^{2N}} \sin ( k r ) \log ( k / \mu_R )
,
\qquad  
n \in \mathbb{N}
\end{equation} 
to study them in a unified way. 
The direct comparison of~\eq{I-quantum} with~\eq{chi-Q-int},~\eq{source-Q-int}, and~\eq{M-de-quant} shows that the one-loop quantum corrections to the effective source, mass function and potential are given, respectively, in terms of $I^{(1)}_{N,\mu}$ by
\begin{equation}
\n{sQ}
\rho_s^{(1)} (r) = 
- \frac{\beta_s M}{\pi^2 r} I^{(1)}_{N_s, \mu_s}(r,3)  
,
\end{equation}
\begin{equation}
\n{mQ}
M_s^{(1)} (r)  = 
\frac{  4 \beta_s M }{\pi}  \left[ r \frac{d}{dr} I^{(1)}_{N, \mu_s}(r,1) - I^{(1)}_{N_s, \mu_s}(r,1) \right]
,
\end{equation}
\begin{equation}
\n{pQ}
\chi_s^{(1)} (r) = 
\frac{4\beta_s GM}{\pi r} I^{(1)}_{N_s, \mu_s}(r,1)
.
\end{equation}

To obtain a solution for~\eq{I-quantum}, we once again employ the Feynman trick of integration. Let 
\begin{equation}
J ( r , \alpha ) \equiv \int_{0}^{\infty} \rd k \, \left( \frac{\mu k}{\mu_R} \right)^{2 \alpha - 1} e^{-2 k^{2N}} \sin ( k r )
.
\end{equation}
Then,
\begin{equation}
\n{eq:IandJ}
I^{(1)}_{N, \mu} (r,n) = \frac{\mu \, \mu_R^n}{2} \frac{\partial }{\partial \al} J ( \mu r, \al ) \Big \vert_{\al = \frac{n+1}{2}}
.
\end{equation}
Using the power series representation of the sine function, we can evaluate $J(r,\al)$ in a way similar to~\eq{basic-I}. By doing so, we find
\begin{equation}
\n{J2}
J ( r, \al ) = \left( \frac{\mu}{\mu_R} \right)^{2 \al - 1} \frac{1}{2^{\frac{\al}{N}}} \, \frac{\sqrt{\pi} \, r}{2^{\frac{1}{2N}+2} N} \sum_{l=0}^{\infty}  \frac{\Gamma \left( \frac{2l + 2 \al + 1}{2N} \right)}{\Gamma \left( \frac{2l+3}{2} \right)} \frac{1}{l!} \left( -\frac{r^2}{2^{\frac{1}{N}+2}} \right)^{l}
,
\end{equation}
where we used the identity $( 2l + 1 ) ! = 2^{2l+1} l! \, \Gamma \left( l + 3/2 \right) / \sqrt{\pi}$ to simplify the factorial $( 2l + 1 ) !$ in the Taylor series of the sine. Formula~\eq{J2} can be cast in a compact form in terms of the Fox--Wright $\Psi$-function (see, e.g., \cite{Wachowski:2018gfr}), 
\begin{equation}
\n{eq:Fox-Wright}
{}_{p} \Psi_{q} [ ( a, A ) ; ( b, B ) ; z ] = \sum_{k= 0}^{\infty} \frac{\prod \limits_{j=1}^{p} \Ga ( a_{j} + A_{j} k )}{\prod \limits_{i=1}^{q} \Ga ( b_{i} + B_{i} k ) } \frac{z^{k}}{k!}
,
\end{equation}
namely,
\begin{equation}
J( r, \al ) =  \left( \frac{\mu}{\mu_R} \right)^{2 \al - 1} \frac{1}{2^{\frac{\al}{N}}} \, \frac{\sqrt{\pi} \, r}{2^{\frac{1}{2N}+2} N} \, {}_{1} \Psi_{1} \left[ \left( \frac{2 \al + 1}{2N}, \frac{1}{N} \right) ; \left( \frac{3}{2}, 1 \right) ;  -\frac{r^2}{2^{\frac{1}{N}+2}}  \right]
.
\end{equation}
Finally, taking the derivative with respect of the parameter $\al$ and using Eq.~\eq{eq:IandJ}, it follows
\begin{equation}
\begin{split} 
\label{eq:Ifinal}
I^{(1)}_{N, \mu} ( r,n ) = & \, \frac{\sqrt{\pi} \mu^{n+2} \, r}{ 2^{\frac{n+2}{2N}+3}N^2} \biggl\{ \left[ N \log \left( \frac{\mu^2}{\mu_R^2} \right) - \log 2 \right] {}_{1} \Psi_{1} \left[ \left( \frac{n+2}{2N}, \frac{1}{N} \right) ; \left( \frac{3}{2}, 1 \right) ; -\frac{\mu^2 r^2}{2^{\frac{1}{N}+2}} \right]
\\
& +\frac{\partial}{\partial a} {}_{1} \Psi_{1} \left[ \left( a, \frac{1}{N} \right) ; \left( \frac{3}{2} , 1 \right) ; - \frac{\mu^2 r^2}{2^{\frac{1}{N}+2}} \right] \bigg \vert_{a=\frac{n+2}{2N}} \biggr\}
.
\end{split}
\end{equation}
Here, the derivative of ${}_{1}\Psi_{1}$ with respect to the parameter $a$ is defined through its Taylor series,
\begin{equation}
\n{eq:PsiWrtA}
\frac{\partial}{\partial a_{}} {}_{1}\Psi_{1} [ ( a_{}, A_{} ) ; ( b_{}, B_{} ) ; z ] = \sum_{k=0}^{\infty} \frac{\Ga ( a_{} + A_{} k ) \psi ( a_{} + A_{} k )}{\Ga ( b_{} + B_{} k )} \frac{z^{k}}{k!},
\end{equation}
where $\psi ( z ) = \rd \ln \Ga(z) /\rd z = \Ga^{\prime} ( z )/\Ga ( z )$ is the digamma function.

\begin{figure}[t]
\centering
\begin{subfigure}{.5\textwidth}
\centering
\includegraphics[width=7.0cm]{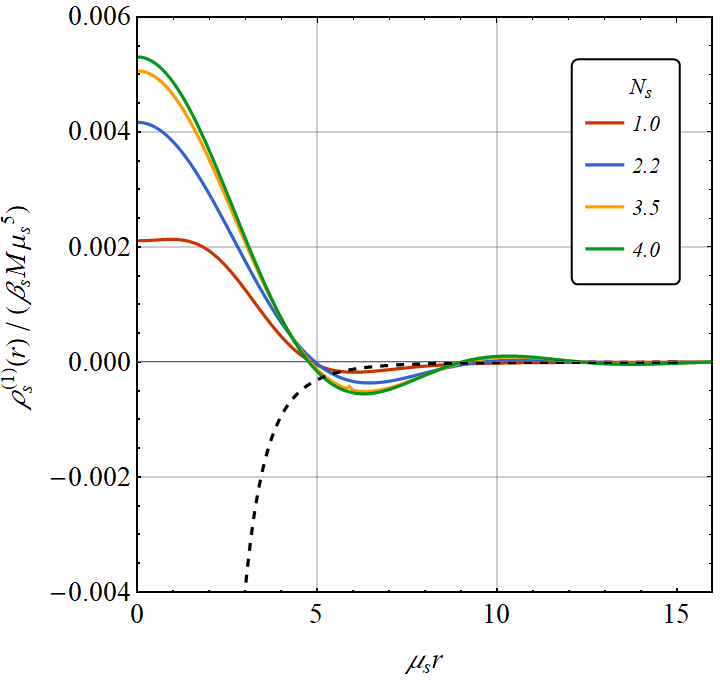}
\end{subfigure}%
\begin{subfigure}{.5\textwidth}
\centering
\includegraphics[width=7.0cm]{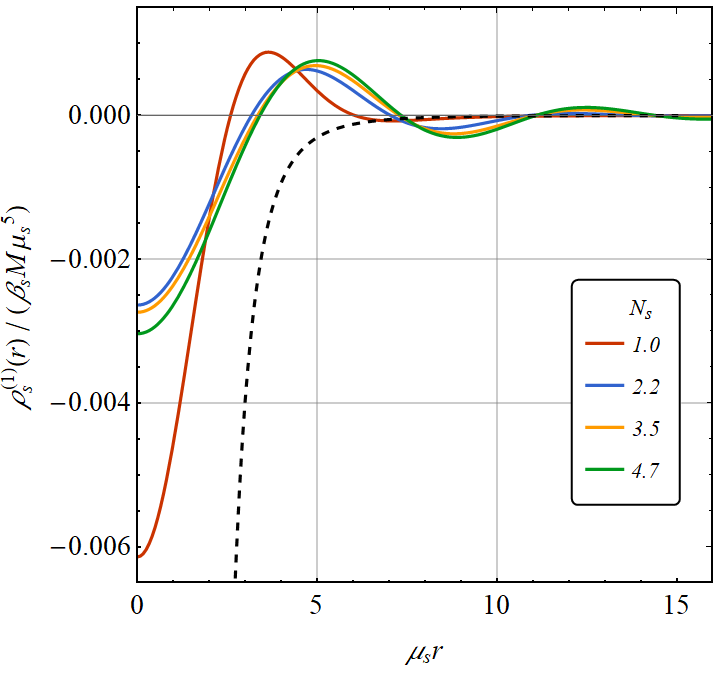}
\end{subfigure}
\caption{$ \rho_s^{(1)} (r) / (\beta_s M \mu_s^5)$ as a function of $\mu_s r$. Left plot: $\mu_R = 1.2 \mu_s$. Right plot: $\mu_R = 0.6 \mu_s$.}
\label{Fig15} 
\end{figure}

\begin{figure}[t]
\begin{subfigure}{.5\textwidth}
\centering
\includegraphics[width=7.0cm]{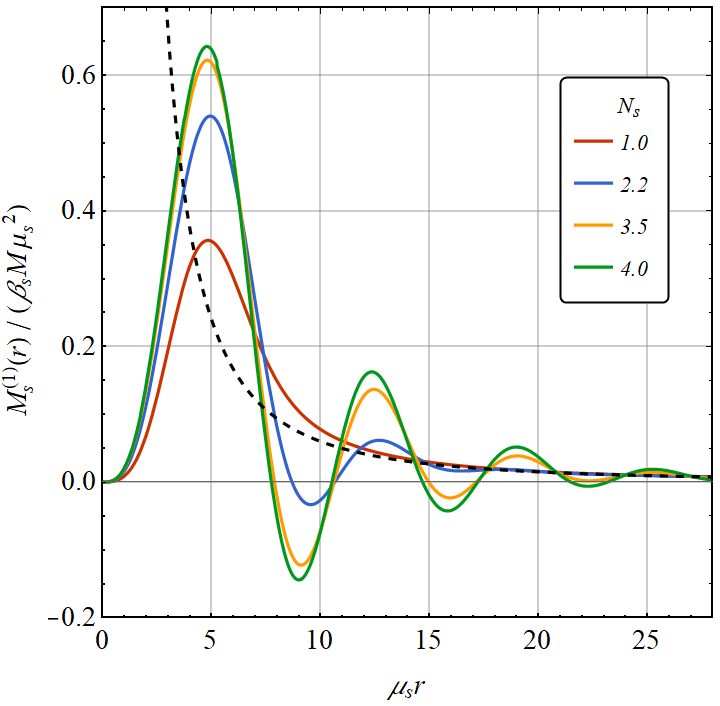}
\end{subfigure}%
\begin{subfigure}{.5\textwidth}
\centering
\includegraphics[width=7.0cm]{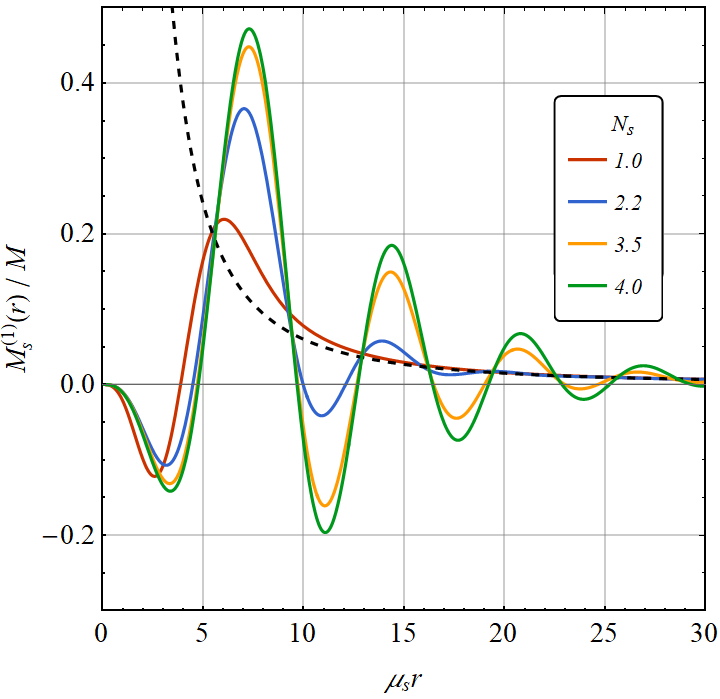}
\end{subfigure}
\caption{$M_s^{(1)} (r) / (\beta_s M \mu_s^2)$ as a function of $\mu_s r$. Left plot: $\mu_R = 1.2 \mu_s$. Right plot: $\mu_R = 0.6 \mu_s$.}
\label{Fig16} 
\end{figure}

\begin{figure}[t]
\begin{subfigure}{.5\textwidth}
\centering
\includegraphics[width=7.0cm]{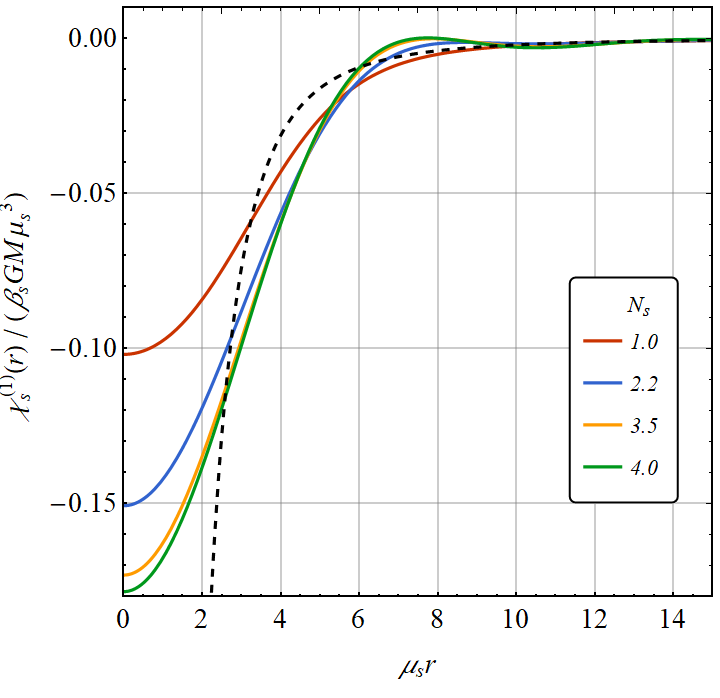}
\end{subfigure}%
\begin{subfigure}{.5\textwidth}
\centering
\includegraphics[width=7.0cm]{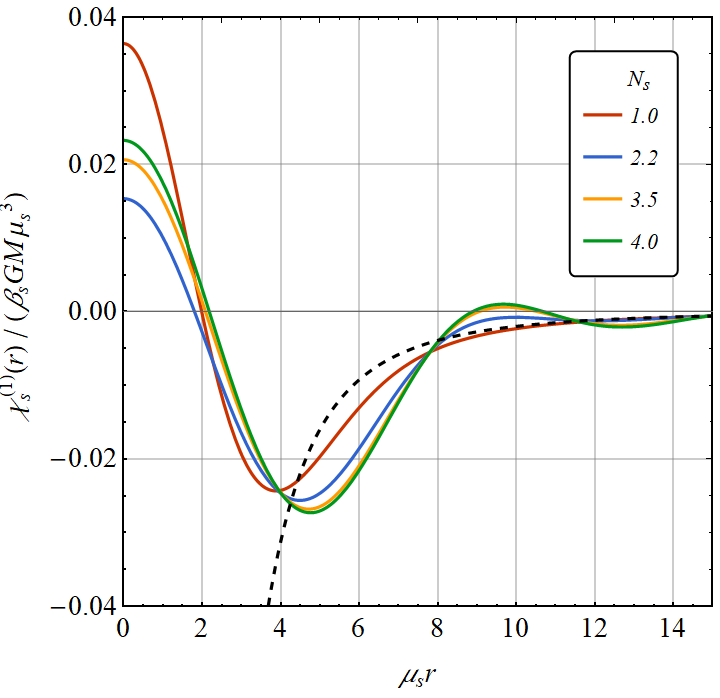}
\end{subfigure}
\caption{$\chi_s^{(1)} (r) / (\beta_s G M \mu_s^3)$ as a function of $\mu_s r$. Left plot: $\mu_R = 1.2 \mu_s$. Right plot: $\mu_R = 0.6 \mu_s$.}
\label{Fig17} 
\end{figure}

With the solution for the integral~\eq{I-quantum} we can obtain the one-loop quantum correction for the effective source, mass function and potential by means of Eqs.~\eq{sQ},~\eq{mQ}~and~\eq{pQ}. In fact,
\begin{equation}
\begin{split} 
\n{rho_quan}
\rho_s^{(1)} (r) = &  \, -\frac{\beta_s M \mu^5_s }{ 2^{\frac{5}{2N_s}+3} \, N^2_s \, \pi^{3/2}} \biggl\{ \left[ N_s \log \left( \frac{\mu_s^2}{\mu_R^2} \right) - \log 2 \right] {}_{1} \Psi_{1} \left[ \left( \frac{5}{2N_s}, \frac{1}{N_s} \right) ; \left( \frac{3}{2}, 1 \right) ; -\frac{\mu^2_s r^2}{ 2^{\frac{1}{N_s} +2 }} \right]
\\
& +\frac{\partial}{\partial a} {}_{1} \Psi_{1} \left[ \left( a, \frac{1}{N_s} \right) ; \left( \frac{3}{2} , 1 \right) ; - \frac{\mu^2_s r^2}{2^{\frac{1}{N_s} + 2}} \right] \bigg \vert_{a=\frac{5}{2N_s}} \biggr\}
,
\end{split}
\end{equation}
\begin{equation}
\begin{split}
\n{mass_quan} 
M_s^{(1)} (r) =
& -\frac{\beta_s M \mu^5_s r^3}{  2^{\frac{5}{2N_s} + 2}  N_s^2 \sqrt{\pi} }\biggl\{ \left[ N_s \log \left( \frac{\mu_s^2}{\mu_R^2} \right) -  \log 2 \right]  
{}_{1} \Psi_{1} \left[ \left( \frac{5}{2N_s}, \frac{1}{N_s} \right) ; \left( \frac{5}{2}, 1 \right) ; -\frac{\mu^2_s r^2}{ 2^{\frac{1}{N_s} +2 }} \right]
\\
& + \frac{\partial}{\partial a} {}_{1} \Psi_{1} \left[ \left( a, \frac{1}{N_s} \right) ; \left( \frac{5}{2}, 1 \right) ; -\frac{\mu_s^2 r^2}{ 2^{\frac{1}{N_s}+2}}  \right] \bigg \vert_{a = \frac{5}{2N_s}} \biggr\}
,
\end{split}
\end{equation}
and
\begin{equation}
\begin{split}
\n{chi_quan}
\chi_s^{(1)} (r) =& \frac{\beta_s GM \mu^3_s}{  2^{\frac{3}{2N_s}+1}  \, N_s^2 \sqrt{\pi}} 
\biggl\{ \left[ N_s \log \left( \frac{\mu_s^2}{\mu_R^2} \right) -  \log 2 \right]  
{}_{1} \Psi_{1} \left[ \left( \frac{3}{2N_s}, \frac{1}{N_s} \right) ; \left( \frac{3}{2}, 1 \right) ; -\frac{\mu^2_s r^2}{ 2^{\frac{1}{N_s} +2 }} \right]
\\
&+ \frac{\partial}{\partial a} {}_{1}\Psi_{1} \left[ \left( a, \frac{1}{N_s} \right) ; \left( \frac{3}{2}, 1 \right) ; -\frac{\mu^2_s r^2}{ 2^{\frac{1}{N_s}+2}} \right] \bigg \vert_{a = \frac{3}{2N_s}} \biggr\}
.
\end{split}
\end{equation}

In Figs.~\ref{Fig15},~\ref{Fig16}, and~\ref{Fig17} we plot, respectively, the solutions~\eq{rho_quan},~\eq{mass_quan}, and~\eq{chi_quan} for some finite values of $N_s$. The dashed black lines represent the far-IR $(\mu_s r \gg 1)$ behavior of the functions, which according to the effective field theory's common lore are given by
\begin{equation}
\n{EFT}
\rho_s^{(1)} (r) \underset{r \to \infty}{\sim} -\frac{3 \beta_s  M}{\pi  r^5}
,
\quad \quad
M_s^{(1)} (r) \underset{r \to \infty}{\sim} \frac{6 \beta_s  M}{r^2}
,
\quad \quad
\chi_s^{(1)} (r) \underset{r \to \infty}{\sim} - \frac{2 \beta_s  G M}{r^3}
.
\end{equation}
Note that these asymptotic behaviors are universal in the sense that they do not depend on the underlying nonlocal model, that is,~\eq{EFT} is independent of~$N_s$, $\mu_s$, and $\mu_R$.\footnote{To derive the expressions in~\eq{EFT} it is enough to make the change of variables $k = q/r$ in the integral~\eq{I-quantum} and notice that $e^{-2(q/\mu r)^{2N}} \longrightarrow  1$, for $\mu r \gg 1$. Also, the term proportional to $\log (\mu_R)$ does not contribute in this limit, as the result of the corresponding integral is a delta function that vanishes for $r \neq 0$. For more details see, e.g,~\cite{Nos6der} and references therein.} 

The plots illustrate that the sign of the one-loop quantum corrections for small values of $\mu_s r$ can change depending on the ratio $\mu_s/\mu_ R$ between the renormalization and nonlocality scales. This change of sign happens roughly for $\mu_s \approx \mu_R$, where $\log(\mu_s^2/\mu_R^2)$ flips sign. To be more precise, for $\mu_s r \ll 1$ we have that
\begin{equation}
\rho_s^{(1)} (r) = - \frac{\beta_s M \mu_s^5}{2^{\frac{5}{2 N_s}+2}  N_s^2 \pi^2 } \Gamma \left(\frac{5}{2 N_s}\right)   \left[N_s \log \left(\frac{\mu_s^2}{\mu_R^2} \right) - \log 2 + \psi \left(  \frac{5}{2 N_s} \right) \right] 
+ O  ( \mu_s^2r^2 )
,
\end{equation}
\begin{equation}
M_s^{(1)} (r) = - \frac{3\beta_s M \mu_s^2}{2^{\frac{5}{2 N_s}}  N_s^2 \pi } \Gamma \left(\frac{5}{2 N_s}\right)   \left[N_s \log \left(\frac{\mu_s^2}{\mu_R^2} \right) - \log 2 + \psi \left(  \frac{5}{2 N_s} \right) \right] (\mu_s r)^3
+ O  (\mu_s^5 r^5 )
\end{equation}
and
\begin{equation}
\chi_s^{(1)} (r) =  \frac{\beta_s G M \mu_s^3}{2^{\frac{3}{2 N_s}}  N_s^2 \pi } \Gamma \left(\frac{3}{2 N_s}\right)  \left[N_s \log \left(\frac{\mu_s^2}{\mu_R^2} \right) - \log 2 + \psi \left(  \frac{3}{2 N_s} \right) \right] 
+ O  ( \mu_s^2r^2 )
.
\end{equation}
Therefore, the one-loop quantum correction to the source and mass function are negative close to $r=0$ if
\begin{equation}
\n{mass_sign}
\log \left( \frac{\mu_s^2}{\mu_R^2} \right) > \frac{1}{N_s} \left[ \log 2 - \psi \left(  \frac{5}{2 N_s} \right) \right]
,
\end{equation}
while the quantum correction to the potential is positive when
\begin{equation}
\log \left( \frac{\mu_s^2}{\mu_R^2} \right) > \frac{1}{N_s} \left[ \log 2 - \psi \left(  \frac{3}{2 N_s} \right) \right]
.
\end{equation}
Also, the plot in Fig.~\ref{Fig16} reveals an interesting new effect: Although the classical mass function [given by Eq.~\eq{mass-GEF}] is positive definite (see Theorem 2 in Sec.~\ref{Sec4}) its quantum counterpart changes sign a finite number of times. The graphs suggest that, at least for small integers $N_s$, the number of negative local minima in $M_s^{(1)} (r)$ is $N_s$ if~\eq{mass_sign} is satisfied, and $N_s - 1$ otherwise.

We point out that given the domain of validity of the Fox--Wright $\Psi$-function, the solutions~\eq{rho_quan},~\eq{mass_quan}, and~\eq{chi_quan} hold only for $N_s > 1/2$. To obtain the solution for $N_s = 1/2$, we directly set this value into \eq{J2}, finding  
\begin{equation}
\begin{split}
J ( r, \al ) &= \left( \frac{\mu}{\mu_R} \right)^{2 \al - 1} \frac{\sqrt{\pi} \,r}{2^{2 \al+2}}  \sum_{l=0}^{\infty}  \frac{\Gamma \left( 2l + 2 \al + 1 \right)}{\Gamma \left( \frac{2l+3}{2} \right)} \frac{1}{l!} \left( - \frac{r^2}{16} \right)^{l}
\\
&= \frac{1}{2^{2 \al + 1} \al} \left( \frac{\mu}{\mu_R} \right)^{2 \al - 1} \left( 1 + \frac{r^2}{4} \right)^{-\al} \Gamma ( 2 \al + 1 ) \sin \left[ 2 \al \arctan \left( \frac{r}{2} \right) \right]
.
\end{split}
\end{equation}
Then,
\begin{equation}
\begin{split}
\rho_s^{(1)} (r) =& - \frac{  2 \beta_s M \mu_s^4}{\pi^2( \mu_s^2 r^2 + 4 )^4 \, r} \biggl\{ 4 \mu_s r ( \mu_s^2 r^2 - 4 ) \left[6 \ga - 11 - 3 \log \left( \frac{\mu_s^2}{\mu_R^2} \right) + 3 \log \left( \mu_s^2 r^2 + 4 \right)  \right]
\\
&+3 ( \mu_s^4 r^4 - 24 \mu_s^2 r^2 + 16 ) \arctan \left( \frac{\mu_s r}{2} \right) \biggr\}
,
\end{split}
\end{equation}
\begin{equation}
\begin{split}
M_s^{(1)} (r) =& 
\frac{ 4 \beta_s M\mu_s^2}{\pi(\mu_s^2 r^2 +4)^3} \biggl\{ 2 \mu_s  r \biggl[ (8 \gamma - 11)  \mu_s^2 r^2 -4 \mu_s^2 r^2 \log \left(\frac{\mu_s^2 }{\mu_R^2}\right) 
\\
&+4 \mu_s^2 r^2 \log \left(\mu_s^2 r^2+4\right)+4\biggr] +\left(3 \mu_s^4 r^4-24 \mu_s^2 r^2-16\right) \arctan \left( \frac{\mu_s r}{2} \right) \biggr\}
,
\end{split}
\end{equation}
\begin{equation}
\chi_s^{(1)} (r) = \frac{ 4 \beta_s GM \mu_s^3 }{\pi(\mu_s^2 r^2 +4 )^2} \left[   
4 - 4 \ga + 2 \log \left( \frac{\mu_s^2}{\mu_R^2} \right) - 2\log ( \mu_s^2 r^2 +4 )  
- \frac{( \mu_s^2 r^2 - 4 )}{\mu_s r} \arctan \left( \frac{\mu_s r}{2} \right)
\right]
,
\end{equation}
where $\ga = 0.57721 \ldots$ is the Euler--Mascheroni constant.

Also, as a consistency check, let us verify if the general results derived here can reproduce $\chi_s^{(1)}(r)$ evaluated in Ref.~\cite{Nos6der} for the particular case $N_s = 1$. First, notice that~\cite{PsiMagico} 
\begin{equation}
{}_{1} \Psi_{1} \left[ \left( a, 1 \right) ; \left( b, 1 \right) ; z \right] 
= \frac{\Ga(a)}{\Ga(b)} \, {}_{1} F_{1} \left( a ; b ; z \right) 
,
\end{equation}
where ${}_1 F_1 (a;b;z) = M(a,b,z)$ is  Kummer's confluent hypergeometric function.
Then, 
\begin{equation}
\begin{split}
{}_{1} \Psi_{1} \left[ \left( \frac{5}{2}, 1 \right) ; \left( \frac{3}{2}, 1 \right) ; -\frac{\mu_s^2 r^2}{8} \right] 
& = \frac32 \, {}_{1} F_{1} \left( \frac{5}{2} ; \frac{3}{2} ; -\frac{\mu_s^2 r^2}{8} \right) 
= \frac32 \, e^{-\frac{\mu_s^2 r^2}{8} } {}_1 F_1 \left( -1; \frac{3}{2}; \frac{\mu_s^2 r^2}{8} \right) 
\\
& = \frac32 \, e^{-\frac{\mu_s^2 r^2}{8} } \left(1 - \frac{\mu_s^2 r^2}{12} \right)
,
\end{split}
\end{equation}
\begin{equation}
{}_{1} \Psi_{1} \left[ \left( \frac{5}{2}, 1 \right) ; \left( \frac{5}{2}, 1 \right) ; -\frac{\mu_s^2 r^2}{8} \right] 
= {}_{1} F_{1} \left( \frac{5}{2} ; \frac{5}{2} ; -\frac{\mu_s^2 r^2}{8} \right) 
= e^{-\frac{\mu_s^2 r^2}{8} }
,
\end{equation}
\begin{equation}
{}_{1} \Psi_{1} \left[ \left( \frac{3}{2}, 1 \right) ; \left( \frac{3}{2}, 1 \right) ; -\frac{\mu_s^2 r^2}{8} \right] 
= {}_{1} F_{1} \left( \frac{3}{2} ; \frac{3}{2} ; -\frac{\mu_s^2 r^2}{8} \right) 
= e^{-\frac{\mu_s^2 r^2}{8} }
.
\end{equation}
For the derivatives of the $\Psi$-function, we use the identities
\begin{small}
\begin{equation}
\begin{split}
\frac{\partial}{\partial a} {}_{1} \Psi_{1} \left[ \left( a, 1 \right) ; \left( \frac{3}{2}, 1 \right) ; -\frac{\mu_s^2 r^2}{8}  \right] \bigg \vert_{a = \frac{5}{2}} 
& = \sum_{k=0}^{\infty} \frac{\Gamma ( \tfrac{5}{2} + k ) \psi \left( \tfrac{5}{2} + k \right) }{\Gamma \left( \tfrac{3}{2} + k \right)} \frac{1}{k!} \left( -\frac{\mu_s^2 r^2}{8} \right)^{k}
\\
& =
\frac{3}{2} \left[ e^ {-\frac{\mu_s ^2 r^2}{8} }  \left( \frac83 - \ga - 2 \log 2 \right)  \left(1 - \frac{\mu_s^2 r^2}{12}\right)
+  \frac{\pa}{\pa a} {}_1 F_1 \left(a,\frac{3}{2},-\frac{\mu_s^2 r^2}{8} \right) \bigg |_{a = \frac{5}{2}}
\right]
,
\end{split}
\end{equation}
\begin{equation}
\begin{split}
\frac{\partial}{\partial a} {}_{1} \Psi_{1} \left[ \left( a, 1 \right) ; \left( \frac{5}{2}, 1 \right) ; -\frac{\mu_s^2 r^2}{8}  \right] \bigg \vert_{a = \frac{5}{2}} 
& = \sum_{k=0}^{\infty} \frac{\Gamma ( \tfrac{5}{2} + k ) \psi \left( \tfrac{5}{2} + k \right) }{\Gamma \left( \tfrac{5}{2} + k \right)} \frac{1}{k!} \left( -\frac{\mu_s^2 r^2}{8} \right)^{k}
\\
&=- e^{- \frac{\mu_s^2 r^2}{8} } \bigl[ \ga - \frac{8}{3} + 2 \log 2
+  \frac{\partial}{\partial a} {}_1 F_1 \left( a; \frac{5}{2}; \frac{\mu_s^2 r^2}{8} \right) \bigg \vert_{a=0} \biggr]
,
\end{split}
\end{equation}
\begin{equation}
\begin{split}
\frac{\partial}{\partial a} {}_{1} \Psi_{1} \left[ \left( a, 1 \right) ; \left( \frac{3}{2} , 1 \right) ; - \frac{\mu_s^2 r^2}{8} \right] \bigg \vert_{a=\frac{3}{2}} 
&=  \sum_{k=0}^{\infty} \frac{\Gamma \left( \tfrac{3}{2} + k \right) \psi \left( \tfrac{3}{2} + k \right)}{\Gamma \left( \tfrac{3}{2} + k \right)} \frac{1}{k!} \left( -\frac{\mu_s^2 r^2}{8} \right)^{k}
\\
&=-e^{-\frac{\mu_s^2 r^2}{8}} \left[\gamma - 2 + 2\log 2 + \frac{\pa}{\pa a} {}_1 F_1 \left( a; \frac{3}{2}; \frac{\mu_s^2 r^2}{8} \right) \bigg |_{a = 0} \right]
.
\end{split}
\end{equation}
\end{small}

Therefore, from~\eq{rho_quan},~\eq{mass_quan}, and~\eq{chi_quan} we get that for $N_s = 1$
\begin{equation}
\begin{split}
\rho_s^{(1)} (r) = &  \, -\frac{3\beta_s \mu^5_s }{ 32(2\pi)^{3/2} } \left\{ e^{- \frac{\mu_s^2 r^2}{8} } \left(1 - \frac{\mu_s^2 r^2}{12} \right)
\left[ \log \left( \frac{\mu_s^2}{\mu_R^2}  \right) - 3  \log 2  + \frac83   -  \gamma \right] +   \frac{\pa}{\pa a} {}_1 F_1 \left( a; \frac{3}{2}; - \frac{\mu_s^2 r^2}{8} \right) \bigg |_{a = \frac52}
\right\}
,
\end{split}
\end{equation}
\begin{equation}
M_{s}^{(1)} ( r ) = -\frac{M \beta_{s} \mu_s^5 r^3}{  16 \sqrt{2\pi}  } e^{- \frac{\mu_s^2 r^2}{8} } \left\{  \log \left( \frac{\mu_s^2}{\mu_R^2} \right) - 3 \log 2 + \frac{8}{3} - \ga - \frac{\partial }{\partial a} {}_1 F_1 \left( a, \frac{5}{2}, \frac{\mu_s^2 r^2}{8} \right) \bigg \vert_{a = 0} \right\}
,
\end{equation}
\begin{equation}
\n{certoorasbolas}
\begin{split}
\chi_s^{(1)} ( r ) = \frac{\beta_s G M \mu_s^3 }{  4  \sqrt{2\pi}}  \,  e^{- \frac{\mu_s^2 r^2}{8}} \left\{ \log \left( \frac{\mu_s^2}{8 \mu_R^2} \right) + 2 - \gamma  - \frac{\partial}{\partial a} {}_1 F_1 \left( a, \frac{3}{2}, \frac{\mu_s^2 r^2}{8} \right) \bigg \vert_{a = 0} \right\}
.
\end{split}
\end{equation}
The solution for the one-loop correction to the potential almost agrees with~\cite{Nos6der}, the only difference being a multiplicative factor in the argument of the logarithm. In fact, there is a typo in Eq.~(100) of~\cite{Nos6der} and the correct result is~\eq{certoorasbolas}.

\begin{figure}[t]
\centering
\begin{subfigure}{.5\textwidth}
\centering
\includegraphics[width=7.5cm]{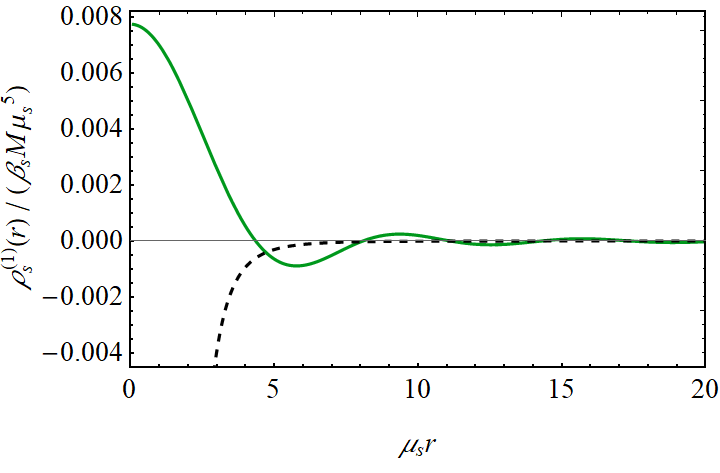}
\end{subfigure}%
\begin{subfigure}{.5\textwidth}
\centering
\includegraphics[width=7.5cm]{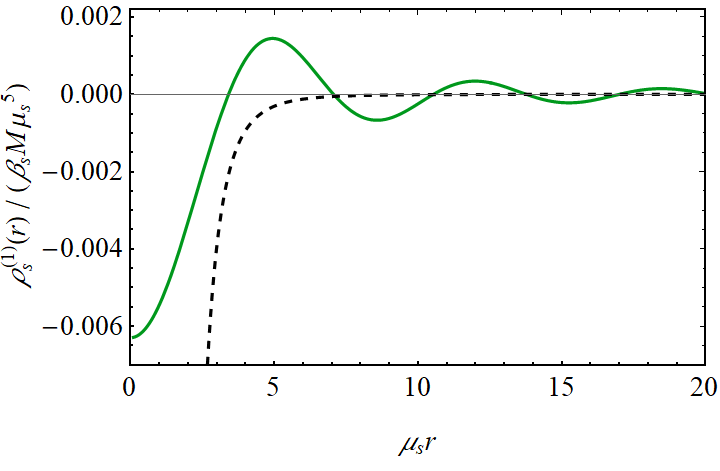}
\end{subfigure}
\caption{Plot of $ \lim_{N_s \to \infty} \rho_s^{(1)} (r) / (\beta_s M \mu_s^5)$ as a function of $\mu_s r$. Left plot: $\mu_R = 1.2 \mu_s$. Right plot: $\mu_R = 0.6 \mu_s$. } 
\label{Fig18}
\end{figure}

\begin{figure}[t]
\begin{subfigure}{.5\textwidth}
\centering
\includegraphics[width=7.5cm]{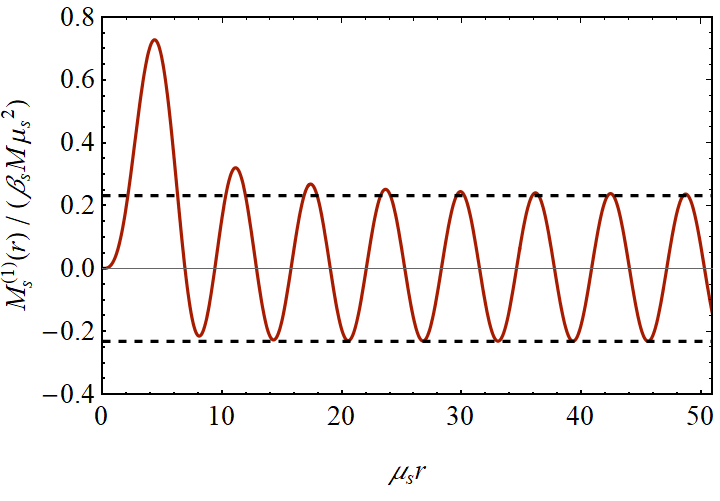}
\end{subfigure}%
\begin{subfigure}{.5\textwidth}
\centering
\includegraphics[width=7.5cm]{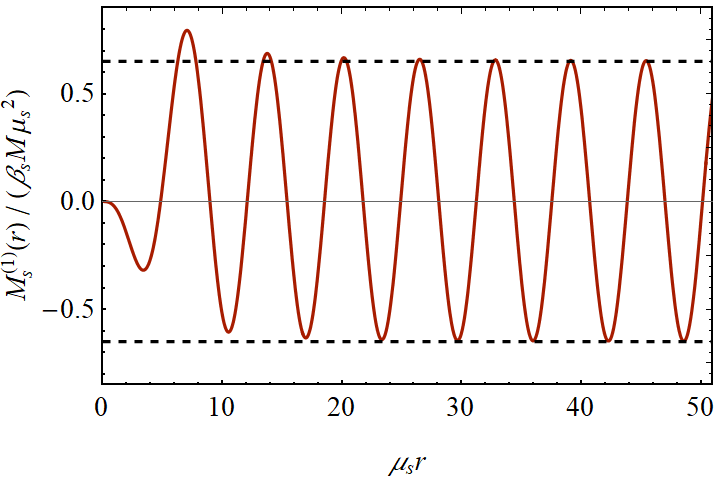}
\end{subfigure}
\caption{Plot of $ \lim_{N_s \to \infty} M_s^{(1)} (r) / (\beta_s M \mu_s^2)$ as a function of $\mu_s r$. The horizontal lines corresponds to $\pm 4 \log \left({\mu_s } / {\mu_R}\right) /\pi $. Left plot: $\mu_R = 1.2 \mu_s$,  $\pm 4 \log \left({\mu_s } / {\mu_R}\right) /\pi = \pm 0.232$. Right plot: $\mu_R = 0.6 \mu_s$, $\pm 4 \log \left({\mu_s } / {\mu_R}\right) /\pi = \pm 0.650$.} 
\label{Fig19}
\vspace{1em} 
\begin{subfigure}{.5\textwidth}
\centering
\includegraphics[width=7.5cm]{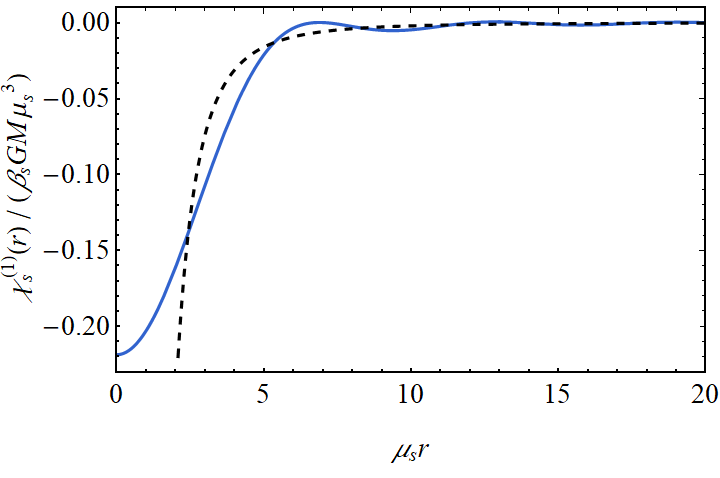}
\end{subfigure}%
\begin{subfigure}{.5\textwidth}
\centering
\includegraphics[width=7.5cm]{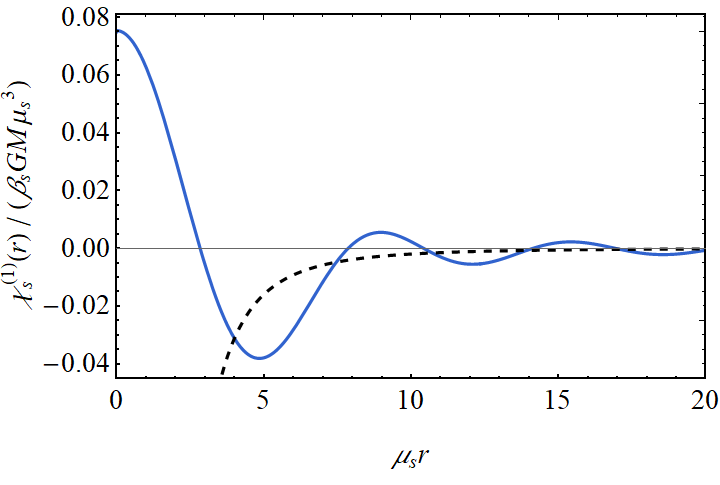}
\end{subfigure}
\caption{Plot of $ \lim_{N_s \to \infty} \chi_s^{(1)} (r) / (\beta_s G M \mu_s^3)$ as a function of $\mu_s r$. Left plot: $\mu_R = 1.2 \mu_s$. Right plot: $\mu_R = 0.6 \mu_s$. } 
\label{Fig20}
\end{figure}

Another case of interest is the limit $N_s \to \infty$. In this case we obtain
\begin{equation}
\begin{split}
\lim_{N \to \infty}    I_{N,\mu} ( r, n ) &= \int_{0}^{\mu} \rd k \,  k^n \sin ( k r ) \log ( k / \mu_R )
,
\end{split}
\end{equation}
which can be directly evaluated for $n \in \left\{ 1, 3 \right\}$, resulting in
\begin{equation}
\begin{split}
\n{rho_inf_Q}
\lim_{N_s \to \infty} \rho_s ^{(1)} (r) = - \frac{\beta_s M}{ \pi^2 r^5} \bigg\{ &   
6 \, \text{Si} ( \mu_s r ) + \left[\mu_s^2 r^2 + 3 (\mu_s^2 r^2 - 2 ) \log \left( \frac{\mu_s}{\mu_R} \right) - 11 \right] \sin ( \mu_s r )  
\\
&
-  \left[(\mu_s^2 r^2 - 6 ) \log \left( \frac{\mu_s}{\mu_R} \right) - 5 \right] \mu_s r \cos ( \mu_s r )
\biggr\}
,
\end{split}
\end{equation}
\begin{equation}
\n{mass_inf_Q}
\begin{split}
\lim_{N_s \to \infty}    M_{s}^{(1)} ( r ) 
= \frac{ 4 M \beta_{s}}{\pi r^2} \biggl\{ & 3 \, \text{Si} \left( \mu_s r \right) + \left[ 3 \mu_s r \cos ( \mu_s r ) - 3 \sin ( \mu_s r ) + \mu_s^2 r^2 \sin ( \mu_s r ) \right] \log \left( \frac{\mu_s}{\mu_R} \right) 
\\
&+ \mu_s r \cos ( \mu_s r )- 4 \sin ( \mu_s r )  \biggr\}
,
\end{split}
\end{equation}
\begin{equation}
\begin{split}
\n{chi_inf_Q}
\lim_{N_s \to \infty} \chi_s^{(1)} (r) = \frac{  4 \beta_s G M}{\pi r^3} \left\{ \sin ( \mu_s r ) - \text{Si} \left( \mu_s r \right) - \left[ \mu_s r \cos ( \mu_s r ) - \sin ( \mu_s r ) \right]  \log \left( \frac{\mu_s}{\mu_R} \right)\right\}
.
\end{split}
\end{equation}

In Figs.~\ref{Fig18},~\ref{Fig19}, and~\ref{Fig20} we plot~\eq{rho_inf_Q},~\eq{mass_inf_Q} and~\eq{chi_inf_Q}, respectively. The asymptotic behavior~\eq{EFT} is violated only for the mass function; the situation is very similar to its classical counterpart, as discussed in Sec~\ref{Sec4}. Indeed, for $\mu_s r \gg 1$ we have
\begin{equation}
\lim_{N_s \to \infty}    M_{s}^{(1)} ( r ) 
\sim \frac{4 \beta_s M \mu_s ^2 }{\pi }  \log \left(\frac{\mu_s }{\mu_R}\right) \sin (\mu_s  r)
,
\end{equation}
showing that the number of sign changes of the mass function tends to infinity as $N_s\to\infty$.

Finally, since the one-loop corrections to the potential evaluated here are even functions of $r$, just like their classical counterpart, they do not spoil the regularity of the curvature invariants at $r=0$, discussed in Sec.~\ref{Sec7}. This happens because the integral~\eq{chi-Q-int} that defines the quantum corrections to the potential is similar to the classical effective source~\eq{eff-sour} but with the effective form factor
\begin{equation}
\tilde{f}_s (k^2) = \frac{[f_s(-k^2)]^2}{\log(k^2 /\mu_R^2 )}.
\end{equation}
Although this effective form factor does not satisfy $\tilde{f}_s(0)=1$, it grows faster than any polynomial, guaranteeing that even when one-loop logarithmic quantum corrections are taken into account, the Newtonian potential is bounded, as well as all the curvature invariants that are polynomial in the Riemann tensor and its derivatives~\cite{Nos6der}. 


\section{Summary and closing remarks}
\label{Sec9}

There exists an infinite number of models under the umbrella of ``nonlocal gravity'' and it is important to identify similarities and differences across this vast family of theories. This point of view motivated the works~\cite{BreTib2,Nos6der}, which focused on the issue of the regularity of the linearized solutions, and also the present paper. Here, however, the main results concern the behavior of the solutions in the Newtonian limit, with possible applications to the weak-field phenomenology of the models. It was noticed long ago that the modified Newtonian potential can oscillate if the form factors of the model are of the type $f_s(\Box)=\exp(-\Box/\mu_s^2)^{N_s}$ with $N_s=2,3,\ldots$~\cite{Edholm_NewPot,Edholm:2017dal,Boos:2018bhd,Perivolaropoulos:2016ucs}, but the explanation of why it does not happen for $N_s=1$ and a detailed comparative analysis of these oscillations for different values of $N_s$ were still pending. These issues were addressed in the present work, together with the following generalizations.

First, our considerations were not restricted to integer values of $N_s$, but assumed that $N_s>0$ could take any real value. Although such fractional operators combined with non-polynomial form factors might look artificial from the point of view of extensions of GR, it allows us to cover other  formulations of quantum gravity phenomenology. For example, effective models from noncommutative geometry often use smeared delta sources with  $N_s=1/2$ (see, e.g.,~\cite{Nozari:2008rc,Liang:2012vx,Kuhfittig:2014iwa}). Therefore, the presentation of our results in terms of the effective source formalism and with generic values of $N_s>0$ can find applications beyond the traditional framework of nonlocal gravity.

Second, we obtained several representations for the effective source $\rho_s(r)$, mass function $M_s(r)$, and potential $\chi_s(r)$, in terms of integrals, a zoo of special functions, and power series. Table~\ref{Taubao} collects the various representations obtained throughout the present work. For the special values $N_s= 1/2$, $N_s= 1$, and in the limiting scenario $N_s\to\infty$ these quantities are known in compact (or even closed) form. The recent mathematical developments of Ref.~\cite{Barvinsky:2019spa} also allowed us to express the solutions for a generic $N_s > 0$ in extremely compact and elegant expressions, simpler than previously known representations in terms of hypergeometric functions~\cite{Boos:2018bhd,Frolov:2015usa,Mo:2022szw}. The main consequence is that the usual formulas for $N_s = 1$ can be directly extended to an arbitrary $N_s > 0$ by simply trading standard exponentials for generalized exponential functions.

\begin{table}[t]
\centering
\begin{tabular}{|c|c|c|c|}
\hline
\hline
\multicolumn{4}{c}{Solutions for special values of $N_s$} \\
\hline
\hline
 & $N_s = 1/2 $ & $N_s=1$ & $N_s \to \infty$ \\
\hline
\rule[-.45cm]{-5pt}{1.1cm} 
$\rho_s(r)$ & $\frac{M \mu^3_s}{\pi^2 } \, \frac{1}{\left(1+ \mu_s^2 r^2\right)^2}$ & $\frac{M \mu^3_s}{8 \pi ^{3/2}} e^{-\frac{\mu^2_s r^2}{4} }$ & $ \frac{M}{2 \pi^2 r^3} \left[ \sin (\mu_s r) - \mu_s r \cos (\mu_s r) \right] $ \\ 
\rule[-.45cm]{-5pt}{1.1cm} 
$M_s (r) $  & $\qquad \, \frac{2M}{\pi} \left[ \arctan ( \mu_s r ) -\frac{\mu_s r}{1 + \mu_s^{2} r^{2}} \right] \qquad\,$  &  $\qquad\, M \left[ {\rm erf} \left( \frac{\mu_s r}{2} \right) - \frac{\mu_s r}{\sqrt{\pi}}  e^{-\frac{\mu^2_s r^2}{4} } \right] \qquad\, $ &  $\qquad\,\, \frac{2M}{\pi} \left[\text{Si}\,  (\mu_s r) - \sin (\mu_s r) \right] \qquad\,\,$  \\ 
\rule[-.45cm]{-5pt}{1.1cm} 
$\chi_s (r) $  & $- \frac{2GM}{\pi r} \arctan ( \mu_s r )$ & $- \frac{GM}{r} \, {\rm erf} \left( \frac{\mu_s r}{2} \right)$  & $- \frac{2GM}{\pi r} \, {\rm Si} \, (\mu_s r) $    \\ 
\hline
\hline
\end{tabular}
\begin{tabular}{|c|c|c|}
\multicolumn{3}{c}{Representations for any $N_s >0$} \\
\hline
\hline
 & \, Generalized exponential function  \, & Integral representations   \\
\hline
\rule[-.45cm]{-5pt}{1.1cm} 
$\rho_s(r)$ & $\frac{M \mu^3_s}{8 \pi^{3/2}} \, \mathcal{E}_{N_s, \frac{3}{2}} \left(-\frac{\mu^2_s r^2}{4} \right)$ & $\frac{M}{2 \pi^2 r^3} \int_0^\infty \rd t \left[ 
\sin ( \mu_s r t^{\frac{1}{2N_s}}) - (\mu_s r t^{\frac{1}{2N_s}}) \cos ( \mu_s r t^{\frac{1}{2N_s}})
\right] e^{-t}$ \\ 
\rule[-.45cm]{-5pt}{1.1cm} 
$M_s (r) $  & $\quad\quad M \Big[ \mathcal{E}{\rm rf}_{N_s,\frac12} \left( \frac{\mu_s r}{2} \right) - \frac{\mu_s r}{\sqrt{\pi}}  {\cal E}_{N_s,\frac12} \left(-\frac{\mu^2_s r^2}{4} \right) \Big]\quad\quad$  & $\frac{2M}{\pi} \int_0^\infty \rd t  \left[\text{Si}\,  (\mu_s r t^{\frac{1}{2N_s}}) - \sin (\mu_s r t^{\frac{1}{2N_s}}) \right] e^{-t}$ \\ 
\rule[-.45cm]{-5pt}{1.1cm} 
$\chi_s (r) $  & $- \frac{GM}{r} \, {\cal E}{\rm rf}_{N_s,\frac12} \left( \frac{\mu_s r}{2} \right)$  & $- \frac{2GM}{\pi r} \int_0^\infty \rd t \,  {\rm Si} \, (\mu_s r t^{\frac{1}{2N_s}}) \, e^{-t}$ \\ 
\hline
\hline
\multicolumn{3}{c}{Representations for any $N_s >1/2$} \\
\hline
\hline
 &  Power series (also for $N_s = 1/2$) & Fox $H$-function \\
\hline
\rule[-.45cm]{-5pt}{1.1cm}  
$\rho_s(r)$ & $\frac{M \mu_s^3}{4 \pi^2 N_s} \sum\limits_{\ell=0}^{\infty}  \frac{(-1)^\ell}{(2\ell + 1)!} \, 
\Ga \left( \frac{2\ell+3}{2N} \right) (\mu_s r)^{2 \ell}$ & $
 \frac{M \mu^3_s}{8 \pi^{3/2} N_s}  H^{1,1}_{1,2} \biggl[
\begin{array}{c}
\left(\frac{2N_s-3}{2N_s}, \frac{1}{N_s} \right) \\
(0,1), ( -\frac{1}{2}, 1 )
\end{array}
\bigg \vert
\frac{\mu^2_s r^2}{4}
\biggr]
$ \\ 
\rule[-.45cm]{-5pt}{1.1cm} 
$M_s (r) $   & $\frac{M }{ \pi N_s } \sum\limits_{\ell=0}^{\infty} \frac{(-1)^\ell}{(2\ell + 1)!} \, 
\Ga \left( \frac{2\ell+3}{2 N_s} \right) \frac{(\mu_s r)^{2 \ell + 3}}{2l + 3}$ & $\frac{4 M}{\sqrt{\pi} N_s} H^{1,2}_{2,3} \biggl[
\begin{array}{c}
(1,2), \left( 1, \frac{1}{N_s} \right) \\
(\frac{3}{2}, 1), \left( 1, 1 \right), ( 0, 2 )
\end{array}
\bigg \vert
\frac{\mu_s^{2} r^{2}}{4}
\biggr]$ \\ 
\rule[-.45cm]{-5pt}{1.1cm} 
$\chi_s (r) $  & $-\frac{ G M \mu_s}{\pi N_s} \sum\limits_{\ell=0}^{\infty} \frac{(-1)^\ell}{(2\ell + 1)!} \, \Ga \left( \frac{2\ell+1}{2N_s} \right) (\mu_s  r)^{2 \ell}$  & $
 -\frac{2 G M}{\sqrt{\pi} r} 
H^{1,2}_{2,3} \biggl[
\begin{array}{c}
(1, 2), \left( 1, \tfrac{1}{N_s} \right) \\
(\tfrac12, 1), \left( 1, 1\right), ( 0, 2 )
\end{array}
\bigg \vert
\frac{\mu_s^2 r^2}{4} 
\biggr]
$     \\ 
\hline
\hline
\multicolumn{3}{c}{Representations for $N_s \in \mathbb{N}$} \\
\hline
\hline
\multicolumn{3}{|c|}{Hypergeometric functions} \\
\hline
\rule[-.45cm]{-5pt}{1.1cm} 
$\rho_s(r)$ & \multicolumn{2}{|c|}{$
\frac{M \mu_s^3}{4 \pi^2 N_s} \sum\limits_{\ell=0}^{N_s - 1}  \frac{(-1)^\ell}{(2\ell + 1)!}  \Ga \left( \frac{2\ell + 3}{2N_s} \right) \, (\mu_s r)^{2\ell} 
{}_0F_{2N_s-2}\left(-; \frac{\ell+1}{N_s}, 
\frac{\ell+2}{N_s},
\ldots, 
\frac{\ell+N_s}{N_s}, 
\frac{2\ell+5}{2N_s},  
\frac{2\ell+7}{2N_s},
\ldots,
\frac{2\ell+2N_s+1}{2N_s}; \left(-\frac{\mu_s^2 r^2}{4N_s}\right)^{N_s}\right)^*
$} \\ 
\rule[-.45cm]{-5pt}{1.1cm} 
$M_s (r) $  & \multicolumn{2}{|c|}{$
\frac{M }{ \pi N_s } \sum\limits_{\ell=0}^{N_s - 1}  \frac{(-1)^\ell}{(2\ell + 1)!} \, 
\Ga \left( \frac{2\ell+3}{2N_s} \right) \frac{(\mu_s r)^{2 \ell + 3}}{2\ell + 3} \,
{}_1 F_{2N_s-1}\left(\frac{2\ell+3}{2N_s}; \frac{\ell+1}{N_s}, 
\frac{\ell+2}{N_s},
\ldots, 
\frac{\ell+N_s}{N_s}, 
\frac{2\ell+5}{2N_s},  
\frac{2\ell+7}{2N_s},
\ldots,
\frac{2\ell+2N_s+3}{2N_s}; \left(-\frac{\mu_s^2 r^2}{4N_s}\right)^{N_s}\right)^*
$} \\ 
\rule[-.45cm]{-5pt}{1.1cm} 
$\chi_s (r) $  & \multicolumn{2}{|c|}{$
 \,\, -\frac{G M \mu_s }{ \pi N_s } \sum\limits_{\ell=0}^{N_s - 1}  \frac{(-1)^\ell}{(2\ell + 1)!} \, 
\Ga \left( \frac{2\ell+1}{2N_s} \right) (\mu_s r)^{2 \ell} \,
{}_1 F_{2N_s-1}\left(\frac{2\ell+1}{2N_s}; 
\frac{2\ell+2}{2 N_s}, 
\frac{2\ell+3}{2 N_s},
\ldots, 
\frac{2\ell+2N_s}{2N_s}, 
\frac{2\ell+2N_s+1}{2N_s}; 
\left(-\frac{\mu_s^2 r^2}{4N_s}\right)^{N_s}\right)^*
$}   \\ 
\hline
\multicolumn{3}{|c|}{Meijer~$G$-function}  \\
\hline
\rule[-.45cm]{-5pt}{1.1cm} 
$\rho_s(r)$ & \multicolumn{2}{|c|}{$
\frac{M \mu_s^3 }{(4\pi N_s)^{3/2}  } \,   G^{N_s,0}_{0,2N_s-1} 
\left(
\left. 
\begin{matrix}
-   &  \\ 
0, \frac{1}{N_s},  \frac{2}{N_s},  \ldots , \frac{N_s-1}{N_s}  ; -\frac{1}{2N_s},  \frac{1}{2N_s},  \ldots,  \frac{2N_s-3}{2N_s} & \\
\end{matrix}
\right|
\left( \tfrac {\mu_s r}{2N_s} \right)^{2N_s}
\right)
$} \\ 
\rule[-.45cm]{-5pt}{1.1cm} 
$M_s (r) $  & \multicolumn{2}{|c|}{$2M \sqrt{\frac{N_s}{\pi}} \,   
G^{N_s,1}_{1,2N_s} 
\left(
\left. 
\begin{matrix}
1   &  \\ 
\frac{3}{2N_s}, \, \frac{5}{2N_s},  \ldots,  \frac{2N_s+1}{2N_s}
; 0, \frac{1}{N_s}, \, \frac{2}{N_s}, \, \ldots , \frac{N_s-1}{N_s}
 & \\
\end{matrix}
\right|
\left( \tfrac {\mu_s r}{2N_s} \right)^{2N_s}
\right) 
$}  \\ 
\rule[-.45cm]{-5pt}{1.1cm} 
$\chi_s (r) $  & \multicolumn{2}{|c|}{$
 -\frac{G M \mu_s }{2 \sqrt{\pi} N_s^{3/2}}\,
G^{N_s,1}_{1,2N_s} 
\left(
\left. 
\begin{matrix}
\frac{2N_s-1}{2N_s}   &  \\  
0, \frac{1}{N_s},  \frac{2}{N_s},  \ldots , \frac{N_s-1}{N_s};
-\frac{1}{2N_s},  \frac{1}{2N_s},  \ldots,  \frac{2N_s-3}{2N_s} 
& \\
\end{matrix}
\right|
\left( \tfrac {\mu_s r}{2N_s} \right)^{2N_s}
\right)
$}     
\\ 
\hline
\end{tabular}
\caption{Collection of the different type of representations of the effective delta source, mass function and potential for different domains of $N_s$.}
\label{Taubao}
\end{table}

With all these representations available, we could obtain various results, for example:
\begin{itemize}

\item[i.] The effective source for the Newtonian potentials is strictly positive for $0<N_s\leqslant 1$ (Theorem 1). This explains the absence of spatial oscillations of the potential in these models.

\item[ii.] Although the effective source oscillates and assumes negative values if $N_s>1$, the effective mass function $M_s$ is always positive (Theorem 2). Therefore, the oscillations in the weak-field solutions of this type of nonlocal gravity theories are very different from those of Lee--Wick gravity~\cite{Accioly:2016qeb,Burzilla:2023xdd}, as discussed in Sec.~\ref{Sec4}. As a physical consequence,  the gravitational force in any GF$_N$ model is always attractive. Moreover, in that section, we obtained an estimate for the radial distance $r_*$ beyond which the oscillations of the gravitational force are suppressed, showing that it grows approximately linearly with $N_s$.

\item[iii.] We also derived approximations for the effective mass function [see Eq.~\eq{pacote}] and the Newtonian potential [Eq.~\eq{chi_ap2}], which can be used in phenomenological applications in a more efficient way than could be done with the exact expressions. In particular, having exact expressions and analytic approximations, in Sec.~\ref{Sec6} we discussed the numerical approximation of~\cite{Perivolaropoulos:2016ucs}, which was used to model laboratory experiments to detect oscillations in the gravitational force.

\item[iv.] All models with $N_s>0$ have a completely regular Newtonian limit, without curvature singularities. This happens because the form factor grows faster than any polynomial, yielding a smooth solution. Therefore, not only the (linearized) curvature invariants but also the ones with covariant derivatives of curvatures are bounded. This differs from the Newtonian limit of any local higher-derivative gravity model and from nonlocal models that have a polynomial behavior in the UV, as in those cases there will always exist curvature-derivative invariants that are singular~\cite{Nos6der}. 

\item[v.] Last but not least, we applied the effective source formalism to calculate the effects of the one-loop logarithmic quantum corrections to the Newtonian-limit solutions. We verified that they recover the behavior predicted by the effective approach to quantum gravity in the limit $\mu_s r \gg 1$, and showed how they can manifest at intermediate scales. For instance, the leading quantum correction to the effective source, mass function, and potential typically oscillates and may change sign for small values of $\mu_s r$, depending on the ratio $\mu_s/\mu_R$. Furthermore, although the classical effective mass $M_s^{(0)}(r)$ is always positive, its one-loop quantum correction $M_s^{(1)}(r)$ can change sign a finite number of times at intermediate values of $\mu_s r$. The physical interpretation of this interesting result could be related to the manifestation of ghost-like degrees of freedom generated by quantum corrections~\cite{Shapiro:2015uxa}, which might assume a tachyonic behavior depending on the value of $N_s$. Nevertheless, as argued in~\cite{Nos6der}, logarithmic or other quantum corrections that are only perturbative cannot affect the spectrum
of the theory in their validity regime. This statement can be understood in our formalism as follows: If we do not break the perturbation theory, then $ |M_s^{(1)} (r)| < |M_s^{(0)} (r) |$, and the total effective mass $M_s(r) = M_s^{(0)} (r) + M_s^{(1)} (r) $ will always remain  positive.
\end{itemize}

The results obtained here, and in particular, the approximations for the effective mass function and potential, can be used to explore the weak-field phenomenology of nonlocal gravity. 
Of course, it would also be interesting to study whether and how the oscillations of the Newtonian-limit solutions manifest in the full nonlinear regime, and if the black hole singularities can really be avoided in these models. However, it is extremely difficult to go beyond the linear approximation in nonlocal models.


\begin{acknowledgments}
\noindent
T.M.S. is grateful to Funda\c{c}\~{a}o de Amparo \`{a} Pesquisa do Estado de Minas Gerais -- FAPEMIG for supporting his MSc project. B.L.G. acknowledges financial support by the Primus grant PRIMUS/23/SCI/005 from Charles University and the support from the Charles University Research Center Grant No. UNCE24/SCI/016. T.P.N.  is grateful to Conselho Nacional de Desenvolvimento Cient\'ifico e Tecnol\'ogico -- CNPq (Brazil) for the financial support.
\end{acknowledgments}


\appendix

\renewcommand{\thesubsection}{\thesection.\arabic{subsection}}


\section{Solutions in terms of generalized hypergeometric functions}
\label{ApA}

The previous works on GF$_N$ gravity theories that obtained Newtonian-limit solutions for $N > 1$ can be summarized as follows: in~\cite{Boos:2018bhd} the effective source was obtained for $N \in \{2,3\}$, while the Newtonian potential was evaluated for $N \in \{2,4\}$ in~\cite{Frolov:2015usa} and~$N = 2$ in~\cite{Mo:2022szw}. The expressions for all these quantities involved a finite sum of generalized hypergeometric functions which becomes more and more complicated for higher values of $N$.\footnote{The power series representation of the potential was obtained in~\cite{Edholm_NewPot}.} So, to be consistent with the previous literature, here we show that for $N_s  \in \mathbb{N}$ the solutions obtained in the main part of the text can be expressed in an equivalent way in terms of generalized hypergeometric functions. 

The essential task is to prove that the basic integral $I_N(r)$ written as the infinite sum~\eq{I-series}, namely
\begin{equation}
\n{I-series-2}
I_N(r) = \frac{1}{2 N} \sum_{p=0}^\infty C_p,
\qquad \qquad
C_p =  \frac{(-1)^p}{(2p + 1)!} \, 
\Ga \left( \frac{2p+3}{2N} \right) r^{2 p}
,
\end{equation}
has as a result, for $N \in \mathbb{N}$, a finite sum of generalized hypergeometric functions. So, using the identity
\begin{equation}
(2p)! = 2^{2p}  p!  \, \frac{ \Ga \left(p+ \tfrac12 \right) }{ \Ga \left( \tfrac12 \right) },
\end{equation}
we can express the factor $(2p+1)!$ in~\eq{I-series-2} in the form 
\begin{equation}
(2p+1)! = (2p + 1)(2p)! =  2^{2p}  p!  \, \left( \frac{p + \tfrac12}{\tfrac12} \right) \frac{ \Ga \left(p+ \tfrac12 \right) }{ \Ga \left( \tfrac12 \right) }
= 2^{2p}  p!  \,  \frac{ \Ga \left(p+ \tfrac32 \right) }{ \Ga \left( \tfrac32 \right) }
,
\end{equation}
which can be rewritten as 
\begin{equation}
\n{maldita}
(2p+1)! = 2^{2p}  (1)_{(p)}  \left( \frac{3}{2} \right)_{(p)},
\end{equation}
where
$
(a)_{(n)} = \Ga(a+n)/\Ga(a)
$
denotes the rising factorial (Pochhammer symbol). Since
\begin{equation}
\n{important_eq}
\sum_{p = 0}^\infty C_p = \sum_{q = 0}^\infty \sum_{\ell = 0}^{N-1} C_{Nq + \ell}
,
\end{equation}
for any convergent series, 
we get for~\eq{I-series-2} 
\begin{equation}
\n{cnql}
I_N(r) = \frac{1}{2 N} \sum_{q = 0}^\infty \sum_{\ell = 0}^{N-1} C_{Nq + \ell},
\end{equation}
where
\begin{equation}
\n{cnq2}
C_{Nq+\ell} = \frac{1}{(1)_{({Nq+\ell})} \left( \frac{3}{2} \right)_{({Nq+\ell})}  } \, 
\Ga \left( q + \frac{3 + 2 \ell}{2N} \right) (-1)^{{Nq+\ell}} \left( \frac{r}{2} \right)^{2Nq+2\ell}
.
\end{equation}
The next step is to use the properties of the rising factorial
\begin{equation}
(a)_{(n+k)} = (a)_{(k)} (a+k)_{(n)}
,
\eeq
\begin{equation}
(a)_{(kn)} = k^{kn} \times \prod_{j=0}^{k-1} \left(\frac{a+j}{k} \right)_{(n)}
,
\end{equation}
to get
\begin{equation}
\n{2nq_factorial} 
(1)_{({Nq+\ell})} \left( \frac{3}{2} \right)_{({Nq+\ell})} = 
(1)_{({\ell})} \left( \frac{3}{2} \right)_{({\ell})} \, (N)^{2Nq} \, \prod_{j=2}^{2N+1}\left( \frac{2 \ell + j}{2N} \right)_{(q)} 
.
\end{equation}
The above identity, together with
\begin{equation}
\Ga \left( q + \frac{3 + 2 \ell}{2N} \right) = \left(\frac{3 + 2 \ell}{2N} \right)_{(q)} 
\Ga \left(\frac{3 + 2 \ell}{2N} \right)
,
\end{equation}
gives the result for~\eq{cnq2}: 
\begin{equation}
C_{Nq+\ell} =  \frac{(-1)^\ell }{ (2\ell+1)! } \,
\Ga \left(\frac{3 + 2 \ell}{2N} \right) r^{2\ell} \,
\frac{1}{ \prod\limits_{\substack{j=2\\j \neq 3}}^{2N+1} 
\left( \frac{2\ell+j}{2N} \right)_{(q)}} (-1)^{{Nq}} \left( \frac{r}{2N} \right)^{2Nq}
,
\end{equation}
where we used~\eq{maldita} once more. Therefore,
\begin{equation}
\begin{split}
I_N (r) 
= \frac{1}{2N} \sum_{\ell = 0}^{N-1} \frac{(-1)^\ell }{ (2\ell+1)! } \, 
\Ga \left(\frac{3 + 2 \ell}{2N} \right) \, r^{2\ell}
\sum_{q=0}^\infty
\frac{1}{  \prod\limits_{\substack{j=2\\j \neq 3 \\ j \neq 2(l - N) }}^{2N+1} 
\left( \frac{2\ell+j}{2N} \right)_{(q)} } \frac{(-1)^{{Nq}}}{q!} \left( \frac{r}{2N} \right)^{2Nq}
.
\end{split}
\end{equation}
Since the generalized hypergeometric function ${}_p F_q$ is defined by
\begin{equation}
\n{HG}
{}_p F_q (a_1, \ldots, a_p ; b_1, \ldots, b_q; z) = \sum_{n=0}^\infty \frac{\prod\limits_{i=1}^p (a)_{(n)}}{\prod\limits_{j=1}^q (b)_{(n)}} \frac{z^n}{n!}
,
\end{equation}
we find that
\begin{equation}
\n{IN-geo}
\begin{split}
I_N (r) = & \,\, \frac{1}{2N}  \sum_{\ell = 0}^{N-1} \frac{(-1)^\ell}{(2\ell + 1)!}  
\Ga \left( \frac{2\ell + 3}{2N} \right) \, r^{2\ell} \, 
\times
\\
& \times
{}_0 F_{2N-2} \left(-; \frac{\ell+1}{N}, 
\frac{\ell+2}{N},
\ldots, 
\frac{\ell+N}{N}, 
\frac{2\ell+5}{2N},  
\frac{2\ell+7}{2N},
\ldots,
\frac{2\ell+2N+1}{2N}; \left(-\frac{r^2}{4N}\right)^{N}\right)^*
,
\end{split}
\end{equation}
where, following the standard notation~\cite{Grad}, the star indicates that the term corresponding to $j = 2l - 2N$ is omitted.

Thus, Eq.~\eq{sour-I} implies that 
\begin{equation}
\begin{split}
\rho_{s} (r) 
= &\,\, \frac{M \mu_s^3}{4 \pi^2 N_s} \sum_{\ell = 0}^{N_s-1} \frac{(-1)^\ell}{(2\ell + 1)!}  \Ga \left( \frac{2\ell + 3}{2N_s} \right) \, (\mu_s r)^{2\ell} \, \times 
\\
& \times 
{}_0F_{2N_s-2}\left(-; \frac{\ell+1}{N_s}, 
\frac{\ell+2}{N_s},
\ldots, 
\frac{\ell+N_s}{N_s}, 
\frac{2\ell+5}{2N_s},  
\frac{2\ell+7}{2N_s},
\ldots,
\frac{2\ell+2N_s+1}{2N_s}; \left(-\frac{\mu_s^2 r^2}{4N_s}\right)^{N_s}\right)^*.
\end{split}
\end{equation}
Given the effective source, to obtain the mass function and the Newtonian potential one can use~\eq{massfunction} and~\eq{chi_intg} with the integration formula~\cite{Grad}
\begin{equation}
\int_0^y \rd x \, x^{\alpha-1} \, 
{}_p F_q \left( {\bf a}_p; {\bf b}_q; \omega x^n\right) = \frac{y^\alpha}{\alpha} \, {}_{p+1}F_{q+1}\left( {\bf a}_p, \frac{\alpha}{n}; {\bf b}_q, \frac{\alpha}{n} + 1; \omega y^n\right)
.
\end{equation}
The results are, respectively,
\begin{equation}
\begin{split}
M_s (r)
= &\,\, \frac{M }{ \pi N_s } \sum_{\ell=0}^{N_s-1} \frac{(-1)^\ell}{(2\ell + 1)!} \, 
\Ga \left( \frac{2\ell+3}{2N_s} \right) \frac{(\mu_s r)^{2 \ell + 3}}{2\ell + 3} \,
\times
\\
&\times
{}_1 F_{2N-1}\left(\frac{2\ell+3}{2N_s}; \frac{\ell+1}{N_s}, 
\frac{\ell+2}{N_s},
\ldots, 
\frac{\ell+N_s}{N_s}, 
\frac{2\ell+5}{2N_s},  
\frac{2\ell+7}{2N_s},
\ldots,
\frac{2\ell+2N_s+3}{2N_s}; \left(-\frac{\mu_s^2 r^2}{4N_s}\right)^{N_s}\right)^*
\end{split}
\end{equation}
and
\begin{equation}
\begin{split}
\chi_s (r) 
=& \,\, -\frac{G M \mu_s }{ \pi N_s } \sum_{\ell=0}^{N_s-1} \frac{(-1)^\ell}{(2\ell + 1)!} \, 
\Ga \left( \frac{2\ell+1}{2N_s} \right) (\mu_s r)^{2 \ell} \,
\times
\\
&\times
{}_1 F_{2N-1}\left(\frac{2\ell+1}{2N_s}; 
\frac{2\ell+2}{2 N_s}, 
\frac{2\ell+3}{2 N_s},
\ldots, 
\frac{2\ell+2N_s}{2N_s}, 
\frac{2\ell+2N_s+1}{2N_s}; 
\left(-\frac{\mu_s^2 r^2}{4N_s}\right)^{N_s}\right)^*
.
\end{split}
\end{equation}


\section{Solutions in terms of Meijer $G$-function}
\label{ApB}

The function $f_\beta(t) = \exp ({-t^\beta}) $, which corresponds to the form factor~\eq{formfactor}, is known as the stretched exponential function. Its application are very broad, for example, it is commonly used as a phenomenological description of mechanical relaxation in disordered systems and in the statistics of the complementary cumulative Weibull distribution. For this reason, there is a rich literature on how to deal with integrals of the stretched exponential function. One of the techniques often used with integrals involving $f_\beta (t)$ (see, e.g.,~\cite{vaca} and references therein) is to write every non-polynomial function in terms of the Meijer~$G$-function 
$$
G^{m,n}_{p,q} 
\left(
\left. 
\begin{matrix}
a_1, \ldots, a_p &  \\ 
b_1, \ldots, b_q & \\
\end{matrix}
\right|
x
\right)
$$
and apply the integration formula for
\begin{equation}
\n{power}
\int_0 ^\infty \rd x \,
x^{\al-1}
G^{s,t}_{u,v} 
\left(
\left. 
\begin{matrix}
c_1, \ldots, c_u  &  \\ 
d_1, \ldots, d_v  & \\
\end{matrix}
\right|
\si x
\right)
G^{m,n}_{p,q} 
\left(
\left. 
\begin{matrix}
a_1, \ldots, a_p  \\ 
b_1, \ldots, b_q \\
\end{matrix}
\right|
\om x^{l/k}
\right)
,
\end{equation}
which can be found, e.g., in Refs.~\cite{algorithm,magicalbook}. 

In order to apply the reasoning above to the integral~\eq{basic-I} (in the case $N \in \mathbb{N}$), we use
\begin{equation}
e^{-f(x)} = G^{1,0}_{0,1} 
\left(
\left. 
\begin{matrix}
- &  \\ 
0 & \\
\end{matrix}
\right|
f(x)
\right)
\qquad \mbox{and} \qquad
\sin (x) = \sqrt{\pi} \, G^{1,0}_{0,2} 
\left(
\left. 
\begin{matrix}
- &  \\ 
\frac{1}{2}, 0 & \\
\end{matrix}
\right|
\frac{x^2}{4}
\right)
,
\end{equation}
so that
\begin{equation}
\n{basic-I-2}
\begin{split}
I_N (r) &=  
\frac{\sqrt{\pi}}{r} \int_0^\infty \rd k \, k \, 
G^{1,0}_{0,2} \left(\left. 
\begin{matrix}
- &  \\ 
\frac{1}{2}, 0 & \\
\end{matrix}
\right|
\frac{r^4 k^2}{4}
\right)
G^{1,0}_{0,1} 
\left(
\left. 
\begin{matrix}
- &  \\ 
0 & \\
\end{matrix}
\right|
k^{2N}
\right)
\\
& = \frac{\sqrt{\pi}}{2r} 
\int_0^\infty \rd x \,
G^{1,0}_{0,2} \left(\left. 
\begin{matrix}
- &  \\ 
\frac{1}{2}, 0 & \\
\end{matrix}
\right|
\frac{r^4 x}{4}
\right)
G^{1,0}_{0,1} 
\left(
\left. 
\begin{matrix}
- &  \\ 
0 & \\
\end{matrix}
\right|
x^{N}
\right)
.
\end{split}
\end{equation}
Therefore, the integral of our interest has the form~\eq{power} with the parameters: 
$\al = 1$,
$s = 1$,
$t = 0$,
$u = 0$,
$v = 2$,
$m = 1$,
$n = 0$,
$p = 0$,
$q = 1$,
$ \om = 1$,
$ l = N$,
$k = 1$,
$d_1=1/2$,
$d_2=0$,
$b_1=0$,
and
$\si = r^2/4$
(while the $a_p$- and $c_u$-terms are absent).
Thus, using the formula found in~\cite{algorithm,magicalbook}, the result for~\eq{basic-I-2} is
\begin{equation}
I_N(r) = \frac{2N^{3/2} \sqrt{\pi}}{ r^3} \, G^{1,N}_{2N,1} 
\left(
\left. 
\begin{matrix}
 -\frac{1}{2N},  \frac{1}{2N},  \ldots,  \frac{2N-3}{2N}; 0, \frac{1}{N},  \frac{2}{N},  \ldots , \frac{N-1}{N}  ;   &  \\ 
0 & \\
\end{matrix}
\right|
\left( \tfrac{2N}{r} \right)^{2N}
\right).
\end{equation}

The above formula can be further simplified using identities, 
such as~\cite{Grad}
\begin{equation}
G^{mn}_{pq} 
\left(
\left. 
\begin{matrix}
 a_1,  a_2,  \ldots,  a_{p}  &  \\ 
 b_1,  \ldots,  b_{q-1},  a_1   & \\
\end{matrix}
\right|
z
\right) = 
G^{m-1,n}_{p-1,q-1} 
\left(
\left. 
\begin{matrix}
a_2, \ldots, a_p &  \\ 
b_1, b_2, \ldots, b_{q-1} & \\
\end{matrix}
\right|
z
\right),
\qquad n, \, p, \, q \geqslant 1
,
\end{equation}
\begin{equation}
z^\al G^{mn}_{pq} 
\left(
\left. 
\begin{matrix}
 {\bf a}_p   &  \\ 
 {\bf b}_q   & \\
\end{matrix}
\right|
z
\right) = 
G^{mn}_{pq} 
\left(
\left. 
\begin{matrix}
{\bf a}_q + \al &  \\ 
{\bf b}_p + \al & \\
\end{matrix}
\right|
z
\right)
,
\end{equation}
and
\begin{equation}
\n{G_iden_import}
G^{mn}_{pq} 
\left(
\left. 
\begin{matrix}
 {\bf a}_p  &  \\ 
 {\bf b}_q  & \\
\end{matrix}
\right|
z
\right) = 
G^{nm}_{qp} 
\left(
\left. 
\begin{matrix}
1 - {\bf b}_q &  \\ 
1 - {\bf a}_p & \\
\end{matrix}
\right|
\tfrac{1}{z}
\right)
,
\end{equation}
which yield
\begin{equation}
\begin{split}
I_N(r) =
\frac{ \sqrt{\pi}}{4 N^{3/2}  } \,  G^{N,0}_{0,2N-1} 
\left(
\left. 
\begin{matrix}
-   &  \\ 
0, \frac{1}{N}, \frac{2}{N},  \ldots , \frac{N-1}{N}  ; -\frac{1}{2N},  \frac{1}{2N},  \ldots,  \frac{2N-3}{2N} & \\
\end{matrix}
\right|
\left( \tfrac {r}{2N} \right)^{2N}
\right) .
\end{split}
\end{equation}
This expression immediately gives the effective source, while the mass function and the potential can be evaluated applying the integration formula
\begin{equation}
\int_0^y \rd x \, x^{\al-1} \,
G^{mn}_{pq}
\left(
\left. 
\begin{matrix}
{\bf a}_p &  \\ 
{\bf b}_q & \\
\end{matrix}
\right|
\omega x
\right)
= y^\al \, G^{m,n+1}_{p+1,q+1}
\left(
\left. 
\begin{matrix}
a_1, \, \ldots, a_n, \, 1 - \al; \, a_{n+1}, \, \ldots , \, a_p &  \\ 
b_1, \, \ldots, b_m; \al, \, b_{m+1}, \, \ldots, \, b_q & \\
\end{matrix}
\right|
\omega y
\right)
.
\end{equation}
The explicit expressions read
\begin{equation}
\n{rG}
\begin{split}
\rho_s (r) = 
\frac{M \mu_s^3 }{(4\pi N_s)^{3/2}  } \,   G^{N_s,0}_{0,2N_s-1} 
\left(
\left. 
\begin{matrix}
-   &  \\ 
0, \frac{1}{N_s},  \frac{2}{N_s},  \ldots , \frac{N_s-1}{N_s}  ; -\frac{1}{2N_s},  \frac{1}{2N_s},  \ldots,  \frac{2N_s-3}{2N_s} & \\
\end{matrix}
\right|
\left( \tfrac {\mu_s r}{2N_s} \right)^{2N_s}
\right),
\end{split}
\end{equation}
\begin{equation}
\n{mG}
\begin{split}
M_s (r) & =   
2M \sqrt{\frac{N_s}{\pi}} \,   
G^{N_s,1}_{1,2N_s} 
\left(
\left. 
\begin{matrix}
1   &  \\ 
\frac{3}{2N_s}, \, \frac{5}{2N_s},  \ldots,  \frac{2N_s+1}{2N_s}
; 0, \frac{1}{N_s}, \, \frac{2}{N_s}, \, \ldots , \frac{N_s-1}{N_s}
 & \\
\end{matrix}
\right|
\left( \tfrac {\mu_s r}{2N_s} \right)^{2N_s}
\right) 
,
\end{split}
\end{equation}
\begin{equation}
\n{cG}
\begin{split}
\chi_s (r) &= 
-\frac{G M \mu_s }{2 \sqrt{\pi} N_s^{3/2}}\,
G^{N_s,1}_{1,2N_s} 
\left(
\left. 
\begin{matrix}
\frac{2N_s-1}{2N_s}   &  \\  
0, \frac{1}{N_s},  \frac{2}{N_s},  \ldots , \frac{N_s-1}{N_s};
-\frac{1}{2N_s},  \frac{1}{2N_s},  \ldots,  \frac{2N_s-3}{2N_s} 
& \\
\end{matrix}
\right|
\left( \tfrac {\mu_s r}{2N_s} \right)^{2N_s}
\right) 
.
\end{split}
\end{equation}

Finally, it is worth mentioning that the above expressions are  equivalent to those obtained in Appendix~\ref{ApA}, as can be proved using the identity~\cite{Grad}
\begin{align} 
\n{slater}
\begin{split}
G^{m, n}_{p, q} \left(
\begin{array}{c}
{\bf a}_{q} \\
{\bf b}_{q}
\end{array}
\bigg\vert
z \right)
= \sum_{h=1}^{m} & \,\, \tfrac{\prod_{j=1}^{m} \Gamma ( b_{j} - b_{h} )^{*} \prod_{j=1}^{n} \Gamma ( 1 + b_{h} - a_{j} ) }{\prod_{j=m+1}^{q} \Gamma ( 1 + b_{h} - b_{j} ) \prod_{j = n + 1}^{p} \Gamma ( a_{j} - b_{h} )} \, z^{b_{h}} 
\, 
{}_{p} F_{q-1} \left(
1 + b_{h} - {\bf a}_{p} ;
( 1 + b_{h} - {\bf b}_{q} )^{*}
;
(-1)^{p - m -n} z \right).
\end{split}
\end{align}
Also, there is a curious fact related to this equation: although the expressions~\eq{rG}--\eq{cG} have $\left( {\mu_s r}/{2N_s} \right)^{2N_s}$  as argument, this does not mean that they are even functions of $r$. Indeed, the right-hand side of~\eq{slater} shows that the parity of the $G$-function is 
defined by ${\bf b}_{q}$ through the term $z^{b_{h}}$, because in our case ${}_p F_q$ is always an even function. Thus, writing these coefficients in Eqs.~\eq{rG}--\eq{cG} in the form
\begin{equation}
{\bf b}_{q} = \left( \frac{\tilde{b}_1, \ldots, \tilde{b}_{m}}{2N_s}; \frac{\tilde{b}_{m+1}, \ldots, \tilde{b}_{q}}{2N_s} \right)
\end{equation}
one can see that 
\begin{equation}
z^{b_h} = \left(\frac{\mu_s r}{2N_s}\right)^{\tilde{b}_h}
, 
\qquad \text{where} \qquad
1 \leqslant h \leqslant m.
\end{equation}
Therefore, if $(\tilde{b}_1, \ldots, \tilde{b}_{m})$ is a sequence of even (odd) numbers, the corresponding object is an even (odd) function of~$r$. Hence, the effective source and potential are even functions, whereas the mass function is odd---as it should be for a theory in which the form factor $f_s(-k^2)$ grows faster than any polynomial~\cite{Nos6der,BreTibLiv}.


\section{Analysis of the convergence of the series~\eq{I-series}}
\label{ApC}

Here we discuss the convergence of~\eq{I-series}. Let us define
\begin{equation}
a_\ell = \frac{(-1)^\ell}{(2\ell+1)!}  \Gamma \left( \frac{2\ell +3}{2N} \right) 
\end{equation}
for all $\ell \in \mathbb{N} \cup \left\{ 0 \right\}$. This can be rewritten as
\begin{equation}
a_\ell = \frac{(-1)^{\ell}}{(2\ell+1)!} \frac{2N}{2\ell+3}  \Gamma \left( 1 + \frac{2\ell + 3}{2N}  \right)
,
\end{equation}
such that
\begin{equation}
\n{aquela}
\left| \frac{a_{\ell+1}}{a_\ell} \right| = \frac{1}{(2\ell+2)(2\ell+5)} \frac{\Gamma \left( 1 + \frac{2\ell+5}{2N} \right)}{\Gamma \left( 1 + \frac{2\ell+3}{2N} \right)} 
.      
\end{equation}

According to Stirling's formula for the gamma function, for sufficiently large $z$,
\begin{equation}
\Gamma ( 1 + z ) \sim \sqrt{2 \pi z} \biggl( \frac{z}{e} \biggr)^{z}
.
\end{equation}
Therefore,
\begin{equation}
\n{limittest}
\lim_{\ell \to \infty}\left| \frac{a_{\ell+1}\, r^{2(\ell+1)}}{a_\ell \,r^{2\ell}} \right| \sim \frac{r^2}{4 N^{\frac{1}{N}}} \,  \ell^{\frac{1}{N} - 2}.
\end{equation}
Hence, 
it follows that the series is absolutely convergent on $\mathbb{R}$ if $N > 1/2$. Since any power series converges uniformly in every compact interval within the convergence radius, then~\eq{IN-series} is uniformly convergent on $[0, L]$ for all $L > 0$ and $N > 1/2$. On the other hand, if $N < 1/2$, the series converges only at $r = 0$. For the critical value $N = 1/2$, although the series converges for $\vert r\vert<1$, it can be summed analytically, and we know the relevant function in closed form [viz. Eq~\eq{source-nonC}],
\begin{equation}
I_{1/2} (r) =  \frac{2}{\left(1+  r^2\right)^2}
;
\end{equation}
see also Eq.~\eq{gef1/2} and the related discussion.


\section{Useful integral representations and proof of Theorem 2}
\label{ApD}

From Eqs.~\eq{I-GEF} and~\eq{GEF-int} we can construct alternative integral representations for the effective source, mass function, and Newtonian potential, namely, 
\begin{equation}
\n{rho-integral}
\rho_s (r)  = \frac{M}{2 \pi^2 r^3} \int_0^\infty \rd t \left[ 
\sin ( \mu_s r t^{\frac{1}{2N_s}}) - (\mu_s r t^{\frac{1}{2N_s}}) \cos ( \mu_s r t^{\frac{1}{2N_s}})
\right] e^{-t}
,
\end{equation}
\begin{equation}
\n{mass-integral}
M_s (r) = \frac{2M}{\pi} \int_0^\infty \rd t  \left[\text{Si}\,  (\mu_s r t^{\frac{1}{2N_s}}) - \sin (\mu_s r t^{\frac{1}{2N_s}}) \right] e^{-t}
,
\end{equation}
\begin{equation}
\n{chi-integral}
\chi_s (r) = - \frac{2GM}{\pi r} \int_0^\infty \rd t \,  {\rm Si} \, (\mu_s r t^{\frac{1}{2N_s}}) \, e^{-t}
,
\end{equation}
where $\text{Si}(z)$ is the sine integral function~\eq{Si(z)}. The above representations hold for any real number $N_s > 0$. They are also very useful for proving some general properties, e.g., the positivity of the mass function, and for performing numerical calculations. Notably, the limit $N_s \to \infty$ can be taken directly, resulting in~\eq{sour-inf},~\eq{mass-inf}, and~\eq{chi_in}. This fact makes the numerical computation of~\eq{rho-integral}--\eq{chi-integral} very stable and efficient even for high values of $N_s$ --- which allowed us to extrapolate some considerations of Sec.~\ref{Sec4} up to $N_s = 1000$.

With the integral representation~\eq{mass-integral} it is straightforward to prove Theorem 2 in Sec.~\ref{Sec4}. In fact, it suffices to show that the integrand of Eq.~\eq{mass-integral} is positive, regardless of $N_s$, i.e.,
\begin{equation}
\n{d4}
g(x) \equiv 
\text{Si} (x) - \sin (x) > 0, 
\quad 
x>0
.
\end{equation}
Let us split the analysis into the domains $0<x<\pi$ and $x>\pi$. In the first case, note that $g(0)=0$ and $g'(x) = \tfrac{ \sin x}{x} - \cos x > 0$ for $x\in (0,\pi)$; therefore $g(x)>0$ in this interval. For $x>\pi$ we use that 
\begin{equation}
\n{d5}
x>0 
\quad \Longrightarrow \quad
\frac{\pi}{2} - \frac{1}{x} \leqslant \text{Si}  (x)  \leqslant \frac{\pi}{2} + \frac{1}{x}
,
\end{equation}
to obtain
\begin{equation}
g(x) \geqslant \frac{\pi}{2} - \frac{1}{x} - 1 \geqslant \frac{\pi}{2} - \frac{1}{\pi} - 1 > 0
.
\end{equation}
This proves~\eqref{d4} and the positivity of the effective mass function.


\section{Oscillations of the mass function}
\label{ApE}

Another useful application of the integral representation~\eq{mass-integral} for the mass function is to estimate the point $r_* \gg 1/\mu_s $, where the oscillations are suppressed and beyond which $M_s(r) \approx M$  (see the discussion in Sec.~\ref{Sec4}). To this end, let us study how the mass function oscillates for large values of $r$ and its dependence on $N_s$. First, notice that for $\mu_s r \gg 1$ the integral of the first term in~\eq{mass-integral} approaches $\pi/2$,
\begin{equation}
\n{MD1}
\int_0^\infty \rd t \, \text{Si} ( \mu_s r t^{ \frac{1}{2 N_s} } ) \, e^{-t}  \, \underset{\mu_s r \gg 1}{\sim} \, \int_0^\infty \rd t \, \frac{\pi}{2} \,   e^{-t}  = 
\frac{\pi}{2}, 
\end{equation}
as it can be shown using inequality~\eq{d5}. Since the mass function oscillates only for $N_s > 1$, for the second term in the integral~\eq{mass-integral}, we make the expansion
\begin{equation}
\n{MD12}
t^{\frac{1}{2N_s}} = e^{\frac{1}{2N_s}\log t} = 1 + \frac{\log t}{2N_s} + O(N_s^{-2}) 
,
\end{equation}
so that, for $N_s\gg 1$,
\begin{equation}
\n{MD2}
\int_0^\infty \rd t \, \sin (\mu_s r t^{\frac{1}{2N_s}})  e^{-t} \approx \int_0^\infty \rd t \, \sin \left( \mu_s r + \frac{\mu_s r}{2N_s} \log t \right) e^{-t}  
.
\end{equation}
The above integral has the general form 
\begin{equation}
\n{intDD}
\int_0^\infty  \rd z \, \sin ( \al + \beta \log z ) e^{-z} = -\Im [ e^{-i \al} \Ga ( 1 - i \beta ) ] 
,
\qquad \qquad 
\al, \, \be > 0
.
\end{equation}
Thus, writing the complex gamma function as 
$ 
\Ga (1 - i \beta) = a + ib, 
$
we have
\begin{equation}
\n{MD3}
\int_0^\infty  \rd z \, \sin ( \al + \beta \log z ) e^{-z} = a \sin \al - b \cos \al 
.
\end{equation}
Therefore, Eqs.~\eq{mass-integral},~\eq{MD1},~\eq{MD2} and~\eq{MD3} yield the approximation $M_s(r) \approx {M}_s^{\rm ap}(r)$, where
\begin{equation}
\n{pacote}
\frac{{M}^{\rm ap}_s(r)}{M} \equiv
1 - \frac{2}{\pi} \left\{ \operatorname{Re} \left[ \Gamma \left( 1 - i \frac{\mu_s r}{2 N_s} \right) \right] \sin(\mu_s r)  - \operatorname{Im} \left[ \Gamma \left( 1 - i \frac{\mu_s r}{2 N_s} \right) \right] \cos(\mu_s r)  \right\} 
,
\end{equation}
which gives the form in which the mass function oscillates for $\mu_s r \gg 1$ and $N_s \gg 1$. The comparison of~\eq{pacote} with the exact result for $M_s(r)$ shows that the approximation seems to be reasonable even for relatively small values of $N_s$ and $\mu_s r$, see Fig.~\ref{Fig21}. The reason for that is twofold: First, the sine integral~\eq{MD1} rapidly converges towards its asymptotic value of $\pi/2$ and, secondly, the term~\eq{MD12} quickly tends to 1 as the parameter $N_s$ increases.

\begin{figure}[t]
\centering
\includegraphics[width=7cm]{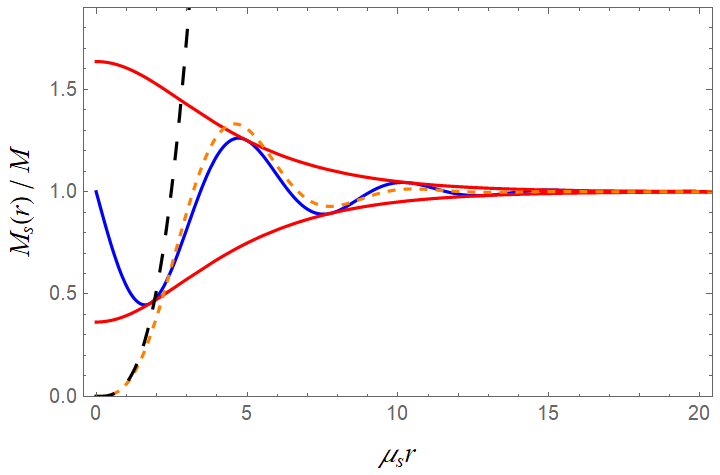}
\includegraphics[width=7cm]{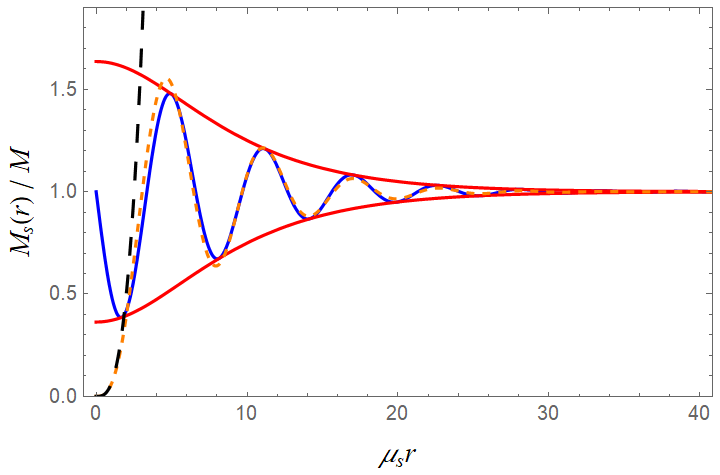}
\includegraphics[width=7cm]{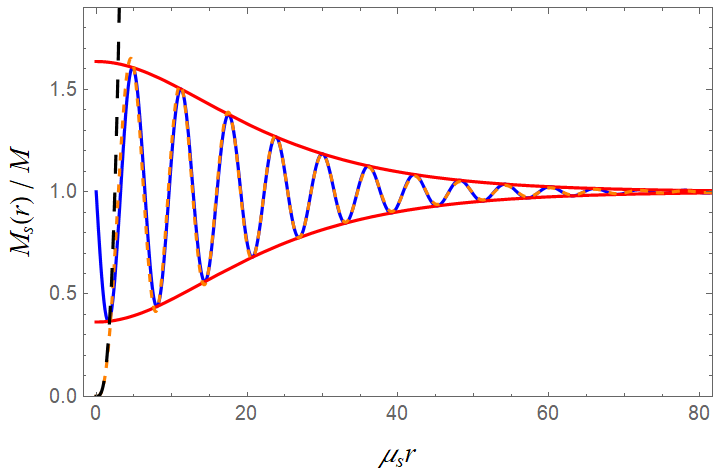}
\includegraphics[width=7cm]{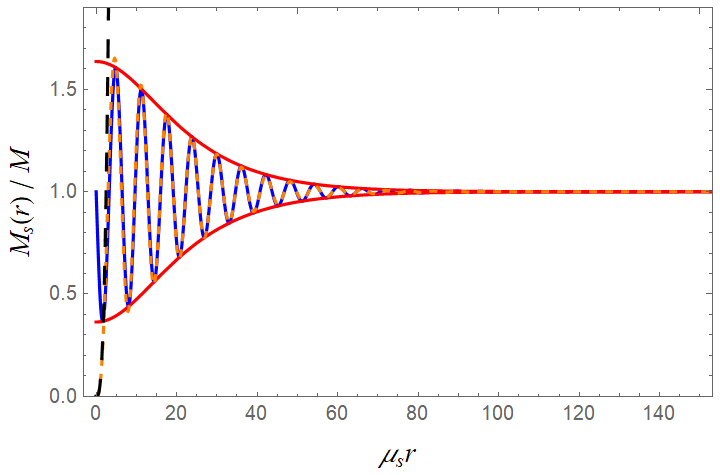}
\caption{Comparison between the exact mass function \eqref{mass-integral} (orange dashed line) and the approximation \eqref{pacote} (solid blue line) for $N_s=2$ (top left), $N_s=4$ (top right), $N_s = 10$ (bottom left), and $N_s = 20$ (bottom right). The red curves represent the wave package enveloping functions $1 \pm \delta$, where $\de$ is given by~\eq{envFunc}. The black dashed line is the $r \to 0$ asymptotic form of the mass function, Eq.~\eq{mass-rto0}.}
\label{Fig21}
\end{figure}

The next step is to find the function that describes the attenuation of the oscillations of Eq.~\eq{pacote}. To this end, we use~\eq{intDD} again, but now we write the complex gamma function in the polar form
$
\Gamma ( 1 - i b ) = \rho \, e^{i \theta}
,\,
$
so that
\begin{equation}
\int_0^\infty  \rd z \, \sin ( \al + \beta \log z ) e^{-z} = \rho \sin (\al - \th)
.
\end{equation}
Since the sine function is limited, we have
\begin{equation}
- \rho  \leqslant \int_{0}^{\infty} \rd z \, \sin ( \al + \be \log z ) e^{-z} \leqslant \rho
,
\end{equation}
which togheter with~\eq{MD1} results in
\begin{equation}
\n{mass-rel-error}
1 - \de  \leqslant  \frac{{M}^{\rm ap}_s (r)}{M} \leqslant 1 + \de
,
\end{equation}
where 
\begin{equation}
\de 
= \frac{\pi}{2} \, \left\vert \Ga \left( 1 - i \frac{\mu_s r}{ 2N_s} \right) \right\vert.
\end{equation} 
The oscillation's enveloping function $1\pm\de(\mu_s r,N_s)$ can be cast in a more useful form using
\begin{equation}
\vert \Gamma(1 \pm i \beta) \vert^2 = \frac{\pi \beta}{\sinh(\pi \beta)}
,
\end{equation}
whence
\begin{equation}
\n{envFunc}
\delta = \sqrt{\frac{2 }{\pi } \frac{\mu_s r}{N_s} \operatorname{csch} \left( \frac{\pi}{2} \frac{\mu_s r}{N_s} \right)}
. 
\end{equation}
Note that $\de\to 0$ as  $\mu_sr\to\infty$. In Fig.~\ref{Fig21} we compare the numerical integration of~\eq{mass-integral} with~\eq{pacote} and~\eq{mass-rel-error} for several values of $N_s$; note that, although technically derived for large values $N_s$ and $\mu_sr$, the approximation works relatively well even when those values are not so large.

\begin{figure}[t]
\centering
\includegraphics[width=7cm]{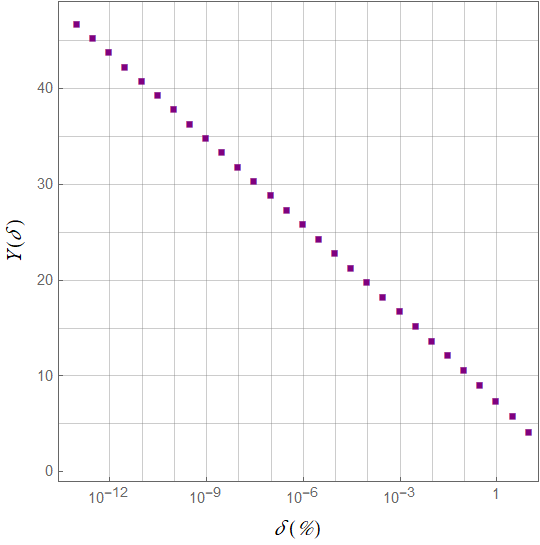}
\caption{Numerical solution for the multiplicative factor $Y$ as a function of the relative error $\delta = |1 - M_s(r)/M|$ (in percentage) in a log-linear scale.} 
\label{Fig22}
\end{figure}

Given an arbitrarily small $\de$, the equation \eqref{envFunc} can be inverted to give the value of $r_*$ such that $r>r_*$ implies $\vert M_s(r)/M\vert<1+\delta$, formally,
\begin{equation}
\frac{\mu_s r_*}{N_s} = Y (\delta)
.
\end{equation}
Since \eqref{envFunc} depends on the combination $\mu_sr/N_s$, it is clear that $\mu_s r_*$ scales linearly with $N_s$, as it was assumed in~\eq{r-estrela}. In Fig.~\ref{Fig22}, we plot the function $Y(\delta)$ obtained by numerically solving the transcendental equation~\eq{envFunc}. As the graph reveals, the $\de$-dependence of $Y$ in a semi-log scale is almost linear. Therefore, it is possible to find an approximation for $Y(\de)$ using a linear regression. To obtain more accurate results, let us divide the values of $Y$ into two domains, namely, $6 \leqslant Y \leqslant 10$ and $10<Y \leqslant 50$. Then, using the numerical data shown in Fig.~\ref{Fig22}, the last squares yield
\begin{equation}
\n{Y_ap_peri}
Y(\de) \approx
\begin{cases}
(-1.385 \pm 0.005) \log (100 \de) + (7.272 \pm 0.006 ) , &\text{for } \quad 6 \leqslant Y \leqslant 10
,
\\ 
(-1.314 \pm 0.002) \log (100 \de) + (7.41 \pm 0.04 ),  & \text{for }\quad  10<Y \leqslant50
.
\end{cases}
\end{equation}
These relations can be used to estimate the point beyond which the nonlocal corrections to the mass function are suppressed, as in Sec.~\ref{Sec4}. For example, for $M_s(r) \approx M$ within $1\%$ it is necessary that $r>r_* \approx 7.3 \,{N_s}/{\mu_s}, \,$ while for a deviation of $0.0001 \%$ we need $r>r_* \approx 20\, {N_s}/{\mu_s}.$



\end{document}